\documentclass[11pt]{article}

\usepackage[margin=1in]{geometry}      
\usepackage{amsmath, amssymb, amsthm}  
\usepackage{mathtools}
\usepackage{bm}                        
\usepackage{graphicx}                  
\usepackage{booktabs}                  
\usepackage{natbib}                    
\usepackage{hyperref}                  
\usepackage{authblk}                   
\usepackage{setspace}                  
\usepackage{enumerate}
\usepackage{algorithm}
\usepackage{algpseudocode}             
\usepackage{float}
\usepackage{subcaption}
\usepackage{multirow}
\usepackage{makecell}
\usepackage{threeparttable}

\hypersetup{
    colorlinks = true,
    urlcolor   = blue,
    citecolor  = blue,
    linkcolor  = black
}

\setcitestyle{authoryear,open={(},close={)}}

\theoremstyle{definition}

\theoremstyle{remark}

\title{ \textbf{Predictor-Informed Bayesian Nonparametric Clustering} }

\author[1]{ Md Yasin Ali Parh}
\author[2]{Jeremy T. Gaskins}
\affil[1]{Department of Biostatistics, University of Michigan, Ann Arbor  \\
\texttt{yasinp@umich.edu}}
\affil[2]{Department of Bioinformatics \& Biostatistics, University of Louisville \\
\texttt{jeremy.gaskins@louisville.edu}}
\date{}

\begin{document}
\maketitle

\begin{abstract}
In this project we are interested in performing clustering of observations such that the cluster membership is influenced by a set of predictors. To that end, we employ the Bayesian nonparameteric Common Atoms Model \citep[CAM;][]{denti2023common}, which is a nested clustering algorithm that utilizes a (fixed) group membership for each observation to encourage more similar clustering of members of the same group. CAM operates by assuming
each group has its own vector of cluster probabilities, which are themselves clustered to allow similar clustering for some groups. We extend this approach by treating the group membership as an unknown latent variable determined as a flexible nonparametric form of the covariate vector. Consequently, observations with similar predictor values will be in the same latent group and are more likely to be clustered together than observations with disparate predictors. We propose a pyramid group model that flexibly partitions the predictor space into these latent group memberships. This pyramid model operates  similarly to a Bayesian regression tree process except that it uses the same splitting rule for at all nodes at the same tree depth which facilitates improved mixing. We outline  a block Gibbs sampler to perform posterior inference from our model. Our methodology is demonstrated in simulation and real data examples. In the real data application, we utilize the RAND Health and Retirement Study to cluster and predict patient outcomes in terms of the number of overnight hospital stays. 
\end{abstract}


\section{Introduction}

Model-based clustering is a collection of methods designed to partition the data into distinct clusters, 
such that cluster members share common features.
This approach typically models the data as a mixture of components, with each cluster associated with a specific (typically, parametric) probability distribution.  
One common example is the Gaussian mixture model \citep{fraley2002model, mclachlan2000finite}.
Traditional approaches to clustering require specifying the number of clusters in advance, which can be restrictive for complex 
data structures. 
The Dirichlet Process (DP) prior \citep{ferguson1973bayesian, ferguson1974prior}
addresses these challenges and is 
the most widely used Bayesian nonparametric model. 
Because draws from a DP are almost surely discrete,
a sequence of values sampled from 
such a  realization has a positive probability of repeating values. 
The pattern of the repeated values induces a partition of the observations into clusters leading to the use of Bayesian nonparametric (BNP) methodology for model-based clustering.
Several studies have demonstrated the application of DP priors to clustering problems
across various field including epidemiology \citep{dunson2005bayesian}, genetics \citep{dunson2007abayesian}, obstetrics \citep{gaskins2023bayesian}, medicine \citep{kottas2002nonparametric}, and econometrics \citep{chib2002semiparametric, hirano2002semiparametric}.

These studies cluster exchangeable samples from a single unknown distribution based only on responses, ignoring any information contained in covariates. However, cluster membership may depend on predictors, such as known groups or covariates, where observations with similar covariate values are believed to be more likely to belong to the same cluster.
Clustering across predefined groups can be modeled using 
Dependent Dirichlet Processes \citep[DDP;][]{maceachern1999dependent, maceachern2000dependent}, an extension of the DP that defines collections of dependent random probability measures, such that the marginal prior of each is itself a DP.
The Hierarchical Dirichlet Process \citep[HDP;][]{teh2004sharing}, a variant of the DDP, partitions data into known groups and forms clusters within each group to capture group-specific latent structures, while allowing clusters to be shared across groups. However, it does not cluster the groups themselves.
The Nested Dirichlet Process \citep[NDP;][]{rodriguez2008nested} facilitates two-layer clustering in this way. First, it clusters similar groups at the distributional level cluster (DC), and then, within each DC, it clusters similar observations at the observational level cluster (OC). 
The NDP does not allow the sharing of OCs across different DCs.  That is, if two groups share the atom $\vartheta_h$ defining an observational cluster, then the groups must be members of the same distributional cluster and would share all observational clusters.  \cite{camerlenghi2019latent} define  this unappealing feature as degeneracy:  if two groups share only a single atom, they cluster together. 
Most recently, the nested Common Atoms Model \citep[CAM;][]{denti2023common} enabled observational level clustering across DCs by the sharing of atoms across the DCs, avoiding the degeneracy property of NDP. A graphical illustration depicting the behavior of HDP, NDP, and CAM is presented in Figure \ref{fig:hdpndpcam} of the Supplementary Materials.

In addition to these studies which consider clustering  that is informed by a known grouping, some studies utilize a general set of predictor variables to determine cluster membership. One approach in this domain is the Product Partition Model with covariates \citep[PPMx;][]{muller2011product}, a model-based clustering algorithm that leverages covariates to compute a weighted average of responses. These weights are based on the similarity of the new patient's covariates to those of existing patients in the cluster. PPMx extends the Product Partition Model \citep[PPM;][]{hartigan1990partition, barry1993bayesian, crowley1997product} by enhancing the cohesion function to increase the likelihood of clustering units with similar covariates.
\cite{rodriguez2011nonparametric}  introduced a novel class of Bayesian nonparametric priors known as Probit Stick-Breaking Process (PSBP), which utilizes stick-breaking constructions with weights derived from probit transformations of normal random variables. This method enables flexible nonparametric regression models that can handle spatial data, time series, and random effect models using predictors.
Additionally, the Ewens-Pitman Attraction (EPA) method, introduced by \cite{dahl2017random}, utilizes a pairwise similarity function among the observations to form clusters.

However, these clustering approaches which use covariates  are not well-suited for data with high dimensionality. 
In particular, the EPA and spatial PSBP approaches use a univariate distance measure to represent the similarity between the covariates of the observations.  While this avoids dimensionality issues, standard implementation will not  learn which predictors are associated with the observational clustering, and if many or most predictors are unrelated to the differences across clusters, the distance measure will be overly determined by the irrelevant features.  
PPMx modifies the cohesion function to account for similarity in the covariate space, but this requires specification of a probability model for the predictor space.  Further, in high dimensional settings, the clustering of the predictors through the cohesion may overwhelm the influence of the outcome, yielding a clustering that provides no information about the response.
While 
\cite{muller2011product} 
proposes variable selection that may mitigate some of these, the CRAN package does not implement this.
Other nested models such as HDP, NDP, and CAM require a fixed group structure to be known, but in most realistic settings, group memberships that stratify the predictor space cannot be specified a priori.
Hence, new methodology for this context is required.


In this project, we consider a clustering approach, similar to the CAM, where a group nesting structure is informed by the set of predictors instead of a known group. To that end, we consider all observations to be partitioned according to an unknown latent group determined by a set of covariates, such that individuals with similar covariates are assigned to the same groups. To determine the latent group structures, we propose a pyramid group model (PGM) that flexibly partitions the predictor space into these latent group memberships. This model is similar to the Bayesian regression tree process \citep{chipman1998bayesian}, except that it uses the same splitting rule for all nodes of the same tree depth. To achieve the clustering, we extended CAM by considering group membership as an unknown latent variable corresponding to the terminal node of the pyramid tree. By combining our PGM with CAM, we cluster outcomes in a nested structure and refer to this model as the Common Atoms Pyramid Group Model (CAPGM). For our proposed model, we introduce a block Gibbs sampler to perform posterior inference.

The remainder of the article is structured as follows. In Section 2, we recall the Dirichlet process and the Common Atoms Model. Section 3 presents our CAPGM model along with its properties. Computational strategies for performing posterior inference are discussed in Section 4. In Section 5, we discuss a simulation study to evaluate the performance of our model, including comparisons with competing methods and estimation accuracy metrics. In Section 6, we apply our model and other competing approaches to the health retirement study (HRS) data. Finally, we conclude with a brief discussion in Section 7.  This manuscript also includes  Supplementary Materials with proofs and additional computational details.

\section{Dirichlet Process and Common Atoms Model}

\subsection{Dirichlet Process}
\label{subsec:DP}
We first recall a few key properties of the DP that are relevant to our approach \citep[for brief review, see][]{li2019tutorial}. 
A random distribution $F(\cdot)$ drawn from a DP is a distribution on the measurable space $(\boldsymbol{\Theta}, \mathcal{F})$ and  is denoted by $F\sim DP (\gamma, \mathcal{G}_0)$, where $\gamma\in \mathbb{R}^+$ is a concentration parameter and the base distribution $\mathcal{G}_0$ is also a probability measure on $(\boldsymbol{\Theta}, \mathcal{F})$. 
Additionally, $F$
is almost surely discrete and puts all probability mass on a countably infinite collections of atoms.
That is,
$
F(\cdot) = \sum_{h = 1}^\infty \pi_h \delta_{\vartheta_h}(\cdot),
$
where $\pi_h$ is the probability associated with atom $\vartheta_h$. \cite{sethuraman1994constructive} provided a constructive definition of the DP through the stick-breaking (SB) representation to describe the probability model for the weights $\pi_1, \pi_2, \ldots$. In this approach a stick with unit-length is divided into an infinite number of segments. The length of the $h^{th}$ segment (weight) is given by
$\pi_h = q_h \prod_{s<h} (1-q_s), $
($h = 1,2, \ldots, H, \ldots$), and $q_h \sim Beta(1, \gamma)$. 
To denote that the probability weights $\boldsymbol{\pi} = \{\pi_1, \ldots, \pi_H, \ldots\}$ come from this stick-breaking model, we write $\boldsymbol{\pi} \sim SB(\gamma)$.

In practice, when using the DP as a part of mixture model, we
have 
$y_i \sim f_y(\cdot\mid\theta_i, \phi)$  ($i = 1, \ldots, n$),
where $f_y(\cdot\mid\theta_i, \phi)$ denotes the distribution of the response $y_i$ given the observation-specific parameter $\theta_i$ and global parameter $\phi$. The $\theta_i$ is drawn from a realization $F(\cdot)$ of the DP. 
To that end, we have
\begin{equation}
  \theta_i \sim F(\cdot) = \sum_{h = 1}^\infty \pi_h \delta_{\vartheta_h}(\cdot).  
  \label{eq:dp_theta_i}
\end{equation}

\noindent
In this formulation, $\theta_i$ takes the value $\vartheta_h$ with probability weight $\pi_h$. Consequently, the samples $\theta_1, \theta_2, \ldots,  \theta_n$ have a positive probability of ties, leading to a clustering or partition of the samples. To identify the cluster memberships of the observations, we define $C_i \in \{1, \ldots, H,\ldots\}$ as the cluster membership indicator for observation $i$, where all observations assigned to cluster $h$ share the common parameter $\vartheta_h$ as the value for their respective $\theta_i$s.

\subsection{Common Atoms Model}
\label{subsec:CAM}
The Common Atoms Model, a variant of the dependent Dirichlet process, is a flexible Bayesian nonparametric clustering method designed for data with a known nested structure \citep{denti2023common}. Unlike the DP, which considers a single distribution for all observations, CAM involves a collection of $G$ related distributions $F_1, F_2, \ldots, F_G $ all on the common space $(\boldsymbol{\Theta}, \mathcal{F})$. CAM addresses the degeneracy issue of NDP by sharing a common set of atoms across all of the group-specific distributions, thereby allowing observations from any group to be clustered together. Within this framework, distributions of the groups that belong to the same distributional cluster (DC) possess the same set of weights for the shared atoms and consequently have the exact same $F_g(\cdot)$ for all groups  in the DC. The distinction between different DCs lies in the different probability weights associated with the common set of atoms.

In this context, let the $i^{th}$ observation be a member of the known and fixed group $g_i \in \{1, \ldots, G\}$. 
Instead of equation (\ref{eq:dp_theta_i}), the distribution for $\theta_i$
is $F_{g_i}(\cdot)$:
\begin{equation}
\theta_{i} \sim F_{g_i}(\cdot) = \sum_{h = 1}^\infty \pi_{g_ih}\delta_{\vartheta_h}(\cdot),
\label{eq:thetagi}
\end{equation}
\noindent
where $\vartheta_h$ ($h = 1, 2, \ldots,H, \ldots$) are the common atoms drawn from $\mathcal{G}_0$, and $\boldsymbol{\pi}_g = \{\pi_{g1}, \ldots, \pi_{gH}, \ldots\}$ is the weight (probability) vector for  group $g$.

If two groups $g$ and $g^\prime$ share the same set of probability weights ($\boldsymbol{\pi}_g = \boldsymbol{\pi}_{g^\prime}$), they are clustered together. That is, they fall in the same distribution cluster. To induce clustering among the groups, the 
vectors $\boldsymbol{\pi}_g$ are drawn from another DP with concentration parameter $\alpha$ and the stick-breaking process $SB(\beta)$ as the base distribution. Consequently, the model is
\begin{equation}
\begin{split}
\boldsymbol{\pi}_g = \{\pi_{g1}, \ldots, \pi_{gH}, \ldots\} & \sim F(\cdot) = \sum_{k=1}^\infty \rho_k \delta_{\boldsymbol{\nu}_k} (\cdot) \sim DP(\alpha, SB(\beta)).
\end{split}
\label{eq:groupprob}    
\end{equation}
\noindent
Here, the DC weights $\boldsymbol{\rho} = \{\rho_1, \ldots, \rho_K,\ldots\}$ come from  $SB(\alpha)$, and they determine the distribution cluster that group $g$ is associated with.  The observation-level clustering probabilities $\boldsymbol{\pi}_g$ are equal to some $\boldsymbol{ \nu}_k = (\nu_{k1}, \nu_{k2}, \ldots., \nu_{kH}, \ldots)$, each of which is drawn independently from $SB(\beta)$.  
If two groups $g$ and $g'$ both select DC $k$ (based on probability $\rho_k$), then they share the same probability vector $\boldsymbol{\nu}_k$, resulting in groups being clustered together, i.e., a distributional cluster.
The concentration parameters $\alpha$ and $\beta$ regulate the number of DCs and OCs, respectively. A larger $\alpha$  encourages the formation of more DCs, while a smaller $\alpha$ favors fewer DCs, as $\boldsymbol{\rho} = \{\rho_1, \ldots, \rho_k, \ldots\}$ will concentrate on few values. Similarly, a larger $\beta$ leads to the creation of more, smaller OCs, whereas a smaller $\beta$ tends to produce fewer, larger OCs.

\section{Common Atoms Pyramid Group Model}
\label{sec:CAPGM_Model}

\subsection{CAPGM Model Specification}

In CAM, clustering is informed by the outcome measurements $Y_i$ and a fixed, known group structure $g_i$, but in many real-world clustering scenarios, there are instead general predictor variables that should impact the clustering,  not an priori known group. To address this, we extend CAM to construct a nested clustering where the group membership is itself a latent variable depending on the set of predictors.
To that end, we propose the Pyramid Group Model (PGM), which flexibly partitions the predictor space to determine the latent groups, similar to the Bayesian classification regression tree model \citep[CART;][]{chipman1998bayesian}, except that PGM uses the same splitting rule at all nodes of the same tree depth.

Let $(Y_i, \boldsymbol{X}_i)$ represents the outcome and the associated predictors for the $i^{th}$ observation, and let $\boldsymbol{\mathcal{X}}$ denote the support for the predictors. In our model, we assume the data is nested according to an unknown group structure determined by the predictors $\boldsymbol{X}_i$. 
That is, we assume there exists some function $G(\cdot)$ that maps the predictor space $\boldsymbol{\mathcal{X}}$ into the space of group labels $\{1,\ldots,G\}$.  With a slight abuse of notation, we refer to $G$ as both the function that provides the group labels and the number of groups.
For individual $i$, $G(\boldsymbol{X}_i)=g_i\in \{1, \ldots, G\}$ provides the covariate group.

The Common Atoms Pyramid Group Model (CAPGM) has two components. The first is analogous to the CAM model described earlier, conditionally on the covariate group specification.
The clustering probabilities of the observations depends on the group through $\boldsymbol{\pi} (\boldsymbol{X}_i) = \boldsymbol{\pi}_{G(\boldsymbol{X}_i)}$, ensuring that observations with similar predictors cluster similarly. Analogous to equation (\ref{eq:thetagi}), the distribution of $\theta_i$  given $\boldsymbol{X}_i$ is represented as 
\begin{align}
  \theta_{i}\mid\boldsymbol{X}_i \sim F_{G(\boldsymbol{X}_i)}(\cdot) = \sum_{h = 1}^\infty \pi_{G(\boldsymbol{X}_i)h}\delta_{\vartheta_h}(\cdot).  
  \label{eq:theta_i|X}
\end{align}

The model defined in equations (\ref{eq:groupprob}) and (\ref{eq:theta_i|X}) can be represented equivalently in terms of latent categorical variables, $D_g$ and $C_i$. Here, $D_g \in \{1, \ldots, K, \ldots\}$ indicates which distributional-level cluster (DC) $k$ is selected for the latent group $g$, and $C_i$ indicates which atom $\vartheta_1, \ldots, \vartheta_h, \ldots$ is selected for observation $i$. Observations $i$ and $i^\prime$ are in the same OC if $C_i = C_{i^\prime}$ ($\theta_i = \theta_{i'} = \vartheta_h$ for some $h$), and observations are in the same predictor group if $G(\boldsymbol{X}_i)=G(\boldsymbol{X}_{i^\prime})$.  Groups $g$ and $g^\prime$ are in the same distribution cluster if $D_g=D_{g^\prime}$, and observations $i$ and $i^\prime$ are in the same DC if $D_{G(\boldsymbol{X}_i)}=D_{G(\boldsymbol{X}_{i^\prime})}$. Thus, $D_{g_i} = k$ and $C_i = h$ indicates that the $i^{th}$ observation is in latent predictor group $g_i = G(\boldsymbol{X}_i)$ and is assigned to $h^{th}$ OC ($\theta_i = \vartheta_h$) and $k^{th}$ DC ($\pi_{g_ih} = \nu_{kh}$, $h = 1, \ldots, H, \ldots$).

The model parameterization that includes these indicator variables is given by the following hierarchy: 
$$
y_i\mid\vartheta_1, \vartheta_2,\ldots, \phi, C_i \sim f_y(\cdot\mid\vartheta_{C_i},\phi),
$$
$$
    C_i\mid g_i=G(\boldsymbol{X}_i), D_g = k \sim \text{Multinomial} (\boldsymbol{\pi}_{G(\boldsymbol{X}_i)} = \boldsymbol{\nu}_k),
$$
 $$
 D_g\mid \boldsymbol{\rho} \sim \text{Multinomial} (\rho_1, \rho_2, \ldots, \rho_K, \ldots),
 $$
 $$
\vartheta_h \sim \mathcal{G}_0(\cdot), \ \ h = 1, \ldots, H, \ldots,
$$
$$
\boldsymbol{\nu}_k \sim SB(\beta) ,  \ \ k = 1, \ldots, K, \ldots,
$$
$$
\boldsymbol{\rho} \sim SB(\alpha).
$$
The hyperparameters $\alpha$ and $\beta$ in the SB distributions can  be assumed known, but it is more common to place conditionally conjugate gamma priors on each:
 $\alpha \sim$ Gamma $(a, b)$ and $\beta \sim$ Gamma $(c, d)$.
Typically, we consider $\boldsymbol{\nu}_k$ and $\boldsymbol{\rho}$ in terms of their SB components:
 $$\rho_k = U_k \prod_{l<k} \left(1-U_l\right), \quad U_k \sim Beta(1, \alpha),$$
 and
 $$\nu_{kh} = q_{kh} \prod_{s<h} (1-q_{ks}), \quad q_{kh} \sim Beta(1,\beta).$$
We use the finite truncation approximation to the infinite summations by setting 
set $U_K = 1$ for large $K$, and for each $k = 1, \ldots, K$, $q_{kH} = 1$ for large $H$ \citep{ishwaran2001gibbs, ishwaran2002dirichlet}. The strategy for selecting the upper bound $H$ and $K$ is discussed in Section \ref{sec:postcomput}.

\subsection{Pyramid Group Model}

\begin{figure}[t]
    \centering
    \includegraphics[width=\linewidth]{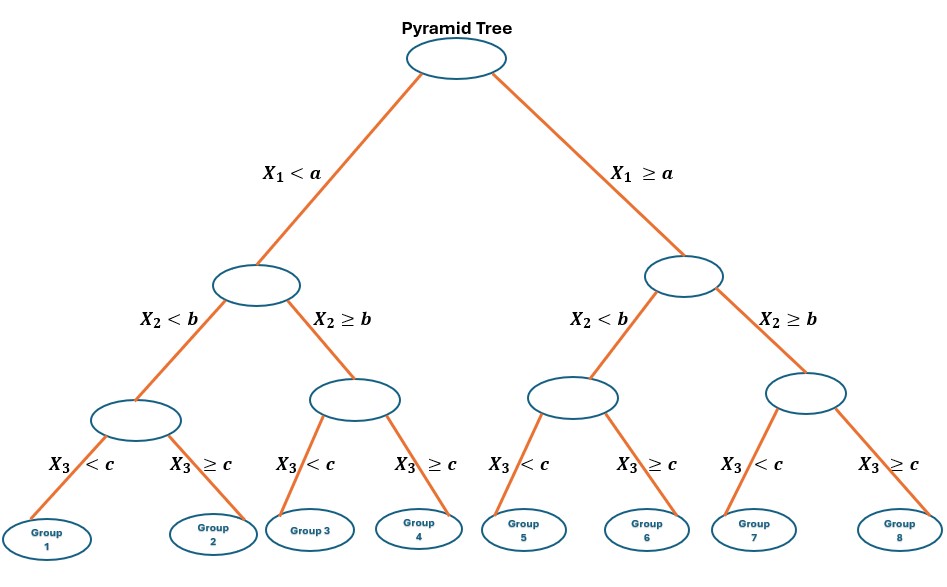}
    \caption{An example pyramid group tree with depth $d=3$.  The pyramid is based on the splitting rules $\{X_1<a\}$, $\{X_2<b\}$, and $\{X_3<c\}$.}
    \label{fig:A PyramidGroupStructure}
\end{figure}

Now we turn to our probability model for the function $G(\boldsymbol{X})$ that flexibly partitions the predictor space $\boldsymbol{\mathcal{X}}$ into groups $1, \ldots, G$.
As noted before, we will model the unknown $G(\boldsymbol{X})$ function using  a type of tree model, such that each group $g\in\{1,\ldots,G\}$ for the CAM is represented  as a terminal node in the tree.
Our model is similar in spirit to the Bayesian classification and regression tree \citep[CART;][]{chipman1998bayesian}, except that we only  consider trees $\mathcal{T}$ 
that use the same splitting rule  for all nodes of a given level.  
We call such choices as pyramid trees.
An example of one such pyramid tree is given in Figure~\ref{fig:A PyramidGroupStructure}. To encourage only moderate complexity to  $\mathcal{T}$ with a reasonable number of groups, we limit $\mathcal{T}$ to split up to a maximum depth $\mathcal{D}$, typically taken as $\mathcal{D} = 10$.

We index the level of the tree \( \mathcal{T} \) as \( \lambda = 1, \ldots, d \), and \( d = d(\mathcal{T}) \leq \mathcal{D} \) is the depth of the tree.
The splitting rule at level $\lambda$ depends on selecting the $j_\lambda\,^{th}$ predictor ($j_\lambda \in \{1, \ldots, P\}$), and a split value $\eta_\lambda$ such that all nodes at level $\lambda$ are split by  determining whether observations are assigned to the left $\{X_{j_\lambda} < \eta_\lambda\}$ or right $ \{X_{j_\lambda} \geq \eta_\lambda\}$.
The probability model for generating the tree $\mathcal{T}$ is governed by two functions: $P_{split} (\lambda)$ and $P_{rule}(j_\lambda, \eta_\lambda)$. 
The function $P_{split} (\lambda)$ is the probability of adding a new level $\lambda$ to  a tree $\mathcal{T}$ of depth $d = \lambda - 1$; that is,  all current terminal nodes will be split. The $P_{rule}(j_\lambda, \eta_\lambda)$ determines the splitting rule that is applied to all the previously terminal nodes.  
The stochastic process for drawing a pyramid tree from the prior can be described in the following recursive manner:

\begin{enumerate}
    \item Start with the zero depth tree $\mathcal{T}$, that is, $d=\lambda = 0$  and $G = 1$, as all the observations are in a single group.
    \item If the current tree has depth $d< \mathcal{D}$, then increase $\lambda$ by one to $d+1$.  We try to add a new level to the bottom of the current pyramid based on the probability 
    \begin{equation}
    P_{split} (\lambda) = a_{\mathcal{T}} \lambda^{-b_{\mathcal{T}}}.
    \label{eq:tree_split}
    \end{equation}
    For $a_{\mathcal{T}}, b_{\mathcal{T}} >0$,  $ P_{split} (\lambda)$ is a decreasing function of  $\lambda$. Large $b_{\mathcal{T}}$  makes deeper nodes less likely to split. \cite{chipman1998bayesian} discussed several combinations of $a_{\mathcal{T}}$ and $ b_{\mathcal{T}}$ for a reasonable number of terminal nodes. 
    \item If a success is obtained, we assign a splitting rule to level $\lambda$ according to the distribution $P_{rule}(j_\lambda, \eta_\lambda)$.
    To choose this rule, we determine the predictor for splitting and the splitting threshold value $\eta_\lambda$. We choose the predictor index $j_\lambda$ based on the discrete uniform distribution (with replacement), 
    \begin{equation}
    j_\lambda \sim \text{uniform}\{1, \ldots, P\},
    \label{eq:tree_var}
    \end{equation}
    and then choose the split value $\eta_\lambda$ uniformly between $Q_{q_1} (X_{j_\lambda})$ and $Q_{q_2} (X_{j_\lambda})$, the $q_1^{th}$ and $q_2^{th}$ quantiles of $X_{j_\lambda}$:
    \begin{equation}
      \eta_\lambda \sim \text{uniform}\left(Q_{q_1} (X_{j_\lambda}), Q_{q_2} (X_{j_\lambda})\right)  .
    \label{eq:tree_splitval}
    \end{equation}
       
While it is possible to sample $\eta_\lambda$ across the full support of $j_\lambda$ by using $q_1 = 0$ and $q_2 = 1$, we typically use ${q_1} =0.05\text{ and } {q_2} = 0.95$ to avoid choosing the split values too close to the edge of the support of $X_{j_\lambda}$. 
\label{step:step_3}
    \item Split all the previous terminal nodes from level $\lambda-1$ by adding branches to each with $X_{j_\lambda}>\eta_\lambda$ and $X_{j_\lambda}\leq \eta_\lambda$.
    Then $\mathcal{T}$ is  the newly created tree with depth $d = \lambda$. We again apply steps 2 and 3 until step $2$ fails or $\lambda = \mathcal{D}$.
    
\end{enumerate}

Each terminal node of the resulting pyramid tree represents a group in the CAM framework. That is, the tree $\mathcal{T}$ represents the $G(\boldsymbol{X})$ function that partitions $\boldsymbol{\mathcal{X}}$ into the groups/nodes $1,\ldots,G$.
 PGM places  individuals with predictors corresponding to the same node in $\mathcal{T}$ into the same latent group.
 From (\ref{eq:groupprob}), individuals belonging to this common predictor group will share the  probability vector $\boldsymbol{\pi}_g$ controlling the membership of the OCs. Predictor groups/nodes with same probability vectors $\boldsymbol{\pi}_g$ will cluster together in the same distribution cluster DC. Individuals assigned to any distributional level cluster $k = D_{G(\boldsymbol{X}_i)}$ have a probability $\nu_{kh}$ of being assigned to observational level cluster $h$, with all observations assigned to cluster $h$ sharing the common parameter $\vartheta_h$. A clear consequence of our CAPGM is that  individuals with similar predictor values have a greater opportunity to be clustered together than observations with desperate predictors.

The growing of the pyramid tree through the stochastic process accommodates continuous, binary, and ordinal variables. While the threshold rule is primarily motivated for continuous predictors, the approach works for binary and ordinal covariates reported a numeric coding. If a binary covariate is selected, it will split the zeros and ones into separate nodes using any $\eta_\lambda \in (0, 1)$. Similarly, for a predictor that is ordinal and numerically coded, the pyramid model will yield a splitting rule that separates a set of numerically lower values from the higher values. Nominal covariates in the model may also be easily incorporated by converting them to a corresponding set of binary dummy variables and allowing each dummy variable to be selected.

By creating a pyramid tree with a depth of $d$, we obtain $2^d$ potential terminal nodes.  We  note that in practice not all nodes are non-empty for the generated tree.  Sampling with replacement in \eqref{eq:tree_var} allows  a predictor to appear at multiple levels, which will necessarily enforce empty nodes.  For instance, consider predictor $j$ to be selected for the first two levels ($j=j_1=j_2$);  at level 2, there will only be three non-empty nodes made of observations with ${X_{j}<\eta_1}, \ {\eta_1 \leq X_{j} < \eta_2}, \ {\eta_2\le X_{j}}$ (assuming $\eta_1<\eta_2$). Additionally, even when we do not repeat a predictor, because we impose the same splitting rule at all nodes in a level, there is no guarantee that the prior nodes have members for each branch of the splitting rule.  However, this is not a concern for our method as such empty nodes do not impact the data likelihood.

Note that the standard Bayesian CART model 
builds a tree by splitting each node according to equation (\ref{eq:tree_split}) instead of splitting the entire level and then generating a splitting rule for the node similar to our step \ref{step:step_3}.  
There are a few benefits to utilizing our PGM structure that imposes the same splitting rule for all nodes of the same level when combining with CAM.  
Firstly, CART wishes to find a sparse tree such that each node has behavior that is unique from the other nodes.  This is not required in CAPGM since the distribution clusters can implicitly recombine nodes.  This means that including additional branches may not impact the behavior since nodes that are placed in the same DC will have the same probability vector $\boldsymbol{\nu}_k$ as if the parent node was unsplit.   For example, if groups 7 and 8 in Figure \ref{fig:A PyramidGroupStructure} were assigned to the same DC, this is the same as the CART tree that does not branch the $4^{th}$ node on level 3.  Similarly, if groups 5-8 were all in the same DC, then $X_1\ge a$ would be effectively serving as a terminal node.  
In fact, as long as $\mathcal{D}$ is large enough, any tree $\mathcal{T^*}$ that can be generated by CART can also be captured in CAPGM by the pyramid $\mathcal{T}$ that includes all splitting rules of any node in $\mathcal{T^*}$ with a set of the distribution clusters that recombine any splits in $\mathcal{T}$ that are not in $\mathcal{T}^*$.  In light of the ability to get equivalent structures from CAPGM and CART, we note that using the CAPGM facilitates the computations and developing the simulation algorithm for inference.  In particular, the levels of the pyramid tree are interchangeable, unlike in CART.

\subsection{CAPGM Properties}
Similar to the properties derived by \cite{denti2023common}, we  consider some theoretical results describing the behavior of the random measures and clusters generated from our CAPGM. We let $P(\boldsymbol{X}_i, \boldsymbol{X}_{i'})$ denote the probability that the two predictor vectors $\boldsymbol{X}_i$ and $\boldsymbol{X}_{i'}$ are assigned to the same terminal node in a randomly selected pyramid tree $\mathcal{T}$. As this will depend on the underlying probability model for the predictors, we do not fully characterize 
the properties of this function.
We can note that $1-a_T \leq P(\boldsymbol{X}_i, \boldsymbol{X}_{i'}) \leq 1$, and as $\boldsymbol{X}_i$ and $\boldsymbol{X}_i$ are more similar (or equal) $P(\boldsymbol{X}_i, \boldsymbol{X}_{i'})$ will approach (equal) $1$.

For observations $i$ and $i'$ with predictors $\boldsymbol{X}_i$ and $\boldsymbol{X}_{i'}$, we have the following properties:
\allowdisplaybreaks
\begin{align*}
    (\text{P-1}) \quad & P\left(F_{G(\boldsymbol{X}_i)}(\cdot) = F_{G(\boldsymbol{X}_{i'})}(\cdot)\right) = \frac{1+\alpha P(\boldsymbol{X}_i, \boldsymbol{X}_{i'})}{1+\alpha} \\
      (\text{P-2}) \quad & \text{Cov} \left(F_{G(\boldsymbol{X}_i)}(A), F_{G(\boldsymbol{X}_{i'})}(B)\right) =  \big[ \mathcal{G}_0 (A\cap B) - \mathcal{G}_0 (A)\mathcal{G}_0 (B)\big] \ \times \\
   & \quad \Bigg(
      \Big[\frac{\alpha}{1+\alpha}\frac{\beta}{1+\beta}\frac{1}{1+2\beta}\Big] P(\boldsymbol{X}_i, \boldsymbol{X}_{i'}) + \Big[\frac{1}{1+\alpha}\frac{1}{1+\beta} + \frac{\alpha}{1+\alpha}\frac{1}{1+2\beta}\Big] 
    \Bigg)  \\
    (\text{P-3}) \quad & \text{Corr} \left(F_{G(\boldsymbol{X}_i)}(A), F_{G(\boldsymbol{X}_{i'})}(A)\right) = 1-\left[1- P(\boldsymbol{X}_i, \boldsymbol{X}_{i'})\right]\frac{\alpha}{1+\alpha}\cdot\frac{\beta}{1+2\beta} \\
    (\text{P-4}) \quad &     P\left(C_i= C_{i'}\right) = \frac{1}{1+\alpha}\left[ \frac{1}{1+\beta}  +\alpha\frac{1}{1+2\beta}\right] + \left[\frac{\alpha}{1+\alpha} \frac{\beta}{1+\beta}\frac{1}{1+2\beta}\right]P(\boldsymbol{X}_i, \boldsymbol{X_{i'}}).
\end{align*}
Proofs for these properties are contained in Section \ref{S2_Proof_Equations} of the Supplementary Materials.

We can clearly see that as $P(\boldsymbol{X}_i, \boldsymbol{X_{i'}})\rightarrow 1$, we have $P\left(F_{G(\boldsymbol{X}_i)}(\cdot) = F_{G(\boldsymbol{X}_{i'})}(\cdot)\right) \rightarrow 1$ from (P-1).  This shows that as observations have more similar predictors, they are more likely to have the same random measure for their $\theta$s.  Similarly, from (P-4), we have $P\left(C_i= C_{i'}\right)$ approaching its maximum value of $\frac{1}{1+\beta}$ as $P(\boldsymbol{X}_i, \boldsymbol{X_{i'}})\rightarrow 1$. 
All of these results are related to their counterparts with the CAM assumption of fixed groups.  
Taking $P(\boldsymbol{X}_i, \boldsymbol{X}_{i'})=1$ for individuals in the same fixed group and $P(\boldsymbol{X}_i, \boldsymbol{X}_{i'})=0$ for individuals in different groups, our results (P-1) through (P-4) coincide with the properties discussed in  \cite{denti2023common}.

\section{Posterior Sampling and Inference}
\label{sec:postcomput}

\subsection{MCMC Sampling}

We  perform posterior inference by using Markov Chain Monte Carlo (MCMC) to generate a collection of samples from the posterior distribution of our proposed CAPGM. The MCMC algorithm consists of a collection of steps that update the CAM parameters and a Metropolis-Hastings (MH) update for the pyramid tree.  The CAM steps are similar to \cite{denti2023common}, except that we  use the finite truncation to the stick-breaking representation \citep{ishwaran2001gibbs, ishwaran2002dirichlet}.  The MH step for $\mathcal{T}$ is similar to the update for the Bayesian CART model \citep{chipman1998bayesian}.  Full details are included in the Supplementary Materials, Section \ref{S3_MCMC_alogorithm}, and we make some comments here.

It is important to note when updating the tree the number of nodes, as well as the clustering of the nodes, will change.  Hence, it is necessary to jointly update the distribution clusters $\textbf{D} = \{D_1, \dots, D_G\}$ with $\mathcal{T}$.  Conditionally on all other parameters, the joint distribution only depends on $n_{gh} = \sum_{i = 1}^n I(G(\boldsymbol{X}_i) = g,  C_i = h)$,  the number of observations assigned to observational level cluster $h$ from the group $g$ of the tree $\mathcal{T}$:
$$
L(\textbf{D}, \mathcal{T}\mid \cdots) = \prod_{g = 1}^G \left( \rho_{D_g}\prod_{h = 1}^H  \pi_{gh}^{n_{gh}}\right)  = \prod_{g = 1}^G \prod_{k = 1}^K \left( \rho_k \prod_{h = 1}^H  \nu_{kh}^{n_{gh}}\right)^{I(D_g = k)} .
$$
Since each $D_g$ is conditionally independent, we can marginalize $\mathbf{D}$ out of this expression to get:
\begin{align}
    L(\mathcal{T}\mid \cdots) = \sum_{\textbf{D}} L(\textbf{D}, \mathcal{T}\mid\cdots ) = \prod_{g = 1}^G \sum_{k=1}^K \left(\rho_k \prod_{h=1}^H \nu_{kh}^{n_{gh}}\right) .
    \label{eq:tree_lik}
\end{align}
Note that the summation in \eqref{eq:tree_lik} requires a finite $K$.  Hence, we need the finite truncation approach to approximating the stick-breaking weights in our model.  For consistency, we also assume an upper bound for $H$ and apply the finite truncation.

Based on the likelihood \eqref{eq:tree_lik} and the tree prior (see equation (\ref{eq:treeprior}) in the Supplementary Materials), we can perform MH sampling for $\mathcal{T}$. For the proposal distribution  we follow the basic approach of sampling the Bayesian CART model \citep{chipman1998bayesian} with slight modifications. Our proposal consists of four possible adjustments to move from the current $\mathcal{T}$ to a proposed $\mathcal{T}'$.
\begin{enumerate}
    \item GROW: Add a new level to $\mathcal{T}$.  Increase the depth $d$ to $d + 1$ and split all previous terminal nodes based on a uniformly selected predictor and a split value $\eta_j$ from \eqref{eq:tree_var} and \eqref{eq:tree_splitval}.
    \item PRUNE: Remove a level from $\mathcal{T}$. Decrease the depth $d$ to $d - 1$ by randomly selecting a level and removing it from the pyramid. $\mathcal{T}'$ is based on the remaining $d-1$ decision rules.
    \item RE-SPLIT: Change the split threshold value $\eta_\lambda$ for a retained predictor.  Randomly chose level $\lambda$, and redraw the split value for $j_\lambda$ from \eqref{eq:tree_splitval}.
    \item CHANGE VARIABLE: Change the variable and its corresponding split value used at one level $\lambda$ in $\mathcal{T}$. Randomly select a level, and draw a new predictor and split value from \eqref{eq:tree_var} and \eqref{eq:tree_splitval}.
\end{enumerate}

Note that if the current tree had depth $d=\mathcal{D}$ and the GROW step is attempted, the proposed $\mathcal{T}'$ will be rejected automatically. Similarly, if the current tree depth is $0$, the PRUNE, RE-SPLIT, and CHANGE VARIABLE steps  will reject automatically. This proposal distribution exhibits several desirable properties. 
Firstly, it ensures a reversible Markov chain, as each transition from 
$\mathcal {T}$ to $\mathcal{T}^\prime$ has a corresponding reverse transition from $\mathcal{T}^\prime$ to $\mathcal {T}$. Specifically, the GROW and PRUNE steps serve as counterparts to each other, while the CHANGE and  RE-SPLIT steps acts as their own counterparts. Movement from the pyramid tree $\mathcal{T}$ to any other valid pyramid tree $\mathcal{T}'$ is possible in a small number of proposals (less than $\mathcal{D}$).  Additionally, the transition densities are all easy to compute as they only involve uniform distributions.  We pre-specify the probabilities of proposing each step type through a vector $p_{MH}=(p_{grow},p_{prune},p_{re-sp},p_{ch var})$, which can be tuned to improve mixing.

Other details about the sampling algorithm can be found in the Supplementary Materials, Section \ref{S3_MCMC_alogorithm}.  We also provide R code to implement our MCMC sampler at the \href{https://github.com/YasinAliParh/CAPGM}{GitHub repository for CAPGM}.

\subsection{Posterior Inference} 
\label{subsec:PosteriorInferance}

As noted, our primary goal in this project is to understand the clustering of observations and how these are impacted by covariates.  However, deriving cluster-specific inferences within a mixture model framework is challenging due to the issues of label-switching and parameter identifiability \citep{stephens2000dealing, jasra2005markov}. These difficulties arise because clusters are exchangeable in MCMC samples, and the likelihood function is invariant to the relabeling of clusters, rendering direct averages across  MCMC samples meaningless. 
One common avenue is to recognize that posterior co-clustering probabilities are identifiable estimands.
Hence, we consider the posterior probabilities that two individuals $i$ and $i'$ are assigned to the same observational cluster $(C_i = C_{i'})$, that they are assigned to the same predictor group $G(\boldsymbol{X}_i) = G(\boldsymbol{X}_{i'})$, and that they assigned to the same distributional cluster ($D_{G(\boldsymbol{X}_i)} = D_{G(\boldsymbol{X}_{i'})}$). These are all  estimated from the MCMC samples by
\begin{align}
 \widehat{\text{Pr}} (C_i = C_{i'}) &= \frac{1}{M} \sum_{m = 1}^M I(C_i^m = C_{i'}^m),
 \label{eq:co-clustering_prob} \\
    \widehat{\text{Pr}}\left(G(\boldsymbol{X}_i) = G(\boldsymbol{X}_{i'})\right) &= \frac{1}{M} \sum_{m = 1}^M I \left(G^m(\boldsymbol{X}_i) = G^m(\boldsymbol{X}_{i'})\right),
    \label{eq:co-grouping_prob}\\
    \widehat{\text{Pr}}(D_{G(\boldsymbol{X}_i)} = D_{G(\boldsymbol{X}_{i'})}) &= \frac{1}{M} \sum_{m = 1}^M I\left(D_{G(\boldsymbol{X}_i)}^m = D_{G(\boldsymbol{X}_{i'})}^m \right),
    \label{eq:co-dist_prob}
\end{align}
 where $M$ is the number of posterior samples, and $C_i^m$, $G^m(\boldsymbol{X}_i)$ and $D_{G(\boldsymbol{X}_i)}^m$ are the sampled indicators for observation $i$  in  iteration $m$ ($m=1,\ldots,M$).

While these pairwise probabilities represent how likely observations are to end up in the same cluster, these values do not correspond to any particular partition of the sample. As we typically desire a point estimate of the cluster, we consider the decision-theoretic strategies proposed by \cite{dahl2006model} and \cite{wade2018bayesian}. In these strategies, an optimal partition is found by minimizing the expected posterior loss under a specified loss function. Following the Dahl recommendation, the optimal clustering $\hat{\boldsymbol {C}}$ is found by minimizing
\begin{align}
   L(\hat{\boldsymbol{C}}, \boldsymbol{C}) = \sum_{i = 1}^n \sum_{i' = 1}^n \left[ I(\hat{C}_i = \hat{C}_{i'}) - \widehat{\text{Pr}} (C_i = C_{i'}) \right]^2,
   \label{eq:dahl_loss_fun}
\end{align}
using the co-clustering matrix determined by  \eqref{eq:co-clustering_prob}.
While \cite{wade2018bayesian} propose a greedy search algorithm to minimize \eqref{eq:dahl_loss_fun}, it is common to only consider $\hat{\boldsymbol{C}}$ among the clusterings observed during the MCMC chain.
Similarly, a point estimate of the predictor groups $\hat{\boldsymbol{G}}$ is obtained by minimizing the corresponding loss with the matrix of probabilities from \eqref{eq:co-grouping_prob}, and a point estimate for the distribution clusters is obtained similarly.

By minimizing the predictor group posterior loss $L(\hat{\boldsymbol{G}}, \boldsymbol{G})$ over the group partitions sampled during MCMC, the optimal predictor group indicators $\hat{\boldsymbol{G}}$ correspond to the samples from one of the iterations.  Hence,  the sampled tree $\mathcal{T}^m$ from that iteration can also be considered as the posterior point estimate of the PGM to provide a single tree that can be used for interpretation.  In addition to this estimator $\hat{\mathcal{T}}$, we may also consider other features of the pyramid tree to provide inference that accounts for the uncertainty in $\mathcal{T}$.
To that end,  we calculate the posterior variable inclusion probability for each predictor, $Pr(X_j \in \mathcal{T})$, as the proportion of times predictor $j$ is included in the tree during MCMC.
Additionally, the most frequent pyramid trees across MCMC iterations, along with the associated covariate combinations, can be identified.
The depth of the tree, the number of non-empty terminal nodes, the number of observations in the most populous terminal node, are all identifiable quantities that can be used to characterize posterior properties of $\mathcal{T}$ and to assess MCMC mixing.

In addition to inference on the clusterings and the pyramid tree, we often are interested in making predictions or performing inference on the individual level.   
To that end, let $\psi(\theta,\phi)$ represent some quantity of interest, depending on cluster-specific parameter $\theta$ and the global parameter $\phi$ from the likelihood $f_y(\cdot\mid\theta,\phi).$  
For instance, if we are interested in the mean of $Y$, $\psi(\theta,\phi)=\int y\,f_y(y\mid\theta,\phi)\,dy$.
This estimand can be identifiably estimated for a predictor combination $\boldsymbol{X}$ by averaging over the posterior samples through 
\begin{align}
\widehat{\psi} (\boldsymbol{X}) &= \frac{1}{M} \sum_{m = 1}^M \hat{\psi}^m(\boldsymbol{X}) = \frac{1}{M} \sum_{m = 1}^M \left[\sum_{h=1}^H Pr(C_i=h\mid G^m(\boldsymbol{X}) )\,\psi(\vartheta_h^m,\phi^m)\right] \nonumber \\
&=\frac{1}{M} \sum_{m = 1}^M \left[\sum_{h=1}^H  \nu_{G(\boldsymbol{X})h}^m \, \psi(\vartheta_h^m,\phi^m)\right].
    \label{eq:fun_theta}
\end{align}
Note that the term in brackets in a weighted average of the estimand $\psi(\vartheta,\phi)$ evaluated at each of the $H$ clusters.  The weights are determined by probability vector associated with the distribution cluster of the node $G(\boldsymbol{X})$ from the pyramid tree $\mathcal{T}^m$.
Uncertainty can be characterized by considering the credible intervals based on the quantiles of the per-iteration estimates $\hat{\psi}^m(\boldsymbol{X}).$

\section{Simulation Study}

\subsection{Data Generation}

Here, we describe one of simulation studies used to investigate the operating characteristics of our methodology.  In this setting, we assume a complete correspondence between the predictors $\boldsymbol{X}$ and the observational cluster $C$, which is a stronger relationship than what is specified by our model.  In Section \ref{S5_Simulation_Study_II} of the Supplementary Materials, we consider a simulation where the OCs are not completely determined by the predictors.

In this simulation study, we considered a sample size of $n = 1000$ patients and $P= 20$ covariates. Each predictor was generated independently from a uniform distribution, $X_{ij} \sim \text{uniform} (-0.5, 0.5)$. 
The eight predictor groups depend on the first three predictors and are determined by all combinations of $X_{i1}\ge 0$, $X_{i2}\ge 0$, and $X_{i3}\ge 0$:
\begin{equation}
G(\boldsymbol{X}_i) = 1+I(X_{i1}\ge 0) + 2\,I(X_{i2}\ge 0) + 4\,I(X_{i3}\ge 0).    \label{eq:TruegroupX}
\end{equation}
These eight groups correspond to four distribution clusters through
\begin{equation}
\label{eq:TruegroupD}
    D_{G(\boldsymbol{X}_i)} = \begin{cases} 
    1& G(\boldsymbol{X}_i)=1,3,5,7;\\ 
    2& G(\boldsymbol{X}_i)=2,6;\\ 
    3& G(\boldsymbol{X}_i)=4;\\ 
    4& G(\boldsymbol{X}_i)=8 \end{cases} \quad =
    \begin{cases}
    1& X_{i1}<0;\\
    2& X_{i1}\ge 0, X_{i2}<0;\\
    3& X_{i1}\ge 0, X_{i2}\ge 0, X_{i3}<0;\\
    4& X_{i1}\ge 0, X_{i2}\ge 0, X_{i3}\ge 0.
    \end{cases}
\end{equation}
As noted previously, we have $D_{G(\boldsymbol{X}_i)} = C_i$ in this simulation.  That is, all members of distribution cluster 1 (those with negative $X_1$) are assigned to observational cluster 1 with $\vartheta_1.$ In terms of the probability vectors $\boldsymbol{\pi}_g$ and $\boldsymbol{\nu}_h$, we have $\nu_{kh}=I(k=h)$.

Conditional on $\boldsymbol{X}_i$ and $D_{G(\boldsymbol{X}_i)} = C_i=h$, we generate the outcome $Y_i$ from $N(\vartheta_h, \phi)$, where all clusters share the equal variance $\phi=1.$
The cluster means are based on four different effect sizes (ES) choices. The atoms $\{\vartheta_1,   \vartheta_2,  \vartheta_3,  \vartheta_4\}$ 
are chosen to be $\{ -\Delta,0, +\Delta, +2\Delta\}$ for ES $\Delta=1,2,3,4.$
Larger $\Delta$ lead to more distinct OCs yielding an easier clustering problem, whereas OCs are less distinct when the ES is small. In the small $\Delta$ cases, the benefit to using the predictors in learning the clustering is more pronounced due to the overlap between the $N(\vartheta_h,\phi)$ distributions. 
With this set up, we generated $100$ independent training datasets and one addition test dataset for each ES.

\subsection{Competitor Methods}

We compared the performance of our model with several existing methods: the standard Dirichlet Process that ignores all information from the covariates as discussed in Section \ref{subsec:DP} and the Common Atoms Model with a fixed and known group membership discussed in \ref{subsec:CAM}. Note that the CAM operators as an oracle estimator, in the sense that it is using the true latent predictor groups from \eqref{eq:TruegroupX},  which is information that other methods do not have. 

We attempted to use the Product Partition Model with Regression on Covariates  \citep[PPMx;][]{muller2011product} through their R package \textit{ppmSuite} \citep{page2023ppmsuite}. However, its implementation provides a modal clustering but not probabilities, so we could not use its output to compute quantities of the form \eqref{eq:fun_theta} for our comparisions.  Additionally, inspection of the output substantially disagreed  from the data generating models suggesting consistent convergence failures. 
Consequently, it was excluded from model comparison.

We include comparison using the probit stick-breaking process  \cite[PSBP;][]{rodriguez2011nonparametric}. PSBP includes many variations, and we considered two of these: dependent PSBPs with Latent Gaussian Process (LGP) and a regression model for distributions. 
The construction of the PSBP model utilizes a stick-breaking process through a  probit transformation of  normals. 
Consequently, instead of model \eqref{eq:theta_i|X}, the PSBP has the form
\begin{equation}
    \theta_i\mid\boldsymbol{X}_i \sim F(\cdot\mid\boldsymbol{X}) = \sum_{h = 1}^H \pi_h(\boldsymbol{X}) \delta_{\vartheta_h}(\cdot),
    \label{eq:psbp}
\end{equation}
where  $\pi_h(\boldsymbol{X}) = \Phi(\alpha_{ih}) \prod_{s<h}[1-\Phi(\alpha_{is})]$, $\Phi(t)$ is the probit function, and $\{\alpha_{ih}\}_{ih}$ are marginally normal and depend on the predictors $\boldsymbol{X}_i$. 
For the regression version of PSBP, we considered $\alpha_{ih} = \boldsymbol{X}_i^T \boldsymbol{\eta}_{h}$ and $\boldsymbol{\eta}_h \sim MVN_P(\boldsymbol{\mu}_P, \mathbf{I}_P)$. 
In the LGP version, the joint distribution of the latent process $\boldsymbol{\alpha}_h=(\alpha_{1h},\ldots,\alpha_{nh})$ comes from a Gaussian process.  The covariances function  $\boldsymbol{\mathcal{S}}(\boldsymbol{X}_i,\boldsymbol{X}_{i^\prime})$ encourages observations with similar predictors to have similar $\alpha_{ih}$ for each $h$.  We implement LGP using $\boldsymbol{\mathcal{S}}(\boldsymbol{X}_i,\boldsymbol{X}_{i^\prime}) = \exp\left\{-{MD (\boldsymbol{X}_i, \boldsymbol{X}_{i'})}/{\sqrt{2P}}\right\}$ with a Mahalonbis distance $$MD(\boldsymbol{X}_i, \boldsymbol{X}_{i'}) = \sqrt{ (\boldsymbol{X}_i - \boldsymbol{X}_{i^\prime})^T \boldsymbol{\Sigma}^{-1} (\boldsymbol{X}_i - \boldsymbol{X}_{i^\prime})}$$ and $\boldsymbol{\Sigma}$ as the sample covariance matrix of the training data.

\subsection{Computational Choices}
 The prior distributions for the model parameters are $\vartheta_h$ is $N(m_0, \tau^2)$ for $m_0=0$ and $\tau^2=10^2$ and $\phi \sim \text{InvGamma} (e, f)$ for $e=f=1$.  The BNP concentration parameters are $\alpha \sim \text{Gamma} (a,b)$, and $\beta\sim \text{Gamma} (c,d)$ with  $a = 2$, $b = 1.5$, $c = 1.5$, $d = 2$. 
 We set the upper bound for the number of DCs at $K = 12$ and for the OCs at $H = 30$.
For the pyramid tree model $\pi (\mathcal{T})$, we use $a_{\mathcal{T}} = 0.95$ and $b_{\mathcal{T}} = 0.5$ \citep{chipman1998bayesian} in \eqref{eq:tree_split}. 
A maximum tree depth of $\mathcal{D}_\lambda = 10$ was considered and split values of the tree levels were drawn between the $q_1=0.05$ and $q_2=0.95$ quantiles. 
For the MH proposal distribution of $\mathcal{T}$, equal probabilities were assigned to the four step types. For each set of training data,  results are based on 5000 posterior samples from a single chain, following a burn-in period of 5000 samples ($\Delta=3,4$) or a burn-in of 15,000 samples ($\Delta=1,2$). Due to the higher computational time required for LGP, its results are derived from 2000 posterior samples after a burn-in period of 3000 or 8000 iterations.
As PSBP is not designed to perform variable selection, we fit both PSBP and CAPGM  models using all $P=20$ predictors and using only the relevant predictors $X_1,X_2,X_3$ to assess the impact of irrelevant covariates.

\subsection{Estimation Accuracy Metrics}
For each method, we obtain the Dahl estimator of the optimal clustering $\hat{\boldsymbol{C}}$ and compare it to the true OC memberships using the Adjusted Rand Index \citep[ARI;][]{rand1971objective, hubert1985comparing}. ARI
measures the degree of agreement between two data partitions.
An ARI of 1 indicates perfect agreement between the estimated and true clusterings, while a value of 0 reflects the level agreement expected under  random assignment.
For CAPGM and CAM, we also compute the ARI for the distribution clusters, and we consider the ARI of the predictor group clusters for CAPGM.  
Additionally, we also compared the methods based on the number of non-empty OCs, the number of reasonably sized  OCs (clusters containing at least $1\%$ of the sample size), the number of DCs, and the size of the largest OC

To evaluate the predictive performance of our model, we consider prediction of the mean of $Y$ as a function of the covariates $\boldsymbol{X}$ using both the training data and the out-of-sample test data.  For each individual, we predict the mean using \eqref{eq:fun_theta} with $\psi(\theta,\phi)=\theta.$ The Root Mean Squared Prediction Error (RMSPE) is then computed as  
$\sqrt{\frac{1}{n} \sum_{i=1}^n [y_i - \hat{\psi}(\boldsymbol{X}_i)]^2}$.

We also compare the goodness of fit of the models using log predictive density  score (LPDS). For each iteration, we estimate the predictive distribution at the observed data using
$$\hat{f}^m (y_i\mid\boldsymbol{X}_i) = \sum_{h= 1}^H Pr(C_i=h \mid G^m(\boldsymbol{X}_i) ) \, f_y(y_i\mid\vartheta_h^m, \phi^m),$$
and compute 
\begin{equation}
\text{LPDS} = \sum_{i = 1}^n \left[ \frac{1}{M} \sum_{m = 1}^M \log \hat{f}^m (y_i\mid\boldsymbol{X}_i) \right].
\end{equation}
The LPDS provides a trustworthy metric for evaluating the model fit for out-of-sample data using parameters estimated from the training sample \citep{gaskins2019hyper, zhou2015spatio}, and higher LPDS scores indicate better model fit.

\subsection{Simulation Results}
\label{Simulation results}

\begin{table}[!t]
\footnotesize
\setlength{\tabcolsep}{2pt}
\caption{RMSPE and LPDS from Simulation Study I. Mean and Standard Error in parentheses are reported based on 100 datasets. RMSPE is calculated according to equations (\ref{eq:fun_theta})}.

\centering
\begin{tabular}{lcccccc}
\hline
&  \multicolumn{3}{c}{Effect Size $\Delta=4$} &  \multicolumn{3}{c}{Effect Size $\Delta=3$} \\
\cmidrule(lr){1-4} \cmidrule(lr){5-7}
&  w/in Sample   & \multicolumn{2}{c}{Out of sample} &  w/in Sample   & \multicolumn{2}{c}{Out of sample} \\
\cmidrule(lr){2-2} \cmidrule(lr){3-4} \cmidrule(lr){5-5} \cmidrule(lr){6-7}
Method     & RMSPE & RMSPE & LPDS & RMSPE & RMSPE & LPDS \\
\cmidrule(lr){1-4} \cmidrule(lr){5-7}
CAPGM                & 1.00 (0.00) & 1.02 (0.00) & -2891 (2)   & 1.04 (0.01) & 1.05 (0.01) & -2920 (9) \\
${X_1, X_2, \& X_3}$ &             &             &             & 1.00 (0.00) & 1.02 (0.00) & -2886 (2) \\ \hline
CAM                  & 1.00 (0.00) & 1.00 (0.00) & -2858 (0.4) & 1.00 (0.00) & 1.00 (0.00) & -2858 (0.4) \\ \hline
DP                   & 4.33 (0.01) & 4.39 (0.00) & -5154 (1)   & 3.31 (0.01) & 3.37 (0.00) & -4881 (1) \\ \hline
PSBP-LGP             & 2.26 (0.01) & 2.70 (0.01) & -4020 (8)   & 1.82 (0.01) & 2.15 (0.01) & -3936 (9) \\
${X_1, X_2,\& X_3}$  &             &             &             & 1.35 (0.00) & 1.41 (0.00) & -3159 (1) \\ \hline
PSBP-Reg             & 1.75 (0.01) & 1.86 (0.01) & -3339 (17)  & 1.47 (0.01) & 1.55 (0.01) & -3285 (9) \\
${X_1, X_2,\& X_3}$  &             &             &             & 1.50 (0.01) & 1.49 (0.01) & -3200 (12) \\ \hline
\midrule
&  \multicolumn{3}{c}{Effect Size $\Delta=2$} &  \multicolumn{3}{c}{Effect Size $\Delta=1$} \\
\cmidrule(lr){1-4} \cmidrule(lr){5-7}
&  w/in Sample   & \multicolumn{2}{c}{Out of sample} &  w/in Sample   & \multicolumn{2}{c}{Out of sample} \\
\cmidrule(lr){2-2} \cmidrule(lr){3-4} \cmidrule(lr){5-5} \cmidrule(lr){6-7}
Method     & RMSPE & RMSPE & LPDS & RMSPE & RMSPE & LPDS \\
\cmidrule(lr){1-4} \cmidrule(lr){5-7}
CAPGM               & 1.07 (0.01) & 1.08 (0.01) & -2974 (7) & 1.03 (0.00) & 1.03 (0.00) & -2904 (1) \\
${X_1, X_2,\& X_3}$ & 1.00 (0.00) & 1.01 (0.00) & -2882 (3) &             &             &               \\ \hline
CAM                 & 1.00 (0.00) & 1.00 (0.00) & -2858 (1) & 1.00 (0.00) & 1.00 (0.00) & -2862 (1) \\ \hline
DP                  & 2.33 (0.00) & 2.38 (0.00) & -4377 (1) & 1.45 (0.00) & 1.48 (0.00) & -3597 (1) \\ \hline
PSBP-LGP            & 1.41 (0.00) & 1.65 (0.00) & -3734 (4) & 1.08 (0.00) & 1.23 (0.00) & -3378 (3) \\
${X_1, X_2,\& X_3}$ & 1.17 (0.00) & 1.21 (0.00) & -3122 (3) &             &             &               \\ \hline
PSBP-Reg            & 1.25 (0.00) & 1.31 (0.00) & -3229 (7) & 1.07 (0.00) & 1.12 (0.00) & -3089 (3) \\
${X_1, X_2,\& X_3}$ & 1.26 (0.00) & 1.26 (0.00) & -3118 (4) &             &             &               \\
\hline
\end{tabular}

\label{tab:sim1_pred}
\end{table}
As shown in Table \ref{tab:sim1_pred}, CAPGM has very competitive RMSPE and LPDS with the oracle estimator CAM across all choices of effect size.  This indicates that the pyramid tree is able to recover the true relationship between the predictors and the predictor groups.  When fitting CAPGM with only the true predictors, the predictions only marginally improve, further confirming the ability of CAPGM to identify which covariates impact the clustering.  
This contrasts to the LGP probit model, which shows substantially worse performance when all $P=20$ covariates are available versus when the model is fit using only the three true predictors.  The regression version of PSBP is robust to the unnecessary predictors, although it has significantly worse performance than CAPGM.  We also note that the PSBP tend to show a large disagreement between the training sample and test sample RMSPEs (especially for $P=20$).  This suggests a level of overfitting to the training data that does not appear to occur in CAPGM.
Finally, without using any information from the predictors, the standard DP model has very poor mean and density recovery.

\begin{table}[!t]
\caption{Cluster information from Simulation Study I, including ARI due to OC. Mean and standard error in parentheses are reported based on the Dahl estimate of the best cluster across the 100 datasets. No.\ OC$_{NE}$ and No.\ OC$_{w/1\% obs}$ represent the number of non-empty observational clusters and the number of OC with a size of at least 1\% of the total sample sizes, respectively.}
\centering
\resizebox{\textwidth}{!}{%
\begin{tabular}{lcccccc}
\hline
& \multicolumn{3}{c}{Effect Size $\Delta=4$} & \multicolumn{3}{c}{Effect Size $\Delta=3$} \\
\cmidrule(lr){1-4} \cmidrule(lr){5-7}
Method & No. OC$_{NE}$ & No. OC$_{w/1\% obs}$ & ARI-OC & No. OC$_{NE}$ & No. OC$_{w/1\% obs}$ & ARI-OC \\
\cmidrule(lr){1-4} \cmidrule(lr){5-7}
CAPGM     & 4.0 (0.0) & 4.0 (0.0) & 1.00 (0.00) & 4.1 (0.0) & 4.0 (0.0) & 0.99 (0.00) \\
${X_1, X_2,\& X_3}$ &     &             &           & 4.0 (0.0) & 4.0 (0.0) & 1.00 (0.00) \\ \hline
CAM       & 4.0 (0.0) & 4.0 (0.0) & 1.00 (0.00) & 4.1 (0.0) & 4.0 (0.0) & 1.00 (0.00) \\ \hline
DP        & 4.9 (0.1) & 4.1 (0.0) & 0.91 (0.00) & 5.5 (0.2) & 4.5 (0.1) & 0.75 (0.00) \\ \hline
PSBP-LGP  & 6.9 (0.2) & 4.6 (0.1) & 0.95 (0.00) & 6.9 (0.2) & 4.7 (0.1) & 0.86 (0.00) \\
${X_1, X_2,\& X_3}$ &     &             &       & 5.4 (0.1) & 4.3 (0.1) & 0.95 (0.01) \\ \hline
PSBP-Reg  & 7.1 (0.3) & 5.0 (0.1) & 0.92 (0.01) & 7.7 (0.3) & 5.2 (0.1) & 0.89 (0.01) \\
${X_1, X_2,\& X_3}$ &     &             &       & 6.0 (0.2) & 4.6 (0.1) & 0.93 (0.01) \\ \hline
\midrule
& \multicolumn{3}{c}{Effect Size $\Delta=2$} & \multicolumn{3}{c}{Effect Size $\Delta=1$} \\
\cmidrule(lr){1-4} \cmidrule(lr){5-7}
Method & No. OC$_{NE}$ & No. OC$_{w/1\% obs}$ & ARI-OC & No. OC$_{NE}$ & No. OC$_{w/1\% obs}$ & ARI-OC \\
\cmidrule(lr){1-4} \cmidrule(lr){5-7}
CAPGM     & 4.2 (0.0) & 4.0 (0.0) & 0.96 (0.00) & 3.4 (0.1) & 3.1 (0.0) & 0.90 (0.00) \\
${X_1, X_2,\& X_3}$  & 4.0 (0.0) & 4.0 (0.0) & 0.99 (0.00)  &     &             &            \\ \hline
CAM       & 4.1 (0.0) & 4.0 (0.0) & 0.99 (0.00) & 4.3 (0.1) & 3.9 (0.0) & 0.91 (0.01) \\ \hline
DP        & 4.4 (0.1) & 3.6 (0.1) & 0.51 (0.00) & 2.4 (0.1) & 2.0 (0.0) & 0.24 (0.00) \\ \hline
PSBP-LGP  & 7.4 (0.2) & 5.2 (0.1) & 0.67 (0.01) & 6.6 (0.1) & 4.7 (0.1) & 0.31 (0.01) \\
${X_1, X_2,\& X_3}$ & 5.7 (0.2) & 4.4 (0.1) & 0.86 (0.00) &     &             &             \\ \hline
PSBP-Reg  &7.5 (0.3) & 5.1 (0.1) & 0.81 (0.01) & 6.9 (0.3) & 4.5 (0.1) & 0.68 (0.01) \\
${X_1, X_2,\& X_3}$  & 6.5 (0.2) & 4.6 (0.1)   & 0.88 (0.01)  &     &             &              \\
\hline
\end{tabular}
}
\label{tab:sim1_clustersize}
\end{table}

In Table \ref{tab:sim1_clustersize}, we evaluate the clustering properties based on the Dahl-estimated partitions.
Recall that the true number of observation clusters should be 4, and they will have expected sizes of 500, 250, 125 and 125.  For $\Delta=2,3,4$, CAPGM obtains approximately the correct number of clusters when considering those with at least 1\% of $n$ (at least 10 members); there is a slight overestimation when counting outlier clusters.  For $\Delta=1$, CAPGM tends to underestimate the number of clusters, which is not unexpected since the two smallest clusters have distributions with substantial overlap: N(1,1) and N(2,1).  CAPGM has excellent ARI across all cases, including 0.90 in the most challenging setting.  
PSBP consistently overestimates the number of clusters and includes many additional small clusters.  Based on ARI, the regression version outperforms the LGP for the more challenging cases with smaller $\Delta$.
Without using the predictors, DP has substantially worse clustering estimates across all $\Delta$, with ARIs that are much lower than the other methods. 
A visual representation of the observational clustering structure, grouping patterns, and distributional clustering of observations for a single dataset, based on pairwise probabilities, is provided in Section \ref{S4_additional_Sim1} of the Supplementary Materials.

\begin{figure}[t]
    \centering
    \includegraphics[width=13cm]{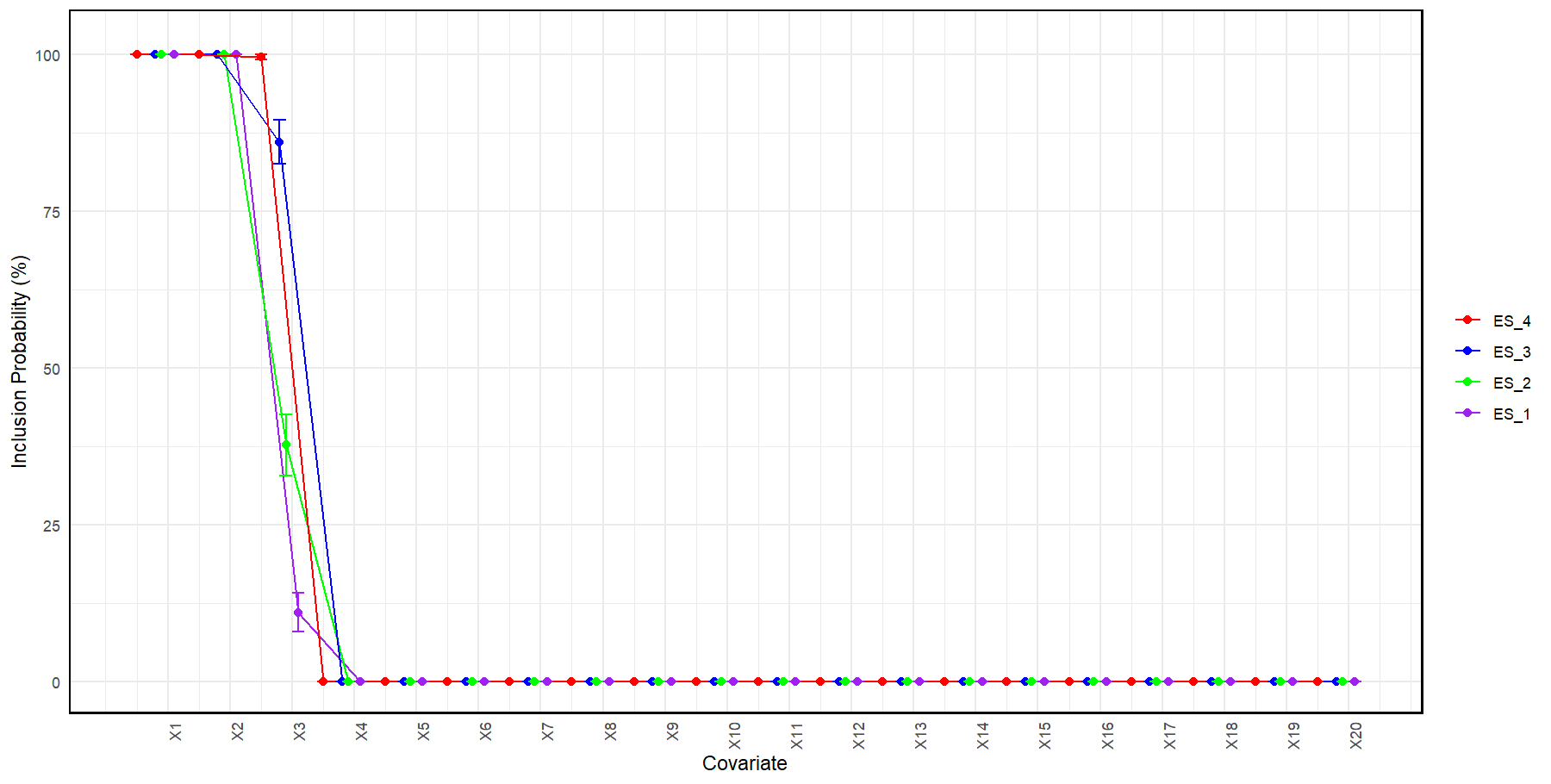}
    \caption{Average inclusion probabilities (as percentages) for each covariate with standard error bars in $\mathcal{T}$ with different effect size. Plot based on 100 datasets.} 
    \label{fig:sim1_x_incprob}
\end{figure}
Having shown that our CAPGM outperforms competitor methods in terms of predictions and clustering, we now  investigate the characteristics specific to our model. We begin with the features of the pyramid tree $\mathcal{T}$, and note
the average inclusion probabilities  illustrated in Figure \ref{fig:sim1_x_incprob}. For ES 4, the inclusion rate of $X_1$, $X_2$, and $X_3$ are $100\%$ with negligible standard error, indicating that throughout MCMC sampling, these covariates were consistently used as partitioning variables. The inclusion probabilities for the remaining (irrelevant) covariates are close to 0. 
The inclusion probabilities of $X_1$ and $X_2$ 
for $\Delta=1,2,3$ are approximately $100\%$, while the inclusion probability of $X_3$ decreases from approximately $90\%$ to $40\%$ to $12\%$ as the effect size decreases. 
Again, its lower inclusion rate  at smaller effect sizes is due to the smaller sample sizes and the high overlap between the distributions of OCs 3 and 4.

\begin{table}[t]
\setlength{\tabcolsep}{2.5pt}
\footnotesize
\caption{Some characteristics of pyramid tree and CAM DC characteristics.  }
\centering
\begin{tabular}{lccccccc}
\hline
&  \multicolumn{5}{c}{Pyramid tree characteristics of CAPGM} &  \multicolumn{2}{c}{CAM DC characteristics} \\
\cmidrule(lr){2-6} \cmidrule(lr){7-8}
Effect Size     & Tree Depth & No. groups  & ARI-Group   & No. DC& ARI-DC & No. DC& ARI-DC\\ \hline
$\Delta=4$            & 3.0 (0.0) & 8.0 (0.0) & 0.99 (0.00) & 4.3 (0.0) & 1.00 (0.00) & 4.1 (0.0) & 1.00 (0.00) \\ \hline
$\Delta=3$            & 2.9 (0.0) & 7.4 (0.1) & 0.94 (0.00) & 3.9 (0.0) & 0.98 (0.00) & 4.0 (0.0) & 1.00 (0.00) \\
$X_1, X_2, \& X_3$    & 3.0 (0.0) & 8.0 (0.0) & 0.99 (0.00) & 4.0 (0.0) & 1.00 (0.00) \\ \hline
$\Delta=2$            & 2.4 (0.1) & 5.5 (0.2) & 0.74 (0.00) & 3.4 (0.1) & 0.95 (0.00) & 4.0 (0.0) & 1.00 (0.00) \\
$X_1, X_2, \& X_3$    & 3.0 (0.0) & 7.9 (0.1) & 1.00 (0.01) & 4.0 (0.0) & 0.99 (0.00) \\ \hline
$\Delta=1$            & 2.1 (0.0) & 4.5 (0.1) & 0.63 (0.01)& 3.2 (0.0) & 0.92 (0.01) & 4.1 (0.0) & 1.00 (0.00) \\

\hline
\end{tabular}

\label{tab:sim1:treeinfo}
\end{table}
Table \ref{tab:sim1:treeinfo} provides additional insight into the sampled pyramid trees. The true tree depth is 3 with 8 true predictor groups.  The depth is accurately recovered for $\Delta=3,4$ and when using the true predictors for $\Delta=2$.  For the more challenging small ES cases, CAPGM can produce shorter (less deep) trees, leading to fewer predictor groups and lower predictor group ARI. 
However, recall that in CAPGM the predictor groups are clustered into DCs that are associated with the observational clustering.  Hence, accurately recovering the DCs is more important for the clustering of the response variables.  Even in these challenging settings, we obtain high ARI for these distribution clusters, with an average ARI of 0.92 even in the $\Delta=1$ case.

We consider another simulation with a different setting in Section \ref{S5_Simulation_Study_II} of the Supplementary Materials.  In that example we consider 4 predictor groups, 3 DCs, 4 OCs, and 3 different ESs, with  observations from multiple DCs being assigned into the same observational cluster.
Under this set up, CAPGM again outperforms all competing methods in terms of both prediction accuracy and recovery of the true clusters.

\section{Analysis of Health Retirement Study Data}

\subsection{Health Retirement Study Data}
In this section, we analyze the RAND Health and Retirement Study (HRS) data to investigate  patterns in hospitalization and how these are impacted by various demographic and medical characteristics. The HRS is a national longitudinal survey conducted in the United States, targeting individuals aged 50 and older, along with their spouses.
The dataset comprises 15 waves 
of bi-yearly interviews,
starting from 1992 \citep{bugliari2023rand, juster1995overview}.

We consider overnight hospital stay (ONHS) at the second wave of measurement as the response $Y$, a count variable reporting the number of nights spent in the hospital during the previous two years.
Our objectives are to observe the clustering patterns of ONHS and by using the  predictors obtained from the baseline survey to predict ONHS during this second wave of data collection, two years after the baseline wave.
Our dataset is restricted to patients who had observed values of the predictors in wave 1 and the outcome variable in wave 2. 
After data cleaning, 10,916 individuals were retained for our analysis, with $P=26$ baseline predictors used. 
An overview of the predictors along with their descriptive statistics is presented in Section \ref{S6_HRS_Data_Details} of the Supplementary Materials. The data were randomly divided into training and test sets. We considered three different sample sizes for the training set: 500 (small), 1000 (moderate), and 2500 (large). Results for the moderate sample size are reported here, while the results for the small and large sample sizes are reported in Sections \ref{S8_HRS_Data_Small_Sample} and \ref{S9_HRS_Data_Big_Sample} of the Supplementary Materials.

The response $Y$ is a count variable characterized by a long-tailed distribution, 
with $83.7\%$ zeros.  Among the non-zero values the five number summaries are $1, 2, 5, 10$, and $350$.
Given the potential for overdispersion, we employ a negative binomial distribution as the response variable PMF, $y\sim NB (r_h, p_h)$. 
The cluster specific parameter is determined by the atoms $\vartheta_h=(r_h,p_h)$, and 
we consider a uniform prior for $p_h$ and $\text{Gamma} (1, 1)$ distribution for $r_h$. Since the posterior distribution of $r_h$ is not conjugate, we employed the MH algorithm to generate $r_h$. For the MH, we use a proposal density of $\text{Uniform} (r_h^m - r^*, r_h^m + r^*)$ for  $r^* = 0.5$.

\subsection{Computational Details}
We fit the training datasets using the same methods as the simulation: CAPGM, CAM, DP, and PSBPs with a latent Gaussian process and regression model. To use CAM, we must have a priori defined groups based on the predictors $\boldsymbol{X}_i$. As there is no known group structure here, we define several simple variations and implement CAM with each for illustration and comparison.
Firstly, fixed groups were assigned based on all combinations of the demographic characteristics of age ($< 50$, 50--59, $\ge60$), sex, and race (White and non-White).
Secondly, groups were constructed based on the number of comorbidities (0, 1, 2, 3, 4, more than 4 chronic conditions).
Finally, we considered insurance status for grouping. Five groups were created according to the type of insurance: no insurance, only government insurance, only private insurance, other insurance, and more than one type of insurance.  For the latent Gaussian process of PSBP, we use indicators for each individual comorbidity and excluded the cumulative number of chronic conditions to avoid the singularity in $\Sigma^{-1}$.

During the MCMC sampling of CAPGM, we assign a higher probability to the SWAP step when proposing $\mathcal{T}'$. This is intended to enhance the mixing of the tree by frequently swapping among variables without pruning or growing the tree, which facilitates the exploration of many predictor combinations. We consider the re-split step to be less important for our data as most of the predictors are binary, and in this cases, a new threshold will not change  $\mathcal{T}$. Specifically, we use 
$p_{MH}=(0.3, 0.15, 0.05,0.5).$
As in the simulation study we take $\mathcal{D} =10,$ $K = 12,$ and $H = 30$ as the truncation levels. For each training dataset, we run three MCMC chains with 10,000 iterations. After discarding the first 5000 burn-in iterations from each chain, we combined the MCMC samples from all chains to perform inference. For PSBP-LGP, due to its computational demands, we were only able to ran 5000 iterations per chain and discarded the first 3000 posterior samples.

\subsection{Results: CAPGM}
During MCMC for the  moderate sized training data ($n=1000$), CAPGM forms an average of 5.52 non-empty clusters with an average of 3.35 clusters containing at least 10 members.  The size of the largest cluster was 808 on average and represents a group of respondents that generally had no overnight hospitalizations during the year. Additionally, the model generates an average of 3.40 distributional clusters.

\begin{table}[t]
\caption{Variable inclusion posterior probability of covariates in $\mathcal{T}$, listed in descending order.}
\footnotesize
\centering
\resizebox{\textwidth}{!}{%
\begin{tabular}{rrrrrrrrr}  \hline
 Covariate & SRHS & HLTHLM & Diab & CCs &  Cancer & Stroke & Age & InPov \\ 
 Probability $(\%)$ & 73.2 & 54.0 & 48.9 & 15.1 & 7.6 & 2.9 & 2.8 & 2.2 \\ \hline
 Covariate  & Ethnicity  & HIBP & Partner & Lung & Psych & VigAct & Insurance & Smoking \\
Probability $(\%)$ & 2.1 & 2.1 & 1.3 & 1.3 & 1.3 & 1.1 & 0.8 & 0.7 \\ \hline
 Covariate   & Asset & Sex & Ever Smoke  & Arthr & BMI & Race  & Education & Heart \\
 Probability $(\%)$  & 0.7 & 0.7 & 0.5 & 0.5 & 0.3 & 0.1 & 0.1 & 0.1 \\
   \hline
\end{tabular}
}
\label{tab:data:varincprob}
\end{table}
The inclusion probabilities of the covariates in the pyramid tree $\mathcal{T}$, as shown in Table \ref{tab:data:varincprob}, indicate that Self-Reported Health Status (SRHS) is the most relevant covariate, being involved in the tree during $73\%$ of iterations. HLTMLM (health problems limiting work) and diabetes also emerge as important covariates, demonstrating substantial influence in partitioning patients into latent groups. 
In contrast to these measures of patient health, the socio-demographic covariates, including sex,  age, race, education, and living below the poverty line (InPov) are found to have a minimal impact on defining predictor group structures in our model. This is further reflected in the demographic-based version of the CAM approach that we discuss later, which performs similarly to DP, indicating that demographic factors have minimal contribution to the group assignments.
Additionally, Table \ref{tab:data_1000:combX} in the Supplementary Materials reports the covariate combinations of the top 10 most frequently sampled pyramid trees.

\begin{figure}[t]
    \centering
    \includegraphics[width=1.05\linewidth]{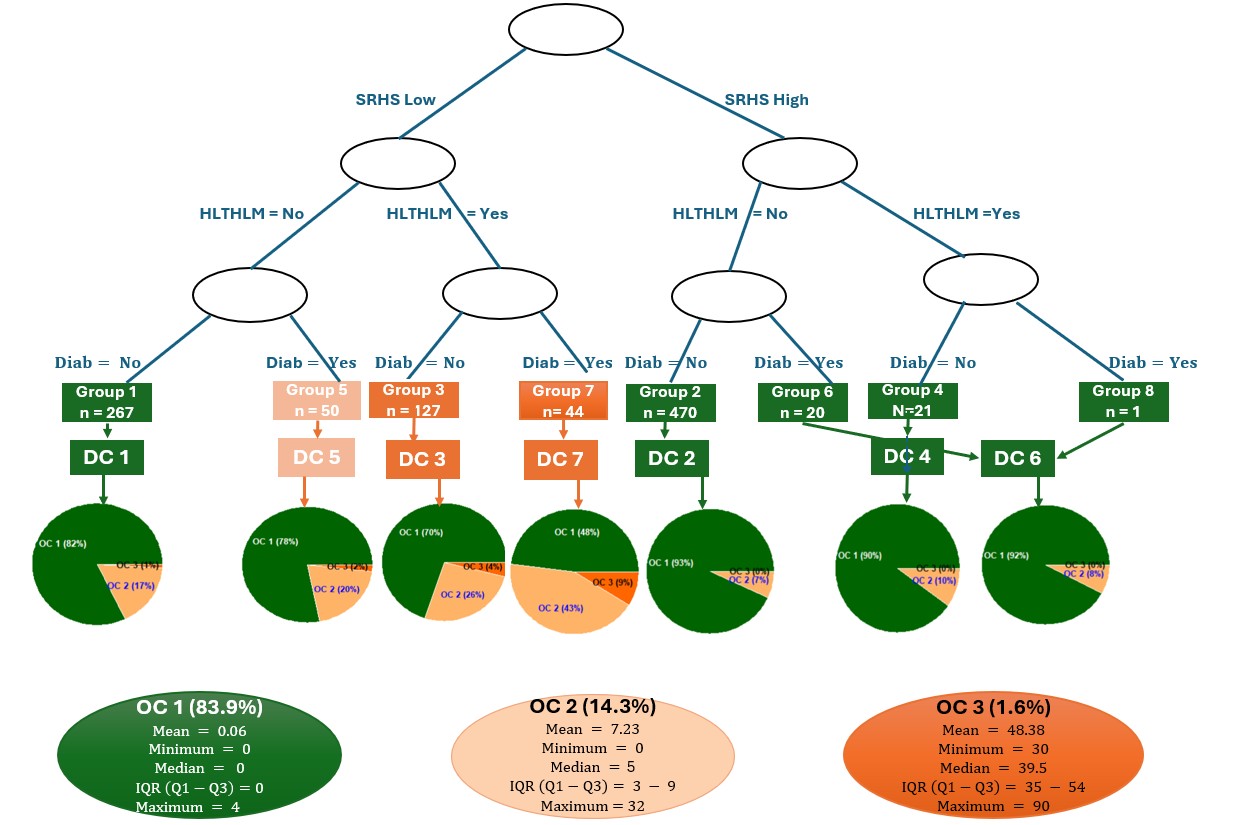}
    \caption{Estimated clustering structures for the HRS data.  The Dahl-estimated predictor groups with the corresponding pyramid tree are connected to seven estimated distribution clusters.  For each distribution cluster, pie charts shows the probabilities that members are assigned to the three observational clusters. 
    }
    \label{fig:capgm.1000.est.clust.strctr}
\end{figure}

We utilize the Dahl approach to obtain a posterior estimate of the OCs, DCs, and predictor groups, and we extract the tree $\mathcal{T}$ from the iteration corresponding to the estimated predictor group clustering.  These are represented in 
Figure \ref{fig:capgm.1000.est.clust.strctr}.
The estimated pyramid tree includes the top three predictors, SRHS, HLTHLM, and diabetes, yielding eight groups.  
The resulting group 2 is expected to be the healthiest, characterized by higher SRHS (very good or excellent, instead of poor, fair, or good), no mobility issues (no HLTHLM), and no diabetes, while group 7 appears to be at the highest risk. 
Six of the predictor groups correspond to their own distribution clusters. The predictor group 8 with high self-reported health but both diabetes and mobility issues contains a single individual, and this group is combined into the same DC as those with diabetes, high reported health, and no HLTMLM.

We obtain three observational clusters 
(excluding two individuals that are placed in outlier clusters).
For summary, the median (and range) of observations within these clusters 0 (0--4), 5 (0--32), and 39.5 (30--90);  this first cluster contains 84\% of observations and is mostly made up of zero counts.
We see that DCs 2, 4, and 6 (associated with high SRHS) have 90\% or higher probability of assigning members to OC1.  The most at-risk group 7 (DC7) assigns over 50\% to OC2 and OC3 representing the higher rates of ONHS.
Further details of the estimated group characteristics are provided in Table \ref{tab:group_est_1000} of the Supplementary Materials, Section \ref{S7_Additional_results_HRS_Data_Ex}, while Figure \ref{fig:est.group.dist.hm} compares the Dahl estimate and pairwise clustering probabilities for the predictor groups and DCs.

\subsection{Results: Comparison Across Methods}

\begin{table}[t]
\setlength{\tabcolsep}{2.5pt}

\caption{RMSPE and LPDS for training and test data from Health and Retirement Study with moderate sample size.} 
\centering
\begin{tabular}{lcccc}
\hline
&  \multicolumn{2}{c}{Within sample} &  \multicolumn{2}{c}{Out of Sample} \\
\cmidrule(lr){2-3} \cmidrule(lr){4-5}
Method    & RMSPE& LPDS & RMSPE & LPDS \\
\hline
CAPGM       & 9.15 & -1076 & 10.33 & -9753\\
CAM (Demo)  & 9.31 & -1097 & 10.50 & -10011\\
CAM (CCs)   & 9.25 & -1081 & 10.43 & -9876\\
CAM (Ins)   & 9.31 & -1094 & 10.43 & -9880\\
DP          & 9.36 & -1102 & 10.50 & -9943\\
PSBP (Reg)  & 8.51 & -1051 & 10.35 & -10236\\
PSBP (LGP)  & 8.58 & -1017 & 10.32 & -10347\\
\hline
\end{tabular}

\label{tab:data_1000_rmspe.lpds}
\end{table}

\begin{figure}[t]
    \centering
    \includegraphics[width=1\linewidth]{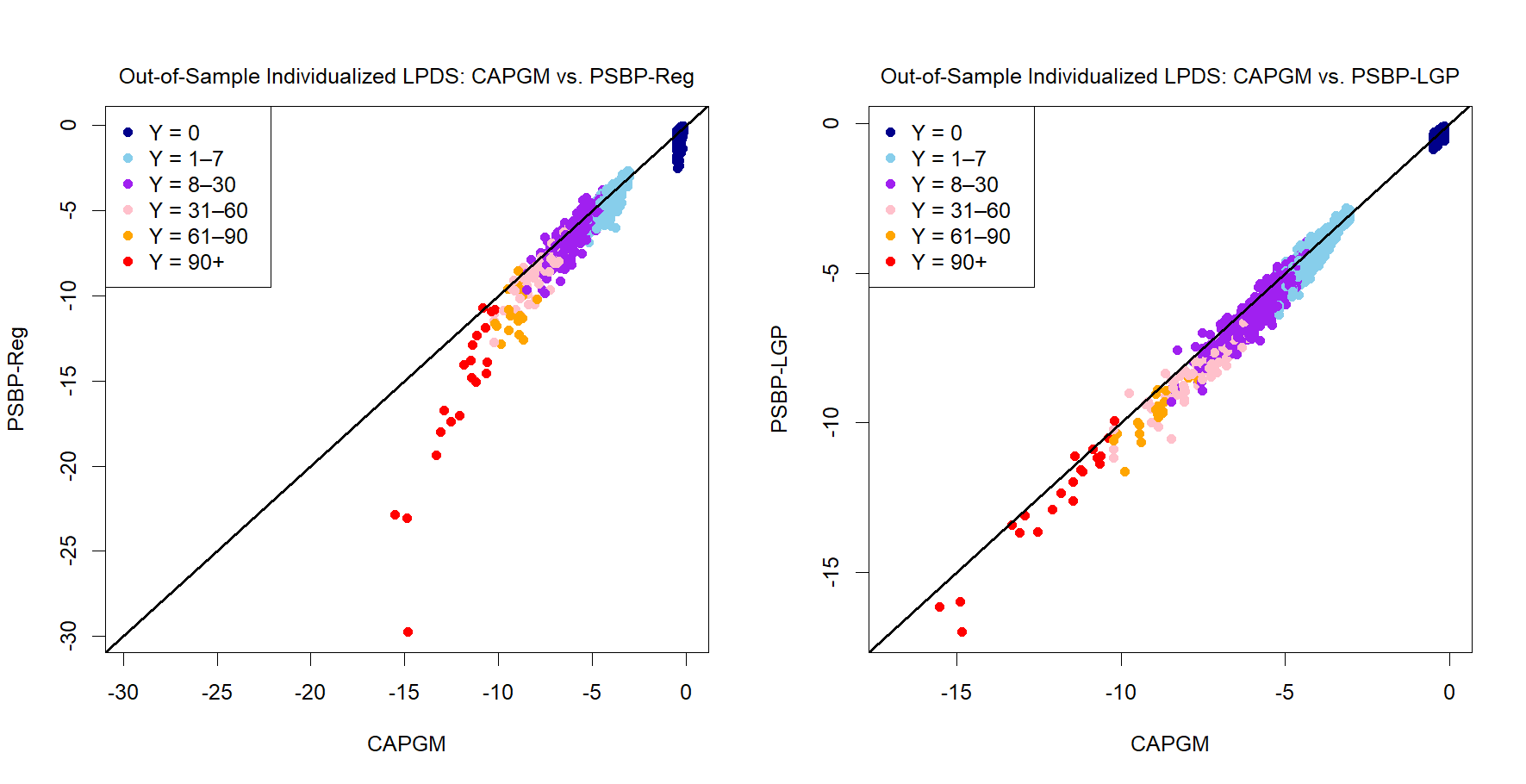}
    \caption{Comparison of model fit between CAPGM, PSBP-Reg and PSBP-LGP. Each point represents the estimate for $E_{post}\left\{ \log {f}(y_i\mid \boldsymbol{X}_i)\right\}$ for an observation in the out-of-sample test data, color-coded by the magnitude of response.}
    \label{fig:capgm.reg.1000.lpds}
\end{figure}

In Table \ref{tab:data_1000_rmspe.lpds}, we compare mean prediction and LPDS across the different competitor models.  
For the training data, the two PSBP model show the lowest mean prediction error within sample and the highest log-score values, indicating the best match to the 1000 within sample respondents' data.  Our CAPGM has the next best performance, surpassing the DP choice that completely ignores the predictors, as well as CAM with groups based on demographic characteristics, number of chronic conditions, and insurance type.
Considering the out-of-sample predictions, CAPGM and the two PSBP models have equivalent mean prediction, while CAPGM has vastly superior LPDS.  The disparity in the log-density is related to the same tendency from the simulation of PSBP to overfit and use too many clusters.  During MCMC sampling, CAPGM uses an average of 5.5 clusters (3.4 with at least 1\% of $n$) compared to 8.3 for PBSP-Reg (4.0) and 6.8 (4.3) for PBSP-LGP.

To further understand the differences in the test data LPDS, we plot the contribution to LPDS for each individual in Figure \ref{fig:capgm.reg.1000.lpds}.  That is, we plot the estimate of $E_{post}\left\{ \log {f}(y_i\mid \boldsymbol{X}_i)\right\}$ for CAPGM against the two PSBP models, color-coded by the magnitude of $y_i$.  For both PSBP models, we see that there are many larger $y_i$s in the test data whose log-density is estimated to be substantially lower under the PSBP models than our CAPGM.  This suggests that PSBP seem to be substantially underestimating the right tail of ONHS.

\begin{figure}[!t]
\centering
\begin{subfigure}{11cm}
  \includegraphics[width=\linewidth]{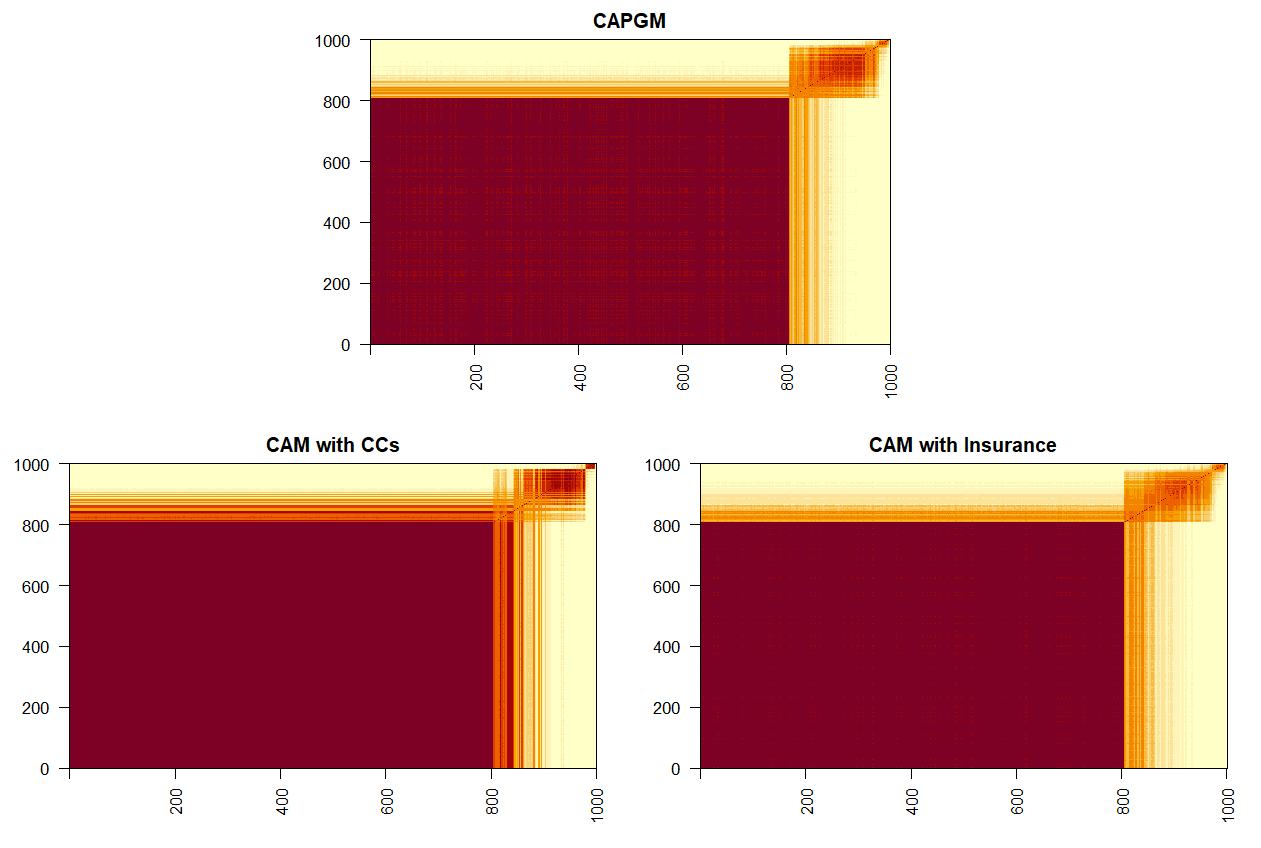}
\end{subfigure} 
\begin{subfigure}{11cm}
  \includegraphics[width=\linewidth]{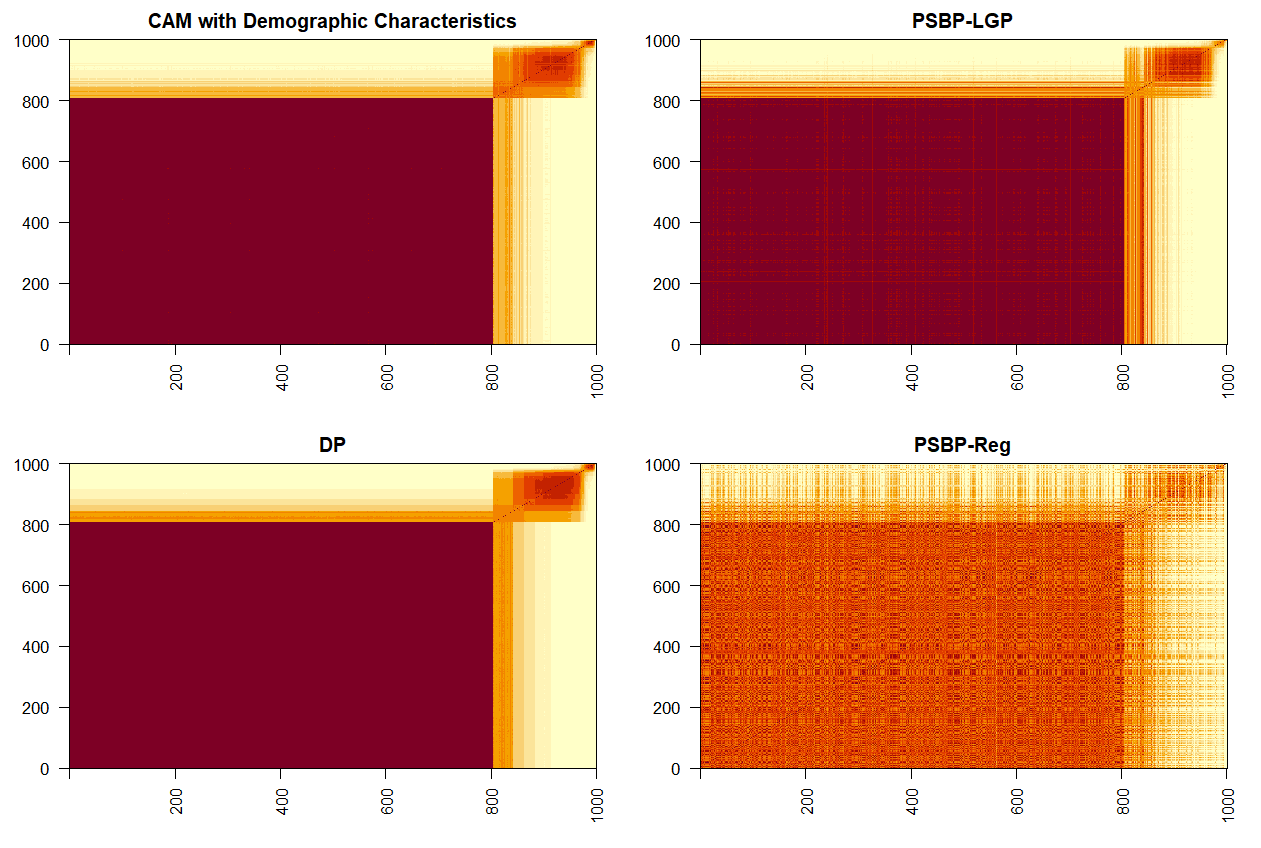}
\end{subfigure}
\caption{Comparison in the  pairwise co-clustering probabilities of ONHS responses across models. Observations are ordered in terms of increasing ONHS.
}
\label{fig:data:heatmap}
\end{figure}

The heat map in Figure \ref{fig:data:heatmap} provides additional insight  about the clustering pattern of the ONHS and reveals the three distinguishable clusters identified by all methods, which generally correspond to the three OCs in our Dahl estimate (see Figure \ref{fig:capgm.1000.est.clust.strctr}).
Of particular interest here is the stark difference in the co-clustering probabilities of the PSBP-Reg model, relative to the other methods.
Consistent with results from the simulation study, 
this method generates more clusters including multiple clusters to explain the zero counts.  Additionally, we note that there are surprisingly high co-clustering probabilities between the zero counts and the positive counts under PSBP-Reg.
Table \ref{tab:data_1000:clustersize} in Section \ref{S7_Additional_results_HRS_Data_Ex} of the Supplementary Materials provides additional clustering information of this data application.  In Sections \ref{S8_HRS_Data_Small_Sample} and \ref{S9_HRS_Data_Big_Sample} of the Supplementary Materials, we also provide results based on using the smaller ($n=500$) and larger ($n=2500$) training datasets, which again demonstrate superior performance of the CAPGM approach.

\section{Discussion}
In this project, we have considered a Bayesian nonparametric model for clustering of observations where the cluster membership is informed by a general set of predictors. To enable predictor-informed clustering, we extended the Common Atoms Model \citep{denti2023common} by treating group membership as an unknown latent variable determined by a set of predictor variables.
To that end, we proposed a pyramid group model to flexibly determine latent group membership, similar to the Bayesian CART model \citep{chipman1998bayesian} but with a key difference: all nodes of the same level in a PGM are split using the same rule at each depth level.
The PGM flexibly partitions the predictor space $\boldsymbol{\mathcal{X}}$ through an easily interpreted structure such  that observations with similar characteristics are assigned to the same terminal node in $\mathcal{T}$. By treating each terminal node as an unknown latent group for CAM, these predictor groups are then clustered based on how their members are assigned to the observational clusters.

To facilitate interpretation of the estimated structures, we partitioned the observations into observational, predictor group, and distribution clusters through a decision-theoretic approach \citep{dahl2006model}. While we utilize this strategy only to get a point estimate for each clustering structure, \cite{wade2018bayesian} show how to obtain uncertainty quantification of these partitions through credible balls.  Due to the relationship between the predictor groups and the PGM, we also use the Dahl approach, restricted to the sampled partitions, to gain point estimation of  $\mathcal{T}$, based on maximizing the pairwise node matching under $\mathcal{T}$.  The usual strategy for finding a point estimate of the tree usually involves maximizing the posterior over the sampled trees \citep{chipman1998bayesian}, but as our PGM is one component of a larger hierarchy, this is not as straightforward.  Just as we leverage the predictor group co-clustering probabilities to obtain the point estimator of the PGM, one could similarly apply the credible ball framework to obtain an uncertainty characterization of the PGM posterior.

A key challenge in implementing the CAPGM model is the risk of the MCMC chain getting stuck in local modes of the pyramid tree, thereby preventing adequate mixing. Running multiple chains at diverse initial trees is critical, and tuning the proposal rates of the four MH step types can also improve performance. In the ONHS example, we assigned higher weight to the CHANGE VARIABLE step to better encourage exploration over pyramids that use different predictors.
More complex MCMC strategies such as parallel tempering 
\citep{earl2005parallel}
or MCMC annealing \citep{geyer1995annealing}
may yield better mixing but will substantially complicate the sampling algorithm, so we have not further investigated their implementation.  In our experiments, we have obtained adequate performance using multiple chains.

Our CAPGM provides a predictor-informed clustering that can be used with any general response distribution $f_y(\cdot\mid \theta,\phi)$.  Our experiments in this manuscript involve univariate normal and count response variables, but this can be extended for more complex response types.  Our future work will consider the use of the CAPGM in the context of a longitudinal model, where we use a set of covariates measured at baseline to cluster longitudinal trends and to predict future observations.  Other types of multivariate responses and clustered data structures could also be investigated within the context of our model.
\bibliographystyle{plainnat}
\bibliography{references}

\clearpage
\appendix

\renewcommand{\thesection}{S.\arabic{section}} 
\renewcommand{\thetable}{S.\arabic{table}} 
\renewcommand{\theequation}{S.\arabic{equation}} 
\renewcommand{\thefigure}{S.\arabic{figure}} 

\setcounter{figure}{0}
\setcounter{table}{0}
\setcounter{equation}{0}

\begin{center}
\vspace{1em}
{\huge Supplementary Material to ``Predictor-Informed Bayesian Nonparametric Clustering''}
\end{center}
\vspace{4em}

\section{A Comparison of Multigroup Bayesian Nonparametric Models}
\label{S1_Multigroup_BNP_comparison}
\begin{figure}[!t]
\centering
\begin{subfigure}{9.5cm}
  \includegraphics[width=\linewidth]{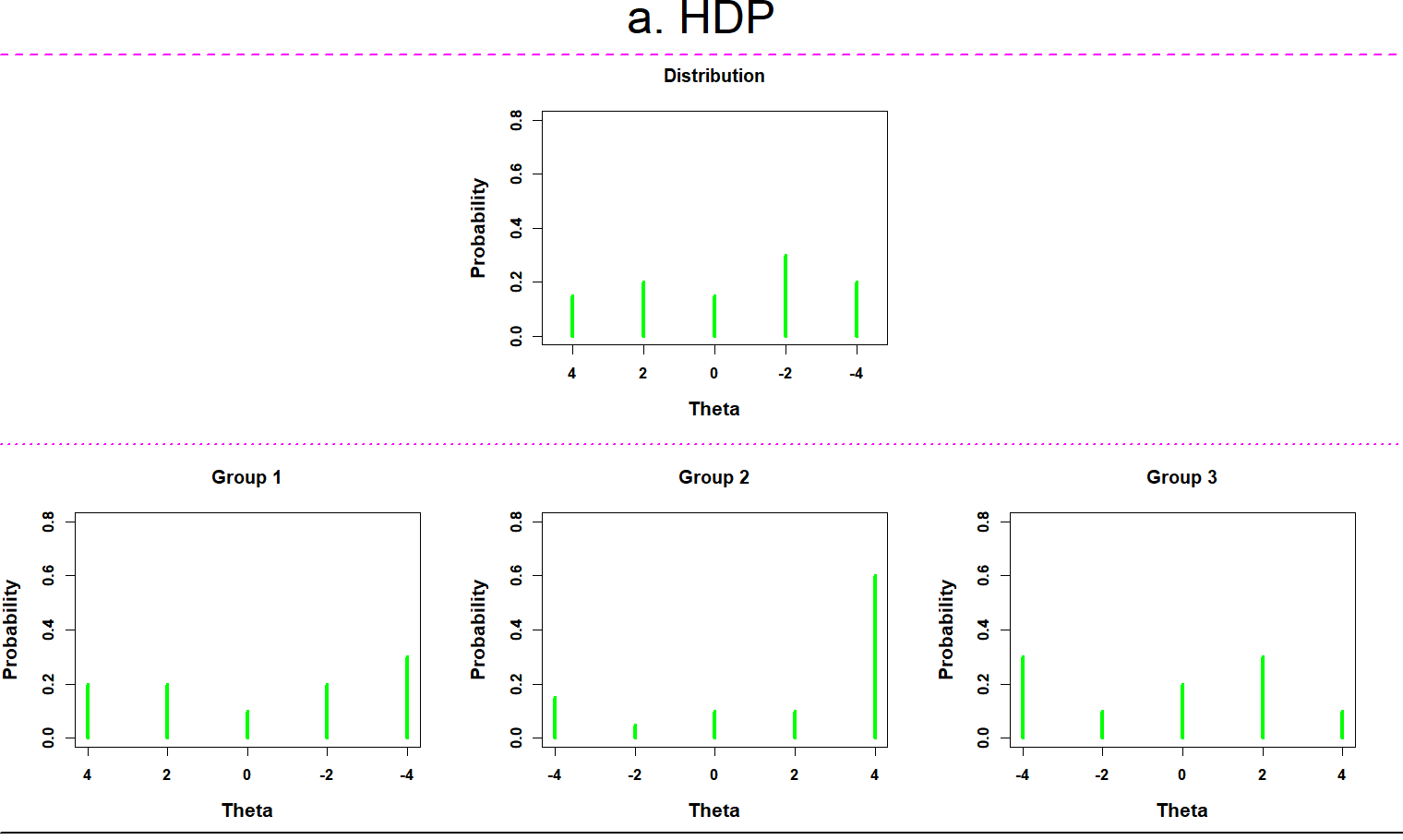}
\end{subfigure}
\begin{subfigure}{9.5cm}
  \includegraphics[width=\linewidth]{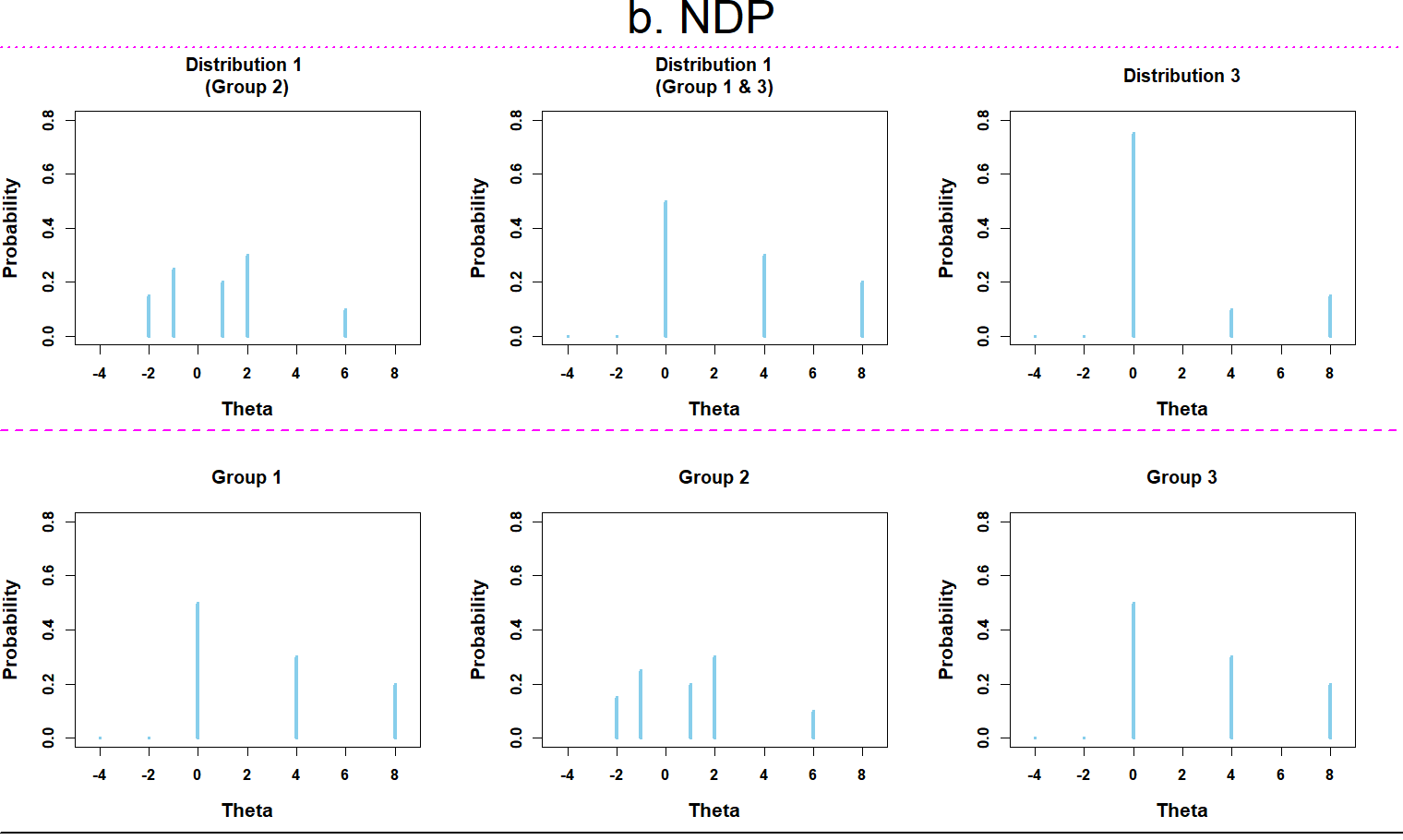}
\end{subfigure}

\begin{subfigure}{9.5cm}
  \includegraphics[width=\linewidth]{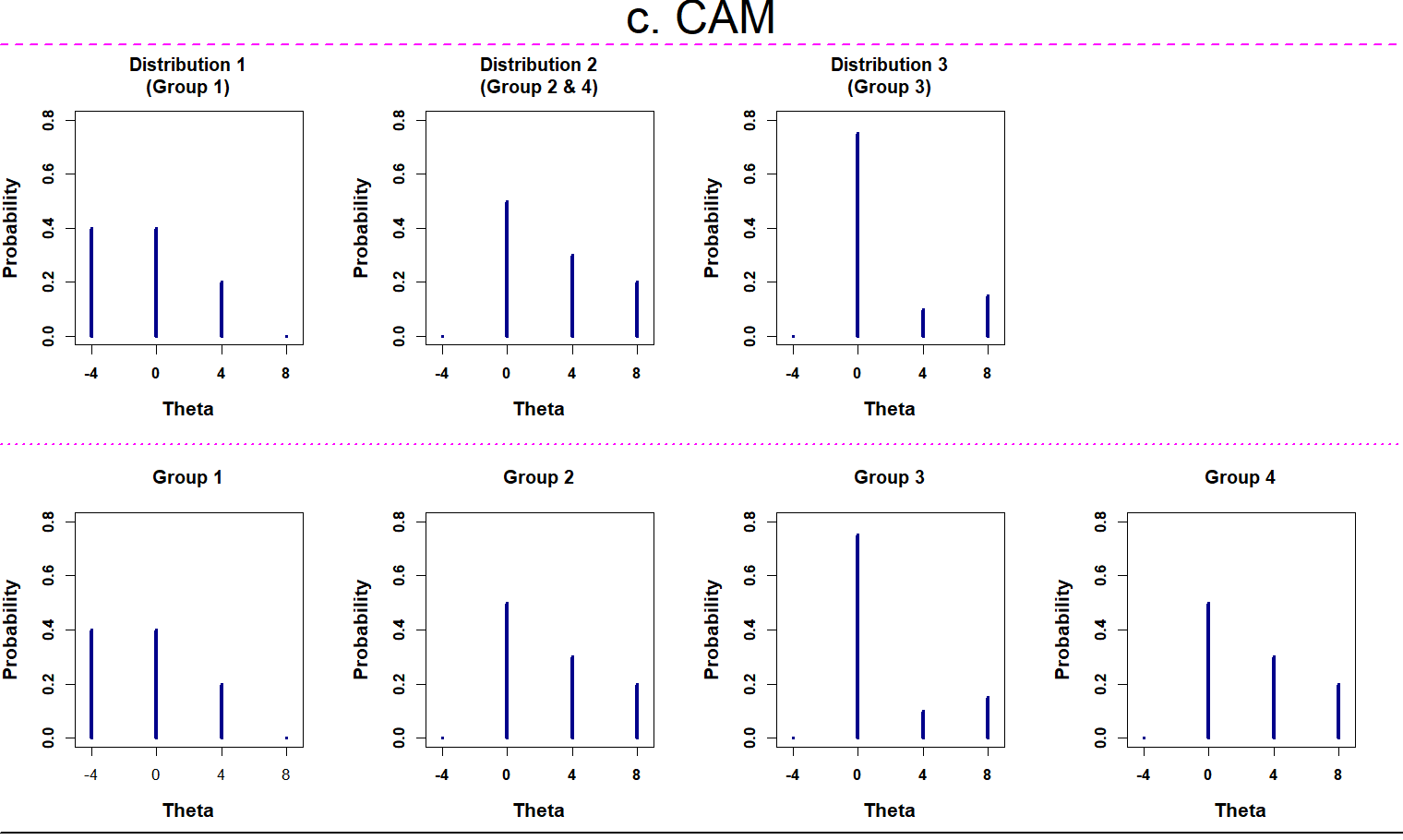}
\end{subfigure}

\caption{Atoms and their corresponding weights for distributions and groups for HDP, NDP, and CAM \citep{teh2004sharing, rodriguez2008nested, denti2023common}.}
\label{fig:hdpndpcam}
\end{figure}

As discussed in the manuscript, the Common Atom Model \citep[CAM;][]{denti2023common} shares similarities with both Hierarchical Dirichlet Process \citep[HDP;][]{teh2004sharing} and Nested Dirichlet Process \citep[NDP;][]{rodriguez2008nested}.
Figure \ref{fig:hdpndpcam} illustrates the similarities and differences between a realization of each of these models.
Like HDP, CAM shares the same set of atoms across groups, mimicking the HDP approach 
As shown in Figure \ref{fig:hdpndpcam}(a),  each group in HDP shares the same set of atoms ($0,\pm 2,\pm4$). However, HDP assigns different weights to the atoms for each group, thus cluster formation occurs only in OC. On the other hand, groups in NDP  either simultaneously share both weights and atoms or they share neither. Figure \ref{fig:hdpndpcam}(b) demonstrates that groups 1 and 3, originating from distribution 2 in the lower level of the hierarchy, share the same atoms and weights, while group 2 has different atoms and weights, leading to OC occurring only within DC. In the construction of CAM, groups sharing the same weights construct DC, and as atoms are shared across all groups, OC occurs across distributions. Figure \ref{fig:hdpndpcam}(c) illustrates that groups 2 and 4, sharing the same weights, form DC 2, and the shared atoms across DC construct OC.

\section{Proof of CAPGM Properties}

\label{S2_Proof_Equations}

In this section, we provide the details proving the results from Section 3.3.  These proofs primarily follow the same steps as in \cite{denti2023common} with the addition of allowing for the randomness of the group assignments due to the PGM.

Before considering our properties (P-1) through (P-4), we recall a property from equation (6) of \cite{denti2023common}.  For two (fixed) groups $g$ and $g'$ ($g\ne g'$), the probability that these groups are assigned to the same distribution cluster is given by
\begin{align}
   P(D_g = D_{g'}) &= E\left[ P(D_g = D_{g'}\mid F)\right] = E\left[\sum_{k = 1}^\infty P(D_g = D_{g'} = k \mid F)\right] \nonumber \\
& = \sum_{i=1}^\infty E\left[ P(D_g = k\mid F) P(D_{g'} = k\mid F)\right]  =  \sum_{k = 1}^\infty E\left[\rho_k^2\right] \nonumber \\
 &=\sum_{k = 1}^\infty E\left[ U_k^2 \prod_{l<k} (1-U_l)^2\right] = \sum_{k = 1}^\infty\left[ E \left[U_k^2\right] \prod_{l<k} E\left[1-U_l\right]^2\right] \nonumber \\
 &=\sum_{k = 1}^\infty\left[ \frac{\beta(3,\alpha)}{\beta(1, \alpha)} \prod_{l<k}\left(\frac{\beta(1, \alpha+2)}{\beta(1,\alpha)}\right)\right]  = \frac{2}{(1+\alpha)(2+\alpha)}\sum_{k=1}^\infty\left[\frac{\alpha}{2+\alpha}\right]^{k-1}\nonumber \\
 &= \frac{2}{(1+\alpha)(2+\alpha)}\sum_{k=0}^\infty\left[\frac{\alpha}{2+\alpha}\right]^{k}
 = \frac{1}{1+\alpha} \label{eq:dgdg}
\end{align}
Here, we recall that the $F$ in the conditioning statements is the CAM distribution of the observation weights (3): $F(\cdot)=\sum_{k=1}^\infty \rho_k \delta_{\boldsymbol{\nu}_k}(\cdot)$.  We also let $\beta(x,y)=\Gamma(x)\Gamma(y)/\Gamma(x+y)$ be the standard beta function.

\subsection*{Proof of (P-1)}
Recall that we have defined $P(X_i, X_{i'})$ to be the probability that observations $i$ and $i'$ belong to the same terminal node, that is, $P(X_i, X_{i'})=P\{G(\boldsymbol{X}_i) = G(\boldsymbol{X}_{i'})\} $. For notational simplification, we let the groups for observations $i$ and $i'$ be $G_i = G(\boldsymbol{X}_i)$ and $G_{i'} = G(\boldsymbol{X}_{i'})$, and their distributions are $F_{G(\boldsymbol{X}_i)}(\cdot)=F_{G_i}(\cdot)$ and $F_{G(\boldsymbol{X}_{i'})}(\cdot)=F_{G_{i'}}(\cdot)$. The predictor groups $G_i$ and $G_{i'}$ for the two observations may either be the same or different. Then,
\begin{align*}
    P\left\{F_{G_i } (\cdot) = F_{G_{i'}} (\cdot)\right\} &= P\left\{F_{G_i } (\cdot) = F_{G_{i'}} (\cdot)\mid G(\boldsymbol{X}_i) = G(\boldsymbol{X}_{i'})\right\}
    \ P\{G(\boldsymbol{X}_i) = G(\boldsymbol{X}_{i'})\}  \\
    & \quad + P\left\{F_{G_i } (\cdot) = F_{G_{i'}} (\cdot)\mid G(\boldsymbol{X}_i) \neq G(\boldsymbol{X}_{i'})\right\} \ P\{G(\boldsymbol{X}_i) \neq G(\boldsymbol{X}_{i'})\} \\
    &= P(\boldsymbol{X}_i, \boldsymbol{X_{i'}})P\left\{D_{G(\boldsymbol{X}_{i})} = D_{G(\boldsymbol{X}_{i'})}\mid G(\boldsymbol{X}_i) = G(\boldsymbol{X}_{i'}) \right\} \\
    & \quad +  \left[1-P(\boldsymbol{X}_i, \boldsymbol{X_{i'}}) \right] P\left\{D_{G(\boldsymbol{X}_{i})} = D_{G(\boldsymbol{X}_{i'})}\mid G(\boldsymbol{X}_i) \neq G(\boldsymbol{X}_{i'}) \right\} \\
    &= P(\boldsymbol{X}_i, \boldsymbol{X_{i'}}) + \left(1-P(\boldsymbol{X}_i, \boldsymbol{X_{i'}}) \right) \frac{1}{1+\alpha} \\
    & = \frac{1+\alpha P(\boldsymbol{X}_i, \boldsymbol{X_{i'}})}{1+\alpha}
\end{align*}
Note that $P\left\{D_{G(\boldsymbol{X}_{i})} = D_{G(\boldsymbol{X}_{i'})}\mid G(\boldsymbol{X}_i) = G(\boldsymbol{X}_{i'}) \right\} =1$ since observations in the same predictor group are necessarily in the same distributional cluster.  Further, the result $P\{D_{G(\mathbf X_i)} =\allowbreak
D_{G(\mathbf X_{i'})} \mid G(\mathbf X_i) \ne G(\mathbf X_{i'})\} = \frac{1}{1+\alpha}$ comes from \eqref{eq:dgdg}.

\subsection*{Proof of (P-2)}

Consider two sets $A, B \in \mathcal{F}$. 
The covariance between $F_{G_i}(A)$ and $F_{G_{i'}}(B))$ defined as 
\begin{align}
    &\text{Cov}\left(F_{G_i}(A), \ F_{G_{i'}}(B)\right) = E\left[F_{G_i}(A) F_{G_{i'}}(B)\right] - E  \left[F_{G_i}(A)\right] E \left[F_{G_{i'}}(B)\right]&&
    \label{eq:cov_start_eq}
\end{align}
First, we focus the expected value of product term.
\begin{align}
&E[F_{G_i}(A) F_{G_{i'}}(B)] =E\left\{ E\left[F_{G_i}(A)  F_{G_{i'}}(B))\mid 
 G_i,G_{i'} \right]    \right\} 
 \nonumber \\ 
& \quad = \quad  E\left[ F_{G_i} (A) F_{G_{i'}} (B) \mid D_{G_i} = D_{G_{i'}},G_i = G_{i'}\right] P\left\{ D_{G_i} = D_{G_{i'}},G_i = G_{i'}\right\}\nonumber \\
& \quad \quad + \quad E\left[ F_{G_i} (A) F_{G_{i'}} (B) \mid D_{G_i} = D_{G_{i'}}, G_i \neq G_{i'}\right] P\left\{ D_{G_i} = D_{G_{i'}},G_i \neq G_{i'}\right\} \nonumber \\
& \quad \quad + \quad E\left[ F_{G_i} (A) F_{G_{i'}} (B) \mid D_{G_i} \neq D_{G_{i'}}, G_i \neq G_{i'}\right] P\left\{ D_{G_i} \ne D_{G_{i'}},G_i \neq G_{i'}\right\} \nonumber \\
& \quad = \quad  E\left[ F_{G_i} (A) F_{G_{i}} (B)\right] P\left\{ D_{G_i} = D_{G_{i'}}\mid G_i = G_{i'}\right\}P\left\{ G_i = G_{i'}\right\}\nonumber \\
& \quad \quad + \quad E\left[ F_{G_i} (A) F_{G_{i}} (B) \right] P\left\{ D_{G_i} = D_{G_{i'}}\mid G_i \ne G_{i'}\right\}P\left\{ G_i \ne G_{i'}\right\}\nonumber \\
& \quad \quad + \quad E\left[ F_{G_i} (A) F_{G_{i'}} (B) \mid D_{G_i} \neq D_{G_{i'}} \right] P\left\{ D_{G_i} \ne D_{G_{i'}}\mid G_i \ne G_{i'}\right\}P\left\{ G_i \ne G_{i'}\right\}\nonumber \\
& \quad = \quad E\left[ F_{G_i}(A) F_{G_i}(B)\right] P(\boldsymbol{X}_i, \boldsymbol{X}_{i'}) \nonumber \\ 
& \quad \quad + \quad E\left[F_{G_i} (A) F_{G_i } (B)\right]\frac{1}{1+\alpha} \left(1-P(\boldsymbol{X}_i, \boldsymbol{X}_{i'})\right)  \nonumber \\
& \quad \quad + \quad E\left[F_{G_i}(A) F_{G_{i'} } (B)\mid D_{G_i} \neq D_{G_{i'}}\right]\frac{\alpha}{1+\alpha}\left(1-P(\boldsymbol{X}_i, \boldsymbol{X}_{i'})\right)  
\label{eq:cov_exp}
\end{align}
In the last equality, we have again used the results from \eqref{eq:dgdg}.  

We now evaluate the expectation when both terms correspond to the same group $G_i=k$.
\begin{align}
  E\left[F_{G_i} (A) F_{G_i}(B)\right] &=  E\left[\left( \sum_{h= 1}^\infty \nu_{kh}\delta_{\vartheta_h} (A) \right)\left(\sum_{h= 1}^\infty \nu_{kh}\delta_{\vartheta_h} (B) \right)\right] \nonumber \\
  & = E\left[\sum_{h= 1}^\infty \nu_{kh}^2\delta_{\vartheta_h} (A\cap B)\right]+ E\left[\sum_{h= 1}^\infty \sum_{s\neq h} \nu_{kh} \nu_{ks} \delta_{\vartheta_h} (A) \delta_{\vartheta_s} (B) \right] \nonumber \\
  &= E\left[\sum_{h= 1}^\infty \nu_{kh}^2\right] \mathcal{G}_0(A\cap B) + \left[1-E\left(\sum_{h= 1}^\infty  \nu_{kh}^2\right)   \right]\mathcal{G}_0(A)\mathcal{G}_0(B)\nonumber \\
  & = \frac{1}{1+\beta} \mathcal{G}_0(A\cap B) + \frac{\beta}{1+\beta}\mathcal{G}_0(A)\mathcal{G}_0(B),
  \label{eq:EFGEFG}
\end{align}
where $\mathcal{G}_0$ is the base distribution of the atoms $\vartheta$ and $ E\left[\sum_{h= 1}^\infty \nu_{kh}^2\right] = \frac{1}{1+\beta}$ from the same argument as $ E\left[\sum_{k= 1}^\infty \rho_{k}^2\right] = \frac{1}{1+\alpha}$ in \eqref{eq:dgdg}.
To prove that 
$E\!\left[\sum_{h=1}^\infty \sum_{s \neq h} \nu_{kh} \nu_{ks}\right] 
= 1 - E\!\left[\sum_{h=1}^\infty \nu_{kh}^2\right]$ in the above equation, consider the square of the sum:  
$$
\left(\sum_{h  =1}^\infty \nu_{kh}\right)^2 
= \sum_{h  =1}^\infty \nu_{kh}^2 
+ \sum_{h  =1}^\infty \sum_{s \neq h} \nu_{kh} \nu_{ks}.
$$
Taking expectations on both sides yields  
\begin{align*}
E\!\left[\sum_{h  =1}^\infty \sum_{s \neq h} \nu_{kh} \nu_{ks}\right] 
&= E\!\left[\left(\sum_{h  =1}^\infty \nu_{kh}\right)^2\right] 
- E\!\left[\sum_{h  =1}^\infty \nu_{kh}^2\right] 
= E\left[(1)^2\right] - E\!\left[\sum_{h  =1}^\infty \nu_{kh}^2\right] \\
&= 1 - \frac{1}{1+\beta} = \frac{\beta}{1+\beta},
\end{align*}
using the fact that $\sum_{h  =1}^\infty \nu_{kh} = 1$ almost surely.

When the groups differ with $G_i=k$ and $G_{i'}=k'$, $E\left[F_{G_i}(A) F_{G_{i'}} (B)\mid D_{G_i} \neq D_{G_{i'}}\right]$ can be computed as follows:
\begin{align}
  & E\left[F_{k}(A) F_{k'}(B) \right] =  E\left[ \left(\sum_{s= 1}^\infty \nu_{ks}\delta_{\vartheta_s} (A) \right) \left(\sum_{h= 1}^\infty \nu_{k'h}\delta_{\vartheta_h} (B) \right) \right] \nonumber \\
  & \quad = \quad E\left[\sum_{s= 1}^\infty \nu_{ks} \nu_{k's}\delta_{\vartheta_s} (A\cap B)\right] +E\left[ \sum_{s\neq h} \nu_{ks} \nu_{k'h} \delta_{\vartheta_s} (A) \delta_{\vartheta_h} (B) \right] \nonumber \\
  & \quad = \quad  E\left[\sum_{s= 1}^\infty \nu_{ks} \nu_{k's}\right] \mathcal{G}_0(A\cap B) + E\left[\sum_{s\neq h}^\infty \nu_{ks} \nu_{k'h}\right]  \mathcal{G}_0(A)\mathcal{G}_0(B) \nonumber \\
  & \quad = \quad \sum_{s=1}^\infty E\left[ \nu_{ks}\right] E\left[\nu_{k's}\right] \mathcal{G}_0(A\cap B) + \left(1- \sum_{s =1}^\infty \left\{E[ \nu_{ks}]\right\}^2\right) \mathcal{G}_0(A)\mathcal{G}_0(B) \nonumber \\
    & \quad = \quad \sum_{s=1}^\infty \left(E\left[ \nu_{ks}\right]\right)^2 \ \mathcal{G}_0(A\cap B) + \left(1- \sum_{s =1}^\infty \left\{E[ \nu_{ks}]\right\}^2\right)  \mathcal{G}_0(A)\mathcal{G}_0(B)
  \nonumber \\
& \quad = \quad \frac{1}{1+2\beta} \mathcal{G}_0(A\cap B) + \frac{2\beta}{1+2\beta} \mathcal{G}_0(A)\mathcal{G}_0(B)
    \label{eq:2nd_exp_res}
\end{align}
The final step uses the result $\sum_{s=1}^\infty \left(E\left[ \nu_{ks}\right]\right)^2  = \frac{1}{1+2\beta}$, which is justified as follows:
\begin{align}
  & \sum_{h = 1}^\infty\left[  E\nu_{kh}\right]^2 = \sum_{h = 1}^\infty \left[E\left\{q_{kh}\prod_{s<h} (1-q_{ks})\right\}\right]^2 = \sum_{h = 1}^\infty \left[ E\left[q_{kh}\right]\prod_{s<h} E\left[(1-q_{ks})\right]\right]^2 \nonumber \\
  & \quad = \quad \sum_{h = 1}^\infty \left[ \frac{1}{1+\beta} \prod_{s<h}\frac{\beta(1, 1+\beta)}{\beta(1,\beta)}\right]^2 
  = \sum_{h = 1}^\infty \left[ \frac{1}{1+\beta} \prod_{s<h} \frac{\beta}{1+\beta}\right]^2
  \nonumber \\
 & \quad = \quad \sum_{h = 1}^\infty\left[\frac{1}{1+\beta}  \left[\frac{\beta}{1+\beta}\right]^{h-1}\right]^2 
  = \frac{1}{[1+\beta]^2}\sum_{h = 0}^\infty \left[\frac{\beta}{1+\beta}\right]^{2h} = \frac{1}{1+2\beta}. \nonumber
\end{align}

To get the expectation of $F_{G_i}(A)$, we see
\begin{align}
    E[F_{G_i}(A)] &= E\left[\sum_{h =1}^\infty \nu_{kh}\delta_{\vartheta_h} (A)\right]  = \sum_{h = 1}^\infty E\left[\nu_{kh} \right] \mathcal{G}_0 (A) = E\left[\sum_{h = 1}^\infty \nu_{kh} \right] \mathcal{G}_0 (A) \nonumber \\
    & = E\left[ 1 \right] \mathcal{G}_0 (A) = \mathcal{G}_0 (A)
    \label{eq:exp_f_A}
\end{align}
Similarly, $ E[F_{G_{i'}}(B)] = \mathcal{G}_0 (B)$.

We now combine \eqref{eq:cov_exp}, \eqref{eq:EFGEFG}, \eqref{eq:2nd_exp_res}, and \eqref{eq:exp_f_A} into \eqref{eq:cov_start_eq} to get the required claim.
\begin{align}
    &\text{Cov}\left(F_{G_i}(A), \ F_{G_{i'}}(B)\right) \nonumber \\
    & \quad = \quad E\left[ F_{G_i}(A) F_{G_i}(B)\right] P(\boldsymbol{X}_i, \boldsymbol{X}_{i'}) \nonumber \\ 
& \quad \quad + \quad E\left[F_{G_i} (A) F_{G_i } (B)\right]\frac{1}{1+\alpha} \left(1-P(\boldsymbol{X}_i, \boldsymbol{X}_{i'})\right)  \nonumber \\
& \quad \quad + \quad E\left[F_{G_i}(A) F_{G_{i'} } (B)\right]\frac{\alpha}{1+\alpha}\left(1-P(\boldsymbol{X}_i, \boldsymbol{X}_{i'})\right)  - E  \left[F_{G_i}(A)\right] E \left[F_{G_{i'}}(B)\right] \nonumber \\
    & \quad = \quad \left[ \frac{1}{1+\beta} \mathcal{G}_0(A\cap B) + \frac{\beta}{1+\beta} \mathcal{G}_0(A)\mathcal{G}_0(B) \right] P(\boldsymbol{X}_i, \boldsymbol{X}_{i'}) \nonumber \\ 
& \quad \quad + \quad \left[ \frac{1}{1+\beta} \mathcal{G}_0(A\cap B) + \frac{\beta}{1+\beta} \mathcal{G}_0(A)\mathcal{G}_0(B) \right] \frac{1}{1+\alpha} \left(1-P(\boldsymbol{X}_i, \boldsymbol{X}_{i'})\right)  \nonumber \\
& \quad \quad + \quad \left[ \frac{1}{1+2\beta} \mathcal{G}_0(A\cap B) + \frac{2\beta}{1+2\beta}\mathcal{G}_0(A)\mathcal{G}_0(B) \right]\frac{\alpha}{1+\alpha}\left(1-P(\boldsymbol{X}_i, \boldsymbol{X}_{i'})\right)  - \mathcal{G}_0(A)\mathcal{G}_0(B) \nonumber \\ 
     & \quad = \quad \left(\left[\frac{\alpha}{1+\alpha}\frac{\beta}{1+\beta}\frac{1}{1+2\beta} \right]P(\boldsymbol{X}_i, \boldsymbol{X}_{i'}) + \left[\frac{1}{1+\alpha}\frac{1}{1+\beta} + \frac{\alpha}{1+\alpha}\frac{1}{1+2\beta}\right]\right)\mathcal{G}_0 (A\cap B) \nonumber \\
 & \quad \quad - \quad \left(\left[ \frac{\alpha}{1+\alpha}\frac{\beta}{1+\beta}\frac{1}{1+2\beta}\right]  P(\boldsymbol{X}_i, \boldsymbol{X}_{i'}) + \left[\frac{1}{1+\alpha}\frac{1}{1+\beta} + \frac{\alpha}{1+\alpha}\frac{1}{1+2\beta} \right]\right) \mathcal{G}_0(A)\mathcal{G}_0(B)
 \nonumber \\
    & \quad = \quad \left(\left[\frac{\alpha}{1+\alpha}\frac{\beta}{1+\beta}\frac{1}{1+2\beta} \right]P(\boldsymbol{X}_i, \boldsymbol{X}_{i'}) + \left[\frac{1}{1+\alpha}\frac{1}{1+\beta} + \frac{\alpha}{1+\alpha}\frac{1}{1+2\beta}\right]\right) \nonumber \\
    & \quad \quad \times \quad \big[ \mathcal{G}_0 (A\cap B) - \mathcal{G}_0 (A)\mathcal{G}_0 (B)\big] .\nonumber 
\end{align}

\subsection*{Proof of (P-3)}

For the correlation in property (P-3), we apply the covariance results from (P-2) with  $A = B \in \mathcal{F}$:
\begin{align}
    &\text{Cov}\left(F_{G_i}(A), \ F_{G_{i'}}(A)\right) \nonumber \\
 &=\left(P(\boldsymbol{X}_i, \boldsymbol{X}_{i'}) \frac{\alpha}{1+\alpha}\frac{\beta}{1+\beta}\frac{1}{1+2\beta} + \left[\frac{1}{1+\alpha}\frac{1}{1+\beta}+\frac{\alpha}{1+\alpha}\frac{1}{1+2\beta}\right]\right)\mathcal{G}_0(A)\left(1-\mathcal{G}_0(A)\right). \nonumber
\end{align}

\noindent
Note that by utilizing \eqref{eq:EFGEFG} for the second moment, we can find the variance of $F_{G_i}(A)$ as
\begin{align}
    \text{Var} (F_{G_i}(A)) &= E\left[F_{G_i}(A)^2\right] - E\left[F_{G_i}(A)\right]^2 = \frac{1}{1+\beta}\mathcal{G}_0(A) + \frac{\beta}{1+\beta}\mathcal{G}_0(A)^2 - \mathcal{G}_0(A)^2 \nonumber \\
    & \quad = \quad \frac{1}{1+\beta}\mathcal{G}_0(A)(1-\mathcal{G}_0(A)).\nonumber
\end{align}

This then provides
\begin{flalign}
    \text{Corr}\left(F_{G_i}(A), \ F_{G_{i'}}(A)\right) &= \frac{\text{Cov} (F_i(A), F_{i'}(A))}{\sqrt{\text{Var}(F_{G_i}(A))\cdot\text{Var}(F_{G_{i'}}(A))}} \nonumber \\
    & = \frac{P(\boldsymbol{X}_i, \boldsymbol{X}_{i'}) \frac{\alpha}{1+\alpha}\frac{\beta}{1+\beta}\frac{1}{1+2\beta} + \left[\frac{1}{1+\alpha}\frac{1}{1+\beta}+\frac{\alpha}{1+\alpha}\frac{1}{1+2\beta}\right]}{\frac{1}{1+\beta}} \nonumber \\
    & = P(\boldsymbol{X}_i, \boldsymbol{X}_{i'}) \frac{\alpha}{1+\alpha}\frac{\beta}{1+2\beta} + \left[\frac{1}{1+\alpha}+\frac{\alpha}{1+\alpha}\frac{1+\beta}{1+2\beta}\right]
    \nonumber \\
    & = P(\boldsymbol{X}_i, \boldsymbol{X}_{i'})\cdot \frac{\alpha}{1+\alpha}\cdot\frac{\beta}{1+2\beta} + \left[1-\frac{\beta}{1+2\beta}\cdot\frac{\alpha}{1+\alpha}\right] \nonumber \\
    & = 1-\left[1- P(\boldsymbol{X}_i, \boldsymbol{X}_{i'})\right]\frac{\alpha}{1+\alpha}\cdot\frac{\beta}{1+2\beta}. \nonumber
\end{flalign}

\subsection*{Proof of (P-4)}
We now consider the property for the observation level clustering probability.
\begin{align}
    & P\{C_i = C_{i'}\} = E\left[P\{C_i = C_{i'}\mid G_i, G_{i'}\}\right] \nonumber \\
    & \quad = \quad E\left[ P\{C_i = C_{i'}\mid G_i = G_{i'}\} \right] P\{G_i = G_{i'}\} \nonumber \\
    & \quad \quad + \quad E\left[ P\{C_i = C_{i'}\mid G_i \neq G_{i'}\} \right]P\{G_i \neq G_{i'}\}
\label{eq:clust_prob_ii'}
\end{align}
To obtain the first expectation, we note
\begin{align}
   & E\left[ P\{C_i = C_{i'}\mid G_i = G_{i'}\} \right] \nonumber \\
 & \quad = \quad E\left[ P\{C_i = C_{i'}\mid D_{G_i} = D_{G_{i'}}\} P\left\{ D_{G_i} = D_{G_{i'}}| G_i = G_{i'}\right\}\right] \nonumber \\
  & \quad = \quad E \left[\sum_{h = 1}^\infty P\left\{C_i= C_{i'} = h \mid D_{G_i} = D_{G_{i'}}\right\}   \right] 
  =
  E\left[ \sum_{h = 1}^\infty \nu_{kh}^2 \right]   = \frac{1}{1+\beta}.
    \label{eq:exp_nu2_kh}
\end{align}

The second term is given by 
\begin{align}
    &E\left[P\{C_i = C_{i'} \mid G_i \neq G_{i'}\}  \right] \nonumber \\
 & \quad = \quad E\left[ P\{C_i = C_{i'}\mid D_{G_i} = D_{G_{i'}}, G_i \neq G_{i'}\} \right]P\left\{ D_g = D_{g'} \mid G_i \neq G_{i'}\right\} + \nonumber \\
    & \quad \quad + \quad E\left[ P\{C_i = C_{i'}\mid D_{G_i} \neq D_{G_{i'}}, G_i \neq G_{i'}\} \right]P\left\{ D_g \neq D_{g'} \mid G_i \neq G_{i'}\right\} \nonumber \\
  & \quad = \quad \sum_{h = 1}^\infty E\left[\nu_{kh}^2 \right] \frac{1}{1+\alpha} + \sum_{h=1}^\infty E\left[\nu_{kh}\ \nu_{k'h}\right]\left[ 1-\frac{1}{1+\alpha}\right] \nonumber \\
  & \quad = \quad \frac{1}{1+\beta} \frac{1}{1+\alpha} +\frac{\alpha}{1+\alpha}\sum_{h = 1}^\infty \left[E(\nu_{kh}) \right]^2 \nonumber \\
  & \quad = \quad \frac{1}{1+\beta} \frac{1}{1+\alpha} +\frac{\alpha}{1+\alpha}\frac{1}{1+2\beta} 
    \label{eq:exp_nu_kh_nu_kh_exp}
\end{align}
Substituting \eqref{eq:exp_nu2_kh} and \eqref{eq:exp_nu_kh_nu_kh_exp}  back into equation (\ref{eq:clust_prob_ii'}), we get the required result:
\begin{align*}
   P\{C_i = C_{i'}\} &= \frac{1}{1+\beta}P(\mathbf X_i, \mathbf X_{i'}) 
   + \left[ \frac{1}{1+\beta}\frac{1}{1+\alpha}
   + \frac{\alpha}{1+\alpha}\frac{1}{1+2\beta} \right]
     (1-P(\mathbf X_i,\mathbf X_{i'})) \\
   &= \frac{1}{1+\alpha}\left[ \frac{1}{1+\beta}
   + \alpha\frac{1}{1+2\beta}\right]
   + \left[ \frac{\alpha}{1+\alpha}\frac{\beta}{1+\beta}
   \frac{1}{1+2\beta}\right] P(\mathbf X_i,\mathbf X_{i'}).
\end{align*}


\section{MCMC Sampling Algorithm Details}
\label{S3_MCMC_alogorithm}
In this section, we provide additional details regarding our MCMC sampling beyond what was described in Section 4 of the main manuscript.

As discussed previously, we use a truncation base stick-breaking representation of the infinite summations in the DPs, following the approach of 
\citet{ishwaran2002dirichlet, ishwaran2001gibbs}.
That is, we select  appropriate upper bounds, denoted as $H$ and $K$, for the infinite sums in equations (3) and (4). 
The upper bounds for $h$ and $k$ were set to at least five more the 95th percentile of the number of non-empty OC and DC, as determined from preliminary MCMC runs.

\subsection*{Pyramid tree prior }
In Section 3.2, we introduce the iterative process for generating a pyramid tree $\mathcal{T}$.  
As noted there, the pyramid tree prior $\pi(\mathcal{T})$ is determined by $P_{split} (\lambda)$ which determines how deep the pyramid goes and $P_{rule}(j_\lambda, \eta_\lambda)$ which determines the splitting rule at level $\lambda$.  As the assignment of splitting rules depends on $\boldsymbol{X}$, the prior $\pi(\mathcal{T})$ will necessarily depend on the predictor space $\boldsymbol{\mathcal{X}}$ \citep{buntine1992learning}. Based on the iterative process, the distribution of pyramid tree prior can be written as follows                        
 \begin{equation}
 \begin{split}
 \pi(\mathcal{T}) & = \left[ \prod_{\lambda =1}^{d} P_{split} (\lambda)\right]  \left(\frac{1}{P}\right)^d \left(1-P_{split}(d+1)\right)\\
 & \quad \times \quad \prod_{\lambda = 1}^d \left[Q_{q_2} (j_\lambda) - Q_{q_1} (j_\lambda)\right]^{-1}. 
 \end{split}
 \label{eq:treeprior}
 \end{equation}
 As we restrict to trees with a maximum depth of $\mathcal{D}$, we have $P_{split}(\mathcal{D} + 1) = 0$.

\subsection*{MCMC Sampling Algorithm}

 After initialization, the MCMC sampling algorithm proceeds through the following steps. We describe the steps for the $m^{th}$ iteration.

 \begin{enumerate}

 \item Sample $\mathcal{T}$ using Metropolis-Hastings  by following the steps below:
\begin{itemize}
    \item Generate a proposal tree $\mathcal{T}^\prime$ from the proposal distribution $\mathfrak{u}(\mathcal{T}^m, \mathcal{T}^\prime)$ by randomly selecting one of the four step types and generating $\mathcal{T}'$ accordingly. See Section 4.1 in the main text.
        
    \item Calculate the MH probability
     $$\alpha (\mathcal{T}^m, \mathcal{T}^\prime) = \text{min} \left\{1, \frac{\mathfrak{u}(\mathcal{T}^\prime, \mathcal{T}^m) L(\mathcal{T}^\prime\mid\cdots) \pi(\mathcal{T}^\prime)}{\mathfrak{u}(\mathcal{T}^m, \mathcal{T}^\prime) L(\mathcal{T}^m\mid\cdots) \pi (\mathcal{T}^m)} \right\},$$
     where the likelihood term $L(\mathcal{T}\mid\cdots) $ is given by (8) of the main manuscript.
        
     \item Accept the proposal tree $\mathcal{T}^\prime$ as the new tree $\mathcal{T}^{m+1}$ with probability $\alpha (\mathcal{T}^m, \mathcal{T}^\prime)$. Otherwise, stay at the current tree, $\mathcal{T}^{m+1} = \mathcal{T}^m$.
 \end{itemize}           

 \item For $ g = 1, \ldots, G$, sample
 $D_g \sim$ Multinomial $(p_1,  \ldots, p_K)$, where 
 $$p_k = P(D_g = k\mid\cdots) \propto \rho_k \prod_h \nu_{kh}^{\sum_{i = 1}^n I(G(\boldsymbol{X}_i) = g_i, C_i = h)}.$$
      
    \item For $i=1, \ldots, n$, sample $C_i \sim$ Multinomial $( 
      \mathfrak{p}_1, \ldots, \mathfrak{p}_H)$, where
    $$\mathfrak{p}_h = P(C_i = h \mid \cdots) \propto \nu_{G(\boldsymbol{X}_i)h} f(y_i\mid\vartheta_h, \phi).$$
    
  \item For $k = 1, \ldots, K - 1,$ sample 
    $$U_k \sim Beta  \left(1 + \sum_{g=1}^G I(D_g = k), \alpha + \sum_{g=1}^G I(D_g>k)\right).$$ 
    After sampling $U_1, \ldots, U_k$, compute  $\rho_k = U_k \prod_{l<k} \left(1-U_l\right)$. Under the truncation approximation, we have fixed $U_K = 1$ so that $\sum_{k = 1}^K \rho_k = 1$.

    \item For each $k = 1, \ldots, K$ and $h = 1, \ldots, H - 1$, sample 
    $$q_{kh} \sim Beta\left(1 + \sum_{i = 1}^n I (D_{g} = k, C_{i} = h), \beta + \sum_{i=1}^n I(D_{g} = k, C_{i}>h)\right).$$
    After sampling $q_{k1}, \ldots, q_{k(H-1)}$, compute $\nu_{kh} = q_{kh} \prod_{s<h} (1-q_{ks})$. For each $k$, we have fixed $q_{kH} = 1$ such that $\sum_{h = 1}^H \nu_{kh} = 1$.

    \item For $h = 1, \ldots, H $,  sample
     $$\vartheta_h \sim \left[\prod_{i:C_i = h} f(y_i\mid\vartheta_h, \phi)\right] \mathcal{G}_0(\vartheta_h).$$
 Depending on the choice of the response likelihood and the prior, this step may or may not be conjugate. For an empty cluster, $\vartheta_h$ is drawn directly from the prior distribution $\mathcal{G}_0(\vartheta_h)$. If $\mathcal{G}_0$ has hyperparameters, they should also be sampled based on $\vartheta_1, \ldots, \vartheta_H$ 

  \item Sample any global parameters from
   $$\phi\sim \left[\prod_{i=1 }^n f(y_{i}\mid \theta_{C_i},\phi)\right] p(\phi).$$
     \item Sample the concentration parameters 
    $$\alpha\sim Gamma \left((K-1) + a, b - \sum_{k=1}^{K-1} log(1-U_k) \right),$$
 $$\beta \sim Gamma\left( c + K(H-1), d -  \sum_{k=1}^K \sum_{h = 1}^{H-1} log(1-q_{kh})\right).$$
 \end{enumerate}

\section{Additional Details from Simulation Study of Section 5}
\label{S4_additional_Sim1}

\begin{figure}[t]
\centering
\begin{subfigure}{11cm}
  \includegraphics[width=\linewidth]{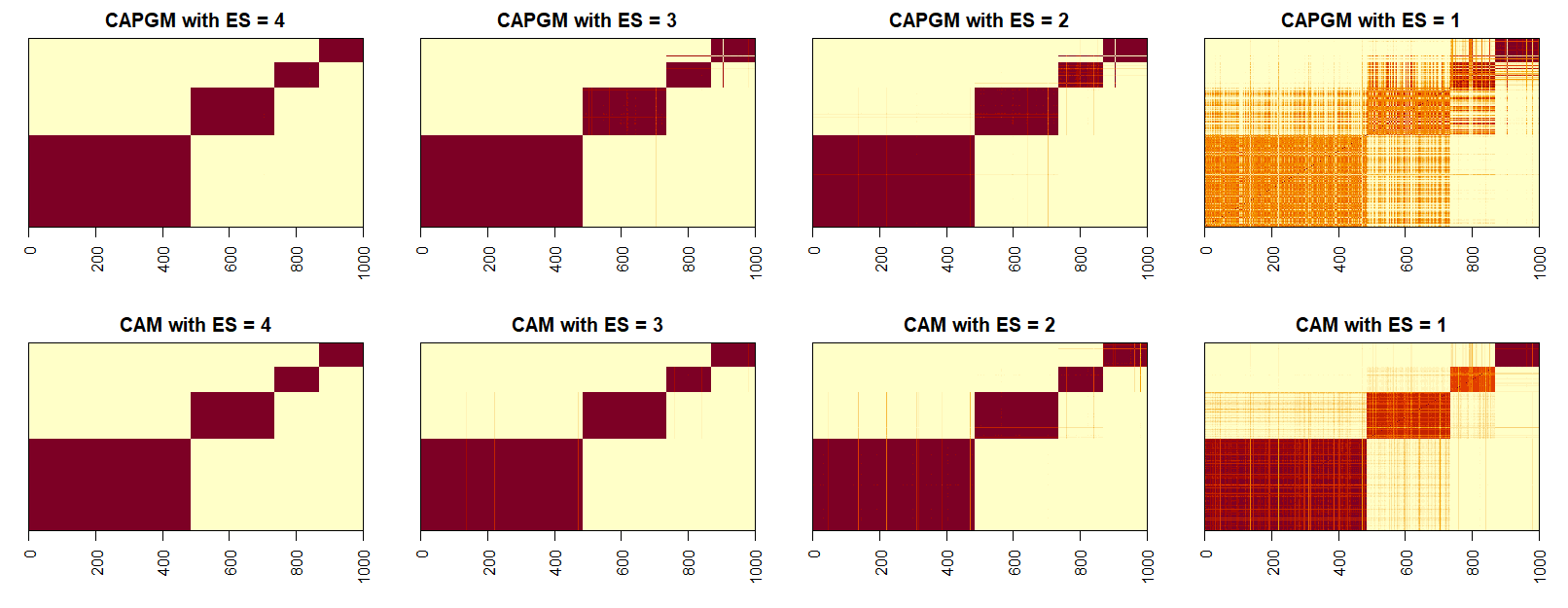}
\end{subfigure} 
\begin{subfigure}{11cm}
  \includegraphics[width=\linewidth]{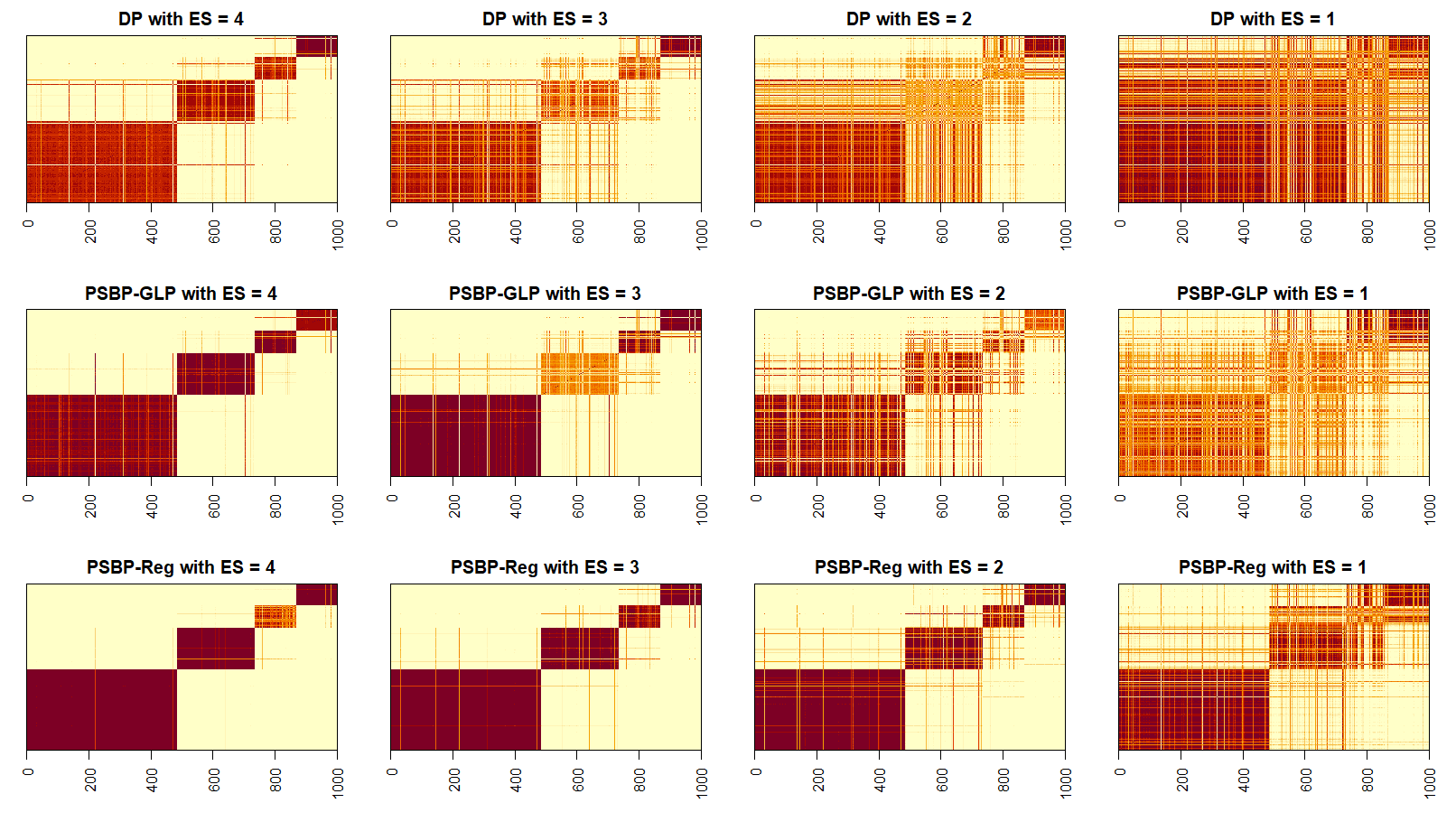}
\end{subfigure}
\caption{Comparison in the pairwise co-clustering probabilities for observations belonging to the same OC across models. This figure produce based a single dataset.}
\label{fig:sim1:oc_heatmap}
\end{figure}

Here, we provide a few additional details from the simulation study discussed in Section 5.
For a visual representation of how the estimated OC structures vary between methods and across the $\Delta$ effect sizes, we display heatmaps of  the co-clustering probability that patients $i$ and $i^\prime$ belong to the same OC, in Figure \ref{fig:sim1:oc_heatmap}.  These heatmaps are produced based on the MCMC samples for a single dataset, 
and high co-clustering probabilities are depicted with deep colors, indicating strong cluster cohesion.  
Observations are ordered such that 1--500, 501--750, 751--875, and 876--1000 are in the same true clusters.

The top two rows of Figure \ref{fig:sim1:oc_heatmap} illustrate four distinct mixture components across different effect sizes (ES) for CAPGM and CAM.  Clearly, both models obtain almost exact recovery for $\Delta=4,3,2$.  For the smaller effect size, the co-clustering probabilities for CAPGM are materially worse than CAM, although they do still generally recover the true structure. Remember that  CAM uses the true predictor grouping structure, while CAPGM must estimate the predictor groups in the presence of significant overlap in the response variable $Y$, leading to a more challenging situation.
With ES 4, ES 3, and ES 2, four distinct clustering blocks are evident in both regression and LGP versions of PSBP. However, for ES 1, the last two components are less distinguishable, with significant overlap, particularly in the LGP version. For DP, there is a significant overlap between clusters, even for the larger effect sizes.

\begin{figure}[t]
    \centering
    \includegraphics[width=13cm]{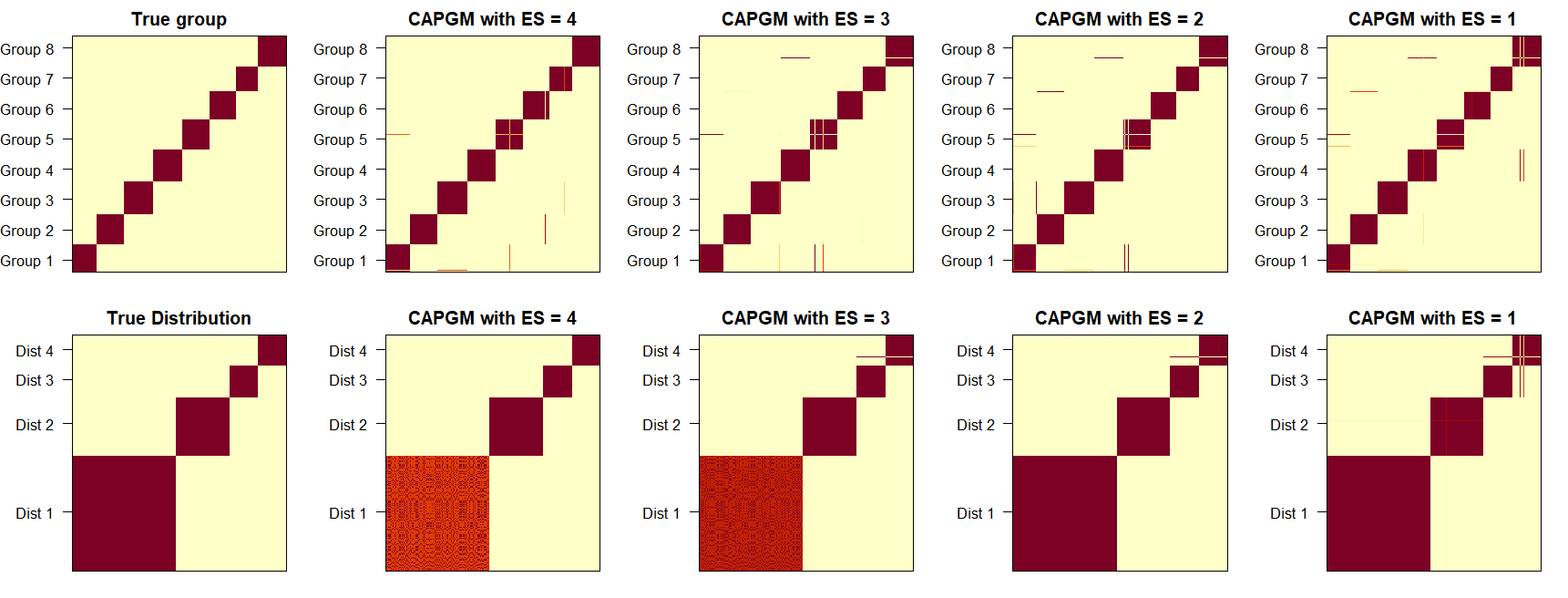}
    \caption{Grouping and distributional structure of the observation based on the pairwise probabilities (CAPGM).} 
    \label{fig:sim1:dc_heatmap}
\end{figure}

We also assess the accuracy of CAPGM in assigning observations to correct (latent) predictor groups and distributional clusters across the MCMC samples using pairwise co-clustering probabilities, as shown in Figure~\ref{fig:sim1:dc_heatmap}. The leftmost heatmaps shows the true predictor group structure (top) and true DC structure (bottom) of the observations. The inferred groups and DCs structures from the MCMC samples closely align with their true counterparts across all ESs. Furthermore, the inferred DC structures closely match the CAPGM's OC assignments (see the top row of Figure~\ref{fig:sim1:oc_heatmap}), which is expected as in our data-generating strategy, we considered $D_{G(\boldsymbol{X}_i)} = C_i$.


\section{Simulation Study II}
\label{S5_Simulation_Study_II}

In this section, we consider an additional simulation study to further investigate the performance of our methodology.  In Section 5 of the manuscript, we considered a case where the DCs correspond exactly to the OCs (i.e., $D_{G(\boldsymbol{X}_i)}=C_i$).  In this setting, we follow the general model framework, where distributional clusters have members within multiple observational clusters.

\begin{table}[t]
\caption{Data generation strategy}
\centering
\begin{tabular}{cccccc}
\hline
\multicolumn{2}{c}{Distribution 1 (Group 1)} & \multicolumn{2}{c}{Distribution 2 (Group 2 \& 4)} & \multicolumn{2}{c}{Distribution 3 (Group 3)} \\
\cmidrule(lr){1-2} \cmidrule(lr){3-4} \cmidrule(lr){5-6}
\makecell{Cluster \\ Probability $\pi_{gh}$} & \multirow{2}{*}{$\vartheta_h$} & \makecell{Cluster \\ Probability $\pi_{gh}$} & \multirow{2}{*}{$\vartheta_h$} & \makecell{Cluster \\ Probability $\pi_{gh}$} & \multirow{2}{*}{$\vartheta_h$} 
\\
\hline
0.40 & $\vartheta_1$  & 0.00 & $\vartheta_1$ & 0.00  & $\vartheta_1$ \\
0.40 & $\vartheta_2$  & 0.50 & $\vartheta_2$  & 0.75 & $\vartheta_2$ \\
0.20 & $\vartheta_3$  & 0.30 & $\vartheta_3$  & 0.10 & $\vartheta_3$ \\
0.00 & $\vartheta_4$  & 0.20 & $\vartheta_4$  & 0.15 & $\vartheta_4$ \\
\hline
\end{tabular}
\label{tab:sim2:datageneration}
\end{table}

Similar to Simulation Study I, we use $n = 1000$, $P = 20$, and $X_{ij} \sim \text{Uniform}(-0.5, 0.5)$. The data generating structure is nested, as illustrated in Table \ref{tab:sim2:datageneration}, featuring four latent predictor groups, three distributional level clusters, and four observational level clusters. Subjects were randomly allocated to four groups with equal probability, determined by the predictors $X_1$ and $X_2$ according to
 \begin{equation*}
\label{eq:sim1:TruegroupX}
    G(\boldsymbol{X}_i) =
    \begin{cases}
        \begin{aligned}
            &1  \ \ \ \  \text{if } -0.5 \leq X_{i1} < 0 \text{ and } -0.5 \leq X_{i2} < 0 \\
            &2  \ \ \ \  \text{if }  \ \ \ 0 \leq X_{i1} \leq 0.5 \ \text{ and }  - 0.5 \leq X_{i2} < 0 \\
            &3  \ \ \ \  \text{if } - 0.5 \leq X_{i1} < 0 \text{ and } \ \ \ \ 0 < X_{i2} \leq 0.5\\
            &4 \ \ \ \  \text{if } \ \ \ \ 0 \leq X_{i1} \leq 0.5  \text{ and } \ \ \ \ 0 \leq X_{i2}  \leq 0.5.
        \end{aligned}
    \end{cases}
\end{equation*} 
Observations from predictor groups 2 and 4 ($X_{i1}\ge 0$) fall into distribution cluster 2, while observations from groups 1 and 3 come from distribution clusters 1 and 3, respectively. 
As in the first simulation the atoms $\{\vartheta_1,   \vartheta_2,  \vartheta_3,   \vartheta_4\}=\{ -\Delta,0,\Delta,2\Delta\}$, and the DC-specific probabilities are given in Table \ref{tab:sim2:datageneration}.
With this set up, we generated 200 independent training datasets and 1 test dataset for ESs $\Delta=2,3,4$.

\begin{table}[t]
\setlength{\tabcolsep}{.5pt}

\caption{RMSPE and LPDS from Simulation Study II. Mean and SE in the parenthesis are reported based on 200 datasets. 
}
\centering
\resizebox{\textwidth}{!}{%
\begin{tabular}{lccccccccc}

\hline
&  \multicolumn{3}{c}{Effect Size $\Delta=4$} &  \multicolumn{3}{c}{Effect Size  $\Delta=3$} &  \multicolumn{3}{c}{Effect Size  $\Delta=2$}\\
\cmidrule(lr){2-4} \cmidrule(lr){5-7} \cmidrule(lr){8-10} 

&  w/in Sample   & \multicolumn{2}{c}{Out of sample} &  w/in Sample   & \multicolumn{2}{c}{Out of sample}
&  w/in Sample   & \multicolumn{2}{c}{Out of sample} \\
\cmidrule(lr){2-2} \cmidrule(lr){3-4} \cmidrule(lr){5-5} \cmidrule(lr){6-7} \cmidrule(lr){8-8} \cmidrule(lr){9-10} 

 Method     &   RMSPE & RMSPE & LPDS &  RMSPE  & RMSPE & LPDS & RMSPE & RMSPE & LPDS \\ \hline
 CAPGM              & 3.20 (0.00) & 3.10 (0.00 )& -4627 (1.68) & 2.49 (0.00) & 2.41 (0.00) & -4360 (0.79) & 1.83 (0.00) & 1.81 (0.00) & -3983 (2.62)\\
 ${X_1 \& X_2}$ & 3.20 (0.01) & 3.10 (0.00) & -4622 (0.81) &   \\
 \hline
  CAM               & 3.20 (0.00) & 3.10 (0.00) & -4609 (0.56) & 2.49 (0.00) & 2.41 (0.00) & -4347 (0.51) & 1.83 (0.00) & 1.81 (0.00) & -3963 (0.70)  \\
  \hline
  DP                & 3.53 (0.01) & 3.43 (0.00) & -4975 (0.45) & 2.73 (0.00) & 2.64 (0.00) & -4651 (0.49) & 1.97 (0.00) & 1.94 (0.00) & -4139 (0.39)  \\
  \hline
  PSBP-LGP          & 2.95 (0.00) & 3.25 (0.00) & -5211 (1.89) & 2.30 (0.00) & 2.52 (0.00) & -4859 (1.59) & 1.71 (0.00) & 1.87 (0.00) & -4269 (2.59)   \\ 
  ${X_1 \& X_2}$& 3.11 (0.01) & 3.14 (0.00) &-4795 (1.04)  \\
  \hline
  PSBP-Reg          & 3.18 (0.01) & 3.23 (0.00) & -4996 (4.72) & 2.46 (0.00) & 2.51 (0.00) & -4695 (3.73) & 1.79 (0.01) & 1.88 (0.00) & -4182 (5.10)  \\   
  ${X_1 \& X_2}$& 3.29 (0.01) & 3.18 (0.00) & -4742 (0.84)  \\

\hline
\end{tabular}
}
\label{tab:sim2_pred}
\end{table}
 
Similarly to simulation study I, we investigate the root mean squared prediction error (RMSPE) for both the training (within) sample and the test (out of) sample.  To measure the accuracy of density recovery, we measure the  log predictive density score (LPDS) for the test data in Table~\ref{tab:sim2_pred}.
We first note that CAPGM and CAM yield almost exactly the same RMSPE performance, even though CAM has the extra information of the true group structure; CAPGM has slightly worse LPDS than CAM but remains competitive.
As in the prior investigation, the PSBP methods yield better performance for the within sample prediction, but worse performance using the withheld test data.  This again indicates that these methods systematically overfit the data they are trained on.
The regression PSBP approach outperformed LGP on the test data
which is reasonable as the regression structure is closer to the data generative model.
Furthermore, with ES 4, we investigated the performance when only using the two true predictors.  In this case, the predictive accuracy of CAPGM remains consistent whether using the full set of predictors or only \( X_1 \) and \( X_2 \) for both training and test data. In contrast, PSBP-LGP and PSBP-Reg perform worse when using only \( X_1 \) and \( X_2 \) for training but yield lower RMSPE and higher log-likelihood for test data compared to using the full set of predictors.

\begin{table}[t]
\setlength{\tabcolsep}{2.5pt}
\caption{Cluster information from Simulation Study II, including ARI due to OC. Mean and standard error in the parenthesis are reported based on 200 datasets. No. OC$_{NE}$ and No. OC$_{w/1\% obs}$ represents number of non-empty OC and number of OC with at least 1\% observations, respectively.}
\centering
\resizebox{\textwidth}{!}{%
\begin{tabular}{lcccccccccccc}
\hline
&  \multicolumn{3}{c}{Effect Size 4} &  \multicolumn{3}{c}{Effect Size 3} &  \multicolumn{3}{c}{Effect Size 2} \\
\cmidrule(lr){2-4} \cmidrule(lr){5-7} \cmidrule(lr){8-10}

 Method   & No. OC$_{NE}$ & No. OC$_{w/1\% obs}$ & ARI-OC & No. OC$_{NE}$ & No. OC$_{w/1\% obs}$ & ARI-OC & No. OC$_{NE}$ & No. OC$_{w/1\% obs}$ & ARI-OC  \\ \hline
CAPGM               & 5.7 (0.1) & 4.5 (0.1) & 0.88 (0.01) & 5.7 (0.1) & 4.5 (0.1) & 0.73 (0.01) & 5.3 (0.1) & 4.4 (0.1) & 0.44 (0.01)\\
${X_1 \& X_2}$      & 5.4 (0.1) & 5.2 (0.1) &  0.86 (0.01) \\ \hline
  CAM               & 6.3 (0.2) & 5.0 (0.1) & 0.82 (0.01) & 6.0 (0.1) & 4.7 (0.1) & 0.71 (0.01) & 5.7 (0.2) & 4.8 (0.1) & 0.45 (0.01)\\ \hline
  DP                & 5.0 (0.1) & 4.1 (0.0) & 0.87 (0.00) & 5.6 (0.1) & 4.5 (0.1)  & 0.69 (0.00) & 3.2 (0.1) & 2.6 (0.1) & 0.34 (0.00)\\ \hline
  PSBP-LGP          & 11.0 (0.2) & 7.3 (0.1) & 0.58 (0.03)  & 10.3 (0.2) & 7.1 (0.1)& 0.53 (0.02) & 9.5 (0.3) & 6.5 (0.2) & 0.29 (0.01)\\ 
  ${X_1 \& X_2}$    & 9.1 (0.2) & 6.3 (0.1) & 0.73 (0.02) \\ \hline
  PSBP-Reg          & 16.6 (0.4)& 9.9 (0.2) & 0.56 (0.02) & 15.4 (0.3) & 9.6 (0.3) & 0.38 (0.01) & 10.6 (0.5) & 6.5 (0.3) & 0.03 (0.01)\\ 
  ${X_1 \& X_2}$    & 9.3 (0.3) & 6.7 (0.2) & 0.73 (0.01) \\

\hline
\end{tabular}
}

\label{tab:sim2_clust_info}
\end{table}

In Table \ref{tab:sim2_clust_info}, we investigate the Dahl estimates of the observational clusters across the various methods.
CAPGM and CAM produced a similar number of  clusters across different ESs and achieved the highest adjusted rand index (ARI), consistently outperforming other methods.  
For ES 4 and ES 3, CAPGM and CAM demonstrated strong alignment between true and estimated clustering with ARI values of $\{0.88, 0.82\}$ and $\{0.73, 0.71\}$, respectively. However, for ES 2, both models showed moderate clustering performance with ARI values of 0.44 and 0.45, respectively. The lower ARI for this ES is likely due to the closer proximity of atoms, causing significant overlap in the generated response variable $Y$.
While DP produced a comparable number of OCs to CAPGM and CAM for ES 4 and ES 3, it generated significantly fewer clusters for ES 2, resulting in a notably lower ARI (0.34) for ES 2. Similar to Simulation Study I, LGP and the regression approach formed a large number of OCs, 
which led to lower ARI values and poorest performance. Specifically, for ES 2, the regression approach's ARI approached zero, indicating performance that is no better than random clustering.  
When comparing the performance when models are fit to all $P=20$ predictors compared to only using the true predictors $X_1\&X_2$, there are noticeable differences observed in PSBP-LGP and PSBP-Reg while ARI is mostly unchanged for CAPGM.
This indicates that CAPGM successfully identifies the covariates associated with the response variable and clusters accordingly, whereas other methods are less capable of doing so.

\begin{figure}[t]
\centering
\begin{subfigure}{11cm}
  \includegraphics[width=\linewidth]{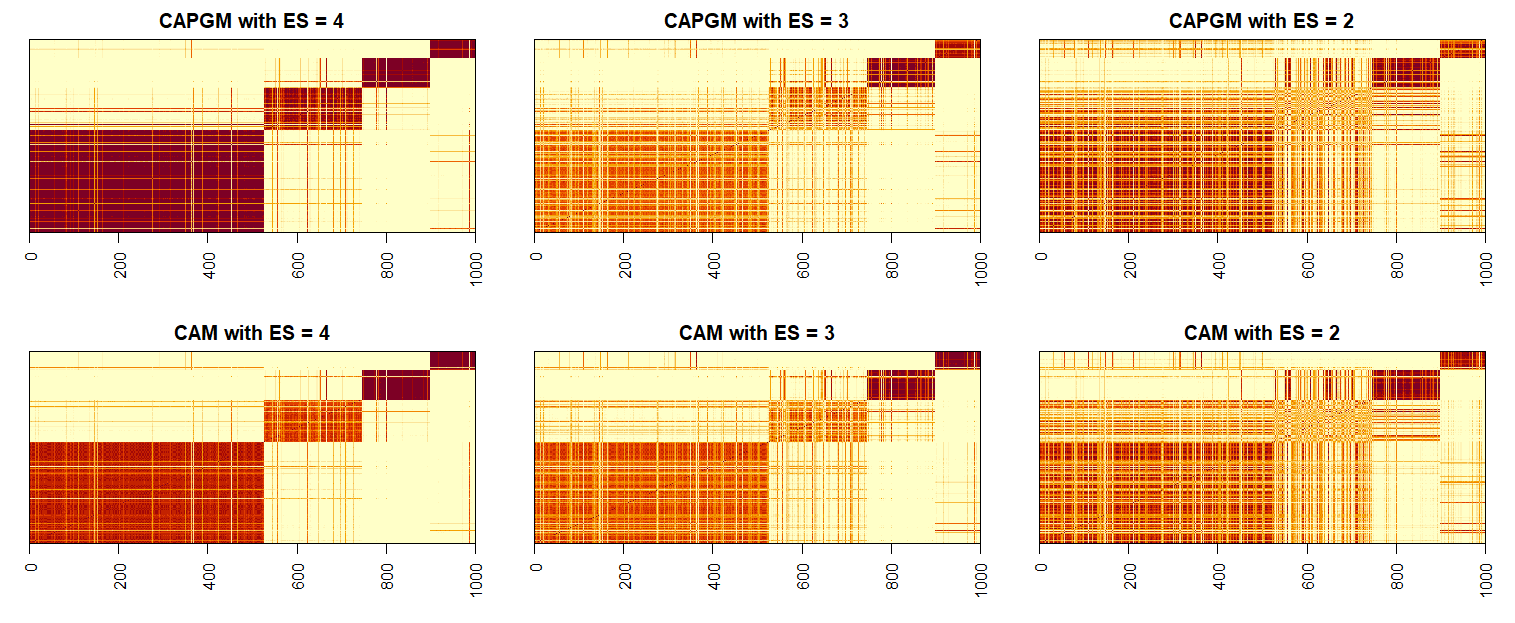}
\end{subfigure} 
\begin{subfigure}{11cm}
  \includegraphics[width=\linewidth]{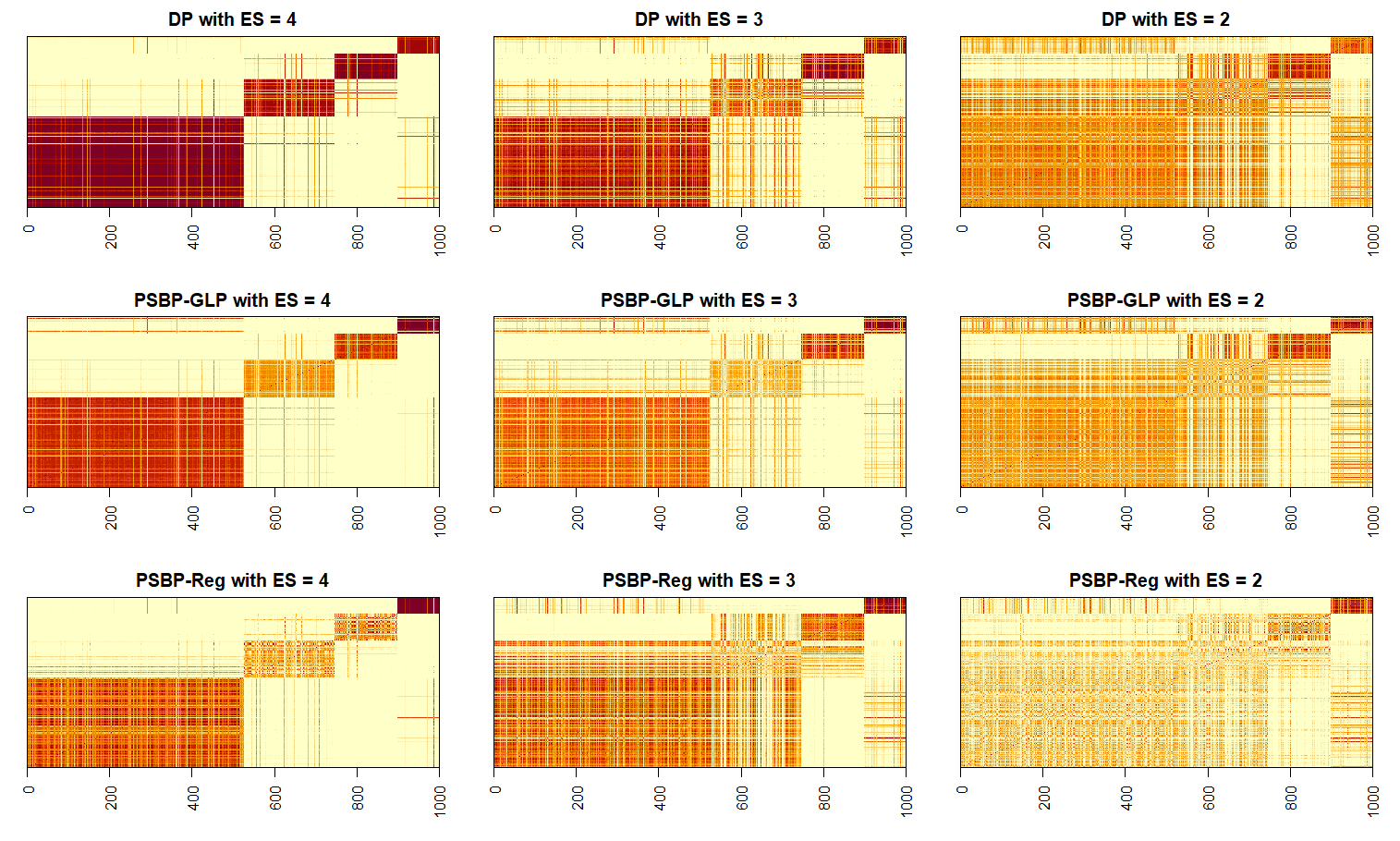}
\end{subfigure}
\caption{Comparison in the pairwise co-clustering probabilities for observations belonging to the same OC across models. This figure produce based on a single dataset.}
\label{fig:sim2:oc_heatmap}
\end{figure}
As in Section \ref{S4_additional_Sim1}, we graphically display the co-clustering probabilities for a representative dataset. 
With ES $\Delta=4$, four distinct observational cluster blocks are clearly identifiable. However, the co-clustering probability that observations $i$ and $i'$ belong to the same cluster is lower for LGP and notably lower for the regression approach. For ES 3, the second clustering block ($\vartheta_3$ = 3) is indistinguishable from the first block in the regression approach, while other methods exhibit clearly identifiable clustering components. For ES 2, the second cluster is hard to identify across all methods, with noticeably lower co-clustering probabilities for DP, LGP, and particularly the regression approach.


\begin{figure}[t]
    \centering
    \includegraphics[width=15cm]{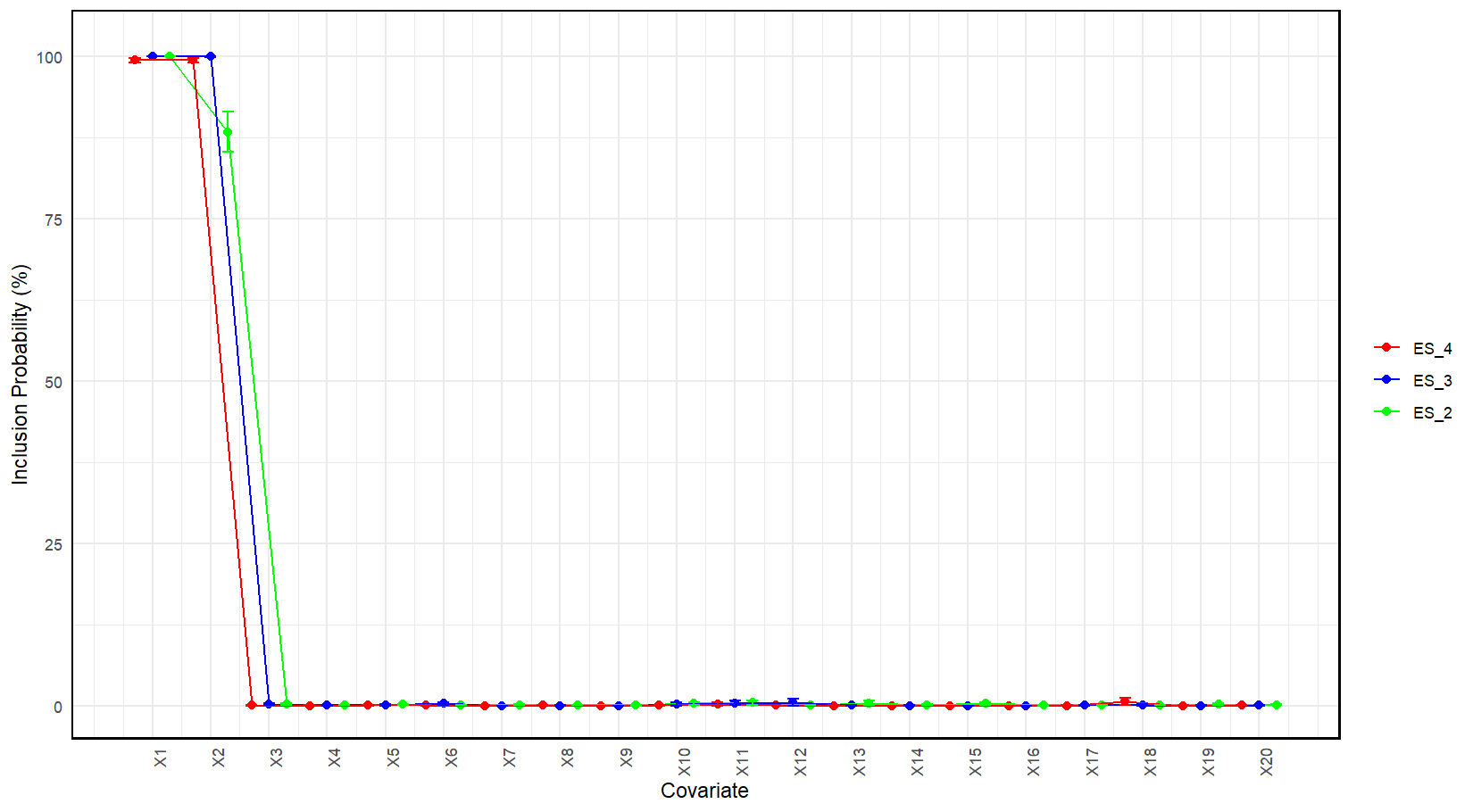}
    \caption{Average inclusion probabilities (in percentage) for each covariate with standard error bars for covariates in $\mathcal{T}$. Plot based on 200 datasets.}
    \label{fig:sim2:x_incprob}
\end{figure}

To investigate the performance of the estimation of the predictor group structure, we begin by considering the variable inclusion probabilities for the PGM model.
As illustrated in Figure \ref{fig:sim2:x_incprob}, the average inclusion probabilities of $X_1$ and $X_2$ in the pyramid tree $\mathcal{T}$ are close to 100\% for ES 4 and ES 3, with negligible standard error. The second most important covariate, $X_2$, was selected less frequently in the MCMC samples for some datasets under ES 2, resulting in a slightly lower inclusion rate. This is likely due to the closer proximity of atoms, causing considerable overlap in the generated response variable $Y$. 
The remaining covariates were not associated with clustering, and correctly have negligible inclusion probabilities. 
This indicates that the pyramid tree effectively identified and prioritized the key predictors for constructing the partitioning structure.

\begin{table}[t]
\setlength{\tabcolsep}{3pt}
\footnotesize
\caption{Some characteristics of pyramid tree and CAM DC characteristics}
\centering
\begin{tabular}{lccccccc}
\hline
&  \multicolumn{5}{c}{Pyramid tree characteristics of CAPGM} &  \multicolumn{2}{c}{CAM DC characteristics} \\
\cmidrule(lr){2-6} \cmidrule(lr){7-8}
Effect Size     & Tree Depth & No. groups  & ARI-Group   & No. DC& ARI-DC & No. DC& ARI-DC\\ \hline
4         & 2.0 (0.0) & 4.0 (0.0) & 0.98 (0.00) & 3.0 (0.0) & 0.98 (0.00) & 3.0 (0.0) & 1.00 (0.00)\\
$X_1\&X_2$& 2.0 (0.0) & 4.0 (0.0) & 0.99 (0.00) & 3.0 (0.0) & 0.99 (0.00)  \\ \hline
3         & 2.0 (0.1) & 4.1 (0.3) & 0.98 (0.01) & 3.0 (0.0) & 0.98 (0.03) & 3.0 (0.0) & 1.00 (0.00) \\ \hline
2         & 1.9 (0.0) & 3.8 (0.1) & 0.89 (0.02) & 2.8 (0.1) & 0.87 (0.02) & 2.9 (0.0) & 0.95 (0.01) \\
\hline
\end{tabular}

\label{tab:sim2:treeinfo}
\end{table}

Table \ref{tab:sim2:treeinfo} provides further insights into the pyramid tree for Simulation Study II. 
These results are based on obtaining the Dahl estimates for the distribution clusters and the predictor group clusters, and obtaining the pyramid tree from the iteration corresponding to the predictor clusters.
The average tree depth remains close to the true tree depth (2) across all ESs.
Similarly, the average number of predictor groups aligns with the true value (4). The ARI values for latent group assignments $G(\boldsymbol{X}_i)$ are close to 1, while a slightly lower performance  for ES $\Delta=2$ (ARI=0.89). The average number of DCs is also close to the true value (3), with ARI values near 1, except for ES 2 (0.87). These results are consistent with CAM for DCs.

\vfill


\section{HRS Data Details}
\label{S6_HRS_Data_Details}

Here we provide a table with details of the baseline summary statistics that are used in the HRS data example form Section 6.



\begin{table}[h]
\caption{Variable Descriptions and Descriptive Statistics for HRS Data}
\small
\centering
\begin{tabular}{cccc}
  \hline
 Variable name &  Level & Mean $\pm$ (SD)/$N (\%)$ \\ 
  \hline
  Age & &55.2 $\pm$ 5.7  \\ \hline
 \multirow{2}{*}{Sex} & Female & 5870 (53.8)\\
  & Male &5046 (46.2)\\ \hline
  \multirow{2}{*}{Race}
  & Non-white & 2181 (20.0)\\
  & White     & 8735 (80.0)\\ \hline
 \multirow{2}{*}{Ethnicity}  & Hispanic & 955 (8.7) \\
 & Non-Hispanic & 9961 (91.3) \\ \hline
  \multirow{2}{*}{Partnership}  & No & 2329 (21.3) \\
  & Yes& 8587 (78.7) \\ \hline
   BMI & & 27.2 $\pm$ 5.2  \\ \hline
 \multirow{3}{*}{Education} & Less than HS & 2838 (26.0) \\
  & GED $\&$ HS grad & 4110 (37.7) \\
  & Some College & 2095 (19.2) \\
  &College $\&$ above & 1873 (17.2) \\ \hline
  \multirow{2}{*}{Household income below the poverty line (InPov)}  &No & 1011 (9.3) \\
  &Yes & 9905 (90.7)\\ \hline

\multirow{5}{*}{Self-reported health status (SRHS)}  & Poor &  803 (7.4) \\
  & Fair & 1533 (14.0) \\
  & Good & 3041 (27.9) \\
  & Very Good & 3078 (28.2)\\
  & Excellent & 2461 (22.5) \\ \hline
\multirow{2}{*}{High BP (HIBP)}  &No & 6729 (61.6)\\
  &Yes & 4187 (38.4) \\ \hline
\multirow{2}{*}{Diabetes (Diab)}  & No & 9732 (89.2)\\
  & Yes & 1184 (10.8)\\ \hline
  \multirow{2}{*}{Cancer}  & No & 10331 (94.6) \\
  & Yes & 585 (5.4) \\ \hline
\multirow{2}{*}{Lung}  & No & 10061 (92.2)\\
  & Yes & 855 (7.8)\\ \hline
\multirow{2}{*}{Heart}  & No  & 9514 (87.2)\\
  & Yes & 1402 (12.8) \\ \hline
\multirow{2}{*}{Stroke}  & No  & 10585 (97.0) \\
  & Yes & 331 (3.0) \\ \hline
\multirow{2}{*}{Psychological condition (Psych)}  &  No  & 9751 (89.3) \\
  &  Yes & 1165 (10.7) \\ \hline
\multirow{2}{*}{Arthritis (Arthr)}  &  No  & 6784 (62.1) \\ 
  &  Yes & 4132 (37.9) \\ \hline
\end{tabular}
\end{table}

\begin{table}[t]
\small
\small
\begin{tabular}{cccc}
  \hline
 Variable name &  Level & Mean $\pm$ (SD)/$N (\%)$ \\ 
  \hline
\multirow{6}{*}{Number of Chronic Conditions (CCs)}& 0 & 3381 (31.0) \\
                                                     & 1 & 3663 (33.6) \\
                                                     & 2 & 2291 (21.0) \\
                                                     & 3 & 965 (8.8) \\
                                                     & 4 & 426 (3.9) \\
                                                     & $>4$ & 190 (1.7) \\ \hline
\multirow{2}{*}{Vigorous physical activity $3+/wk$ (VigAct)}  & No & 8798 (80.6) \\
  & Yes & 2118 (19.4) \\ \hline
 \multirow{2}{*}{Ever smoke} & No & 4020 (36.8) \\
  & Yes & 6896 (63.2) \\ \hline
\multirow{2}{*}{Smoking}  & No & 7984 (73.1) \\
  & Yes & 2932 (26.9) \\ \hline
\multirow{2}{*}{ Health problems limit work (HLTHLM)}  & No & 8649 (79.2) \\
  & Yes & 2267 (20.8) \\ \hline
  Total Asset && 212640 $\pm$ 443249\\ \hline
\multirow{2}{*}{Insurance}  &No & 1494 (13.7)\\
  &Yes & 9422 (86.3)\\ \hline

\end{tabular}
\end{table}

\clearpage
\section{Additional Results from HRS Data Example from Section 6}
\label{S7_Additional_results_HRS_Data_Ex}

In Sections 6.3 and 6.4, we consider the performance of our model on the HRS data using a training sample with size $n=1000$.  Here, we provide additional details regarding the model fit and summaries.
The estimated PGM structure based on the Dahl estimator of the group predictors was displayed in Figure 3.  
In  Table \ref{tab:data_1000:combX}, we report the top ten most probable covariate combinations in the posterior samples of the  pyramid trees.
The results indicate that most of the popular pyramid trees involve the covariate self-reported health status (SRHS). All of the selected  combinations are predominantly associated with health-related covariates, highlighting the significant role of SRHS and its interaction with other health-related factors in the partitioning process. 
Recall that the Dahl-estimated tree involved SRHS, HLTHLM and Diabetes, which was the fifth most visited structure during MCMC sampling.

\begin{table}[t]
\small
\caption{Top 10 combination of $\boldsymbol{X}$ in $\mathcal{T}$ for the moderate sample size training data.}
\centering
\begin{tabular}{lc}
\hline
Combination of $X$ & Posterior probability \\
\hline
SRHS and Diabetes            & 17.7 \\
SRHS                         & 11.7 \\
SRHS and HLTHLM              & 11.1 \\
Diabetes and  HLTHLM         & 8.9 \\
SRHS, Diabetes and HLTHLM    & 6.5 \\
Comorbidity and HLTHLM       & 4.3 \\
HLTHLM                       & 4.2 \\
Diabetes, Comorbidity and HLTHLM          & 4.1 \\
SRHS, Diabetes, and Cancer                & 2.5 \\
SRHS, Comorbidity and HLTHLM              & 2.1 \\
\hline
\end{tabular}

\label{tab:data_1000:combX}
\end{table}

\begin{table}[t]
\scriptsize
\caption{Estimated group characteristics. Group size (n) is shown under each group for moderate sample size.}
\centering
\resizebox{\textwidth}{!}{%
\begin{tabular}{lc|cccccccc}
\hline
& & Group 1 & Group 2 & Group 3 & Group 4 & Group 5 & Group 6 & Group 7 & Group 8\\ 
Covariate & Category&$n=267$ & $n=470$ & $n=127$ &  $n=21$ &  $n=50$ &  $n=20$ &  $n=44$ &  $n=1$ \\ \hline

\multirow{2}{*}{SRHS (\%)}&SRHS Low & 100 & 0 & 100 & 0 & 100 & 0 & 100 & 0 \\ 
  &SRHS High& 0 & 100 & 0 & 100 & 0 & 100 & 0 & 100 \\
\hline
\multirow{2}{*}{HLTHLM (\%)}&  No  & 100 & 100 & 0 & 0 & 100 & 100 & 0 & 0 \\ 
&  Yes & 0 & 0 & 100 & 100 & 0 & 0 & 100 & 100\\ \hline

\multirow{2}{*}{Diabetes (\%)}& No & 100 & 100 & 100 & 100 & 0 & 0 & 0 & 0 \\ 
  &Yes & 0 & 0 & 0 & 0 & 100 & 100 & 100 & 100 \\ \hline

\multirow{2}{*}{CCs (\%)}&CCs $<2$ & 70 & 87 & 36 & 71 & 24 & 30 & 5 & 0 \\
&CCs $\geq 2$ & 30 & 13 & 64 & 29 & 76 & 70 & 96 & 100 \\ \hline

\multirow{2}{*}{Sex(\%)}& Female & 51 & 57 & 54 & 62 & 52 & 60 & 48 & 0 \\
& Male & 49 & 43 & 47 & 38 & 48 & 40 & 52 & 100 \\ \hline

Age (mean) & &55 & 54 & 55 & 55 & 57 & 57 & 57 & 55 \\ \hline
\end{tabular}
}
\label{tab:group_est_1000}
\end{table}

In Figure 3, we show the pyramid tree structure that yields the 8 predictor groups for our interpretation of the model estimates.  
In Table \ref{tab:group_est_1000} we further investigate the characteristics of the eight estimated groups, particularly with respect to some of the other predictors that are not a part of $\hat{\mathcal{T}}$. 
Consistent with the results based only on the pyramid estimate,  group 2 is the healthiest, characterized by higher SRHS, no HLTHLM, no diabetes, and a lower number of CCs, while group 7 is the least healthy. Individuals in groups 1, 4, and 6 generally exhibit only one risk factor—low SRHS, HLTHLM, or diabetes. Groups 3, 5, and 8 are defined by the presence of two risks. The average age is consistent across all groups, with males and females being almost evenly distributed. 

\begin{figure}[t]
    \centering
    \includegraphics[width=13cm]{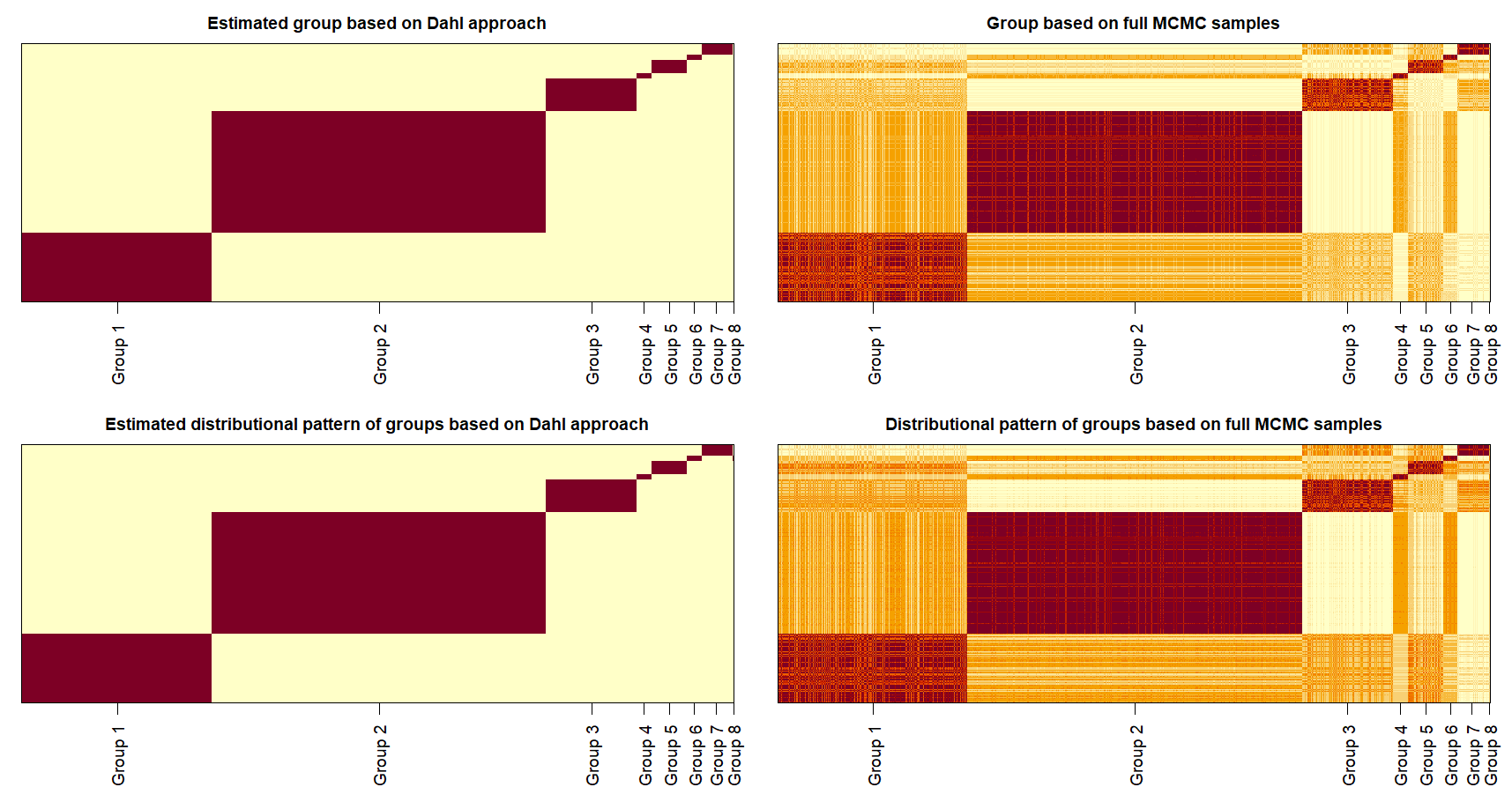}
     \caption{Pairwise co-clustering probabilities for observations belonging to the same group and DC for moderate sample size: Dahl approach (left) vs. entire posterior samples (right)}
    \label{fig:est.group.dist.hm}    
\end{figure}
Figure \ref{fig:est.group.dist.hm} illustrates the predictor group and distributional cluster structure of observations, that is, whether two observations $i$ and $i'$ share the same predictor group ($G(X_i) = G(X_{i'})$) and distributional cluster ($D_{G(X_i)} = D_{G(X_{i'})}$).
The heatmaps (top-left and bottom-left) based on the Dahl estimate reveal equivalent patterns in group and DC assignments in the cluster point estimators---with the exception of the single observation in predictor group 8 who is then classified with group 6 in the DCs.
Displaying the pairwise probabilities that $G(X_i) = G(X_{i'})$ and $D_{G(X_i)} = D_{G(X_{i'})}$ across MCMC samples demonstrate similar grouping and distributional patterns.
We can see that observations in groups 2, 4, and 6 are frequently combined into  a common DCs during MCMC, even as the final Dahl estimate does not include this structure. 
Note that observations in groups 2, 4, and 6 share similar health statuses (fewer risk factors) and are less likely to require hospital stays.

\begin{table}[t]
\caption{For moderate sample size, average number of non-empty OCs, observational clusters with at least 10 observations, distributional clusters, and size of largest observational cluster. 95\% credible intervals are displayed for each.
}
\centering
\begin{tabular}{lcccc}
\hline
Method     & non-empty OC & OC w $\geq 1\%$ Obs & DC   & Largest cluster size \\ \hline
CAPGM      & 5.52 (4, 8)  & 3.35 (3, 4) & 3.40 (2, 5) & 808 (749, 862) \\ 
CAM (Demo) & 5.47 (3, 9) & 3.24 (2, 4) & 2.37 (1, 4) & 804 (626, 851)\\
CAM (CCs)  & 4.71 (3, 7) & 3.08 (2, 4) & 3.72 (2, 5) & 851 (741, 874) \\
CAM (Ins)  & 6.13 (3, 10) & 3.48 (2, 5) & 2.24 (1, 3) & 815 (586, 876) \\
DP         & 5.50 (3, 9) & 3.30 (2, 5) & &818 (710, 853) \\ 
PSBP (Reg) & 8.34 (6, 12) & 4.01 (3, 5) &   & 702 (466, 872)\\
PSBP (LGP) & 6.84 (4, 12) & 4.29 (3, 7) & & 820 (735, 854) \\
\hline
\end{tabular}
\begin{tablenotes}
\footnotesize
\item Demo, CCs and Ins represents Demographics, number of chronic conditions and Insurance, respectively. 
\end{tablenotes}
\label{tab:data_1000:clustersize}
\end{table}

In Section 6.4, we compared the predictive performance across the various methods.  Here, we provide additional details about the response-level clustering across the respective MCMC samples.
The number of non-empty observational clusters, OCs with at least 1\% of the training data sample size, and the size of the largest cluster are shown in Table \ref{tab:data_1000:clustersize}.  This indicates that CAPGM forms fewer clusters than PSBP-LGP and PSBP-regression, but more than DP. CAM with the various a priori selected grouping structures generate a more disperse range of OCs with wider intervals for both the number of clusters and the size of the largest cluster.


\section{Results from HRS Data with Small Sample}
\label{S8_HRS_Data_Small_Sample}

\subsection*{Results: CAPGM}

In Section 6 of the manuscript, we used a training data set of size $n=1000$ to fit CAPGM and the other models.  In this section, we investigate the performance when the smaller training data ($n=500$) is used.
In this case CAPGM finds fewer observational and distributional clusters compared to the moderate sample size. Specifically, it yields an average of 4.68 non-empty observational clusters during MCMC sampling, 3.39 clusters with at least 1\% of the observations, and an average largest cluster size of 415. Additionally, it forms an average of 3.14 distributional clusters.

\begin{table}[t]
\caption{Inclusion probability of covariates in $\mathcal{T}$, listed in descending order, for small sample size training data.}
\footnotesize
\centering
\resizebox{\textwidth}{!}{%
\begin{tabular}{rrrrrrrrr}  \hline
 Covariate& HLTHLM & CCs & Diab & SRHS & InPov & Psych & Lung & HIBP \\ \hline
 Probability$\%$ & 89.4 & 45.4 & 21.8 & 10.7 & 10.1 & 8.0 & 7.0 & 5.2 \\ \hline
 Covariate & Stroke & Partner & Cancr & Arthr & Ever Smoke & Ethnicity & Insurnace & BMI\\
Probability$\%$ &4.6 & 4.5 & 3.9 & 2.1 & 1.8 & 1.7 & 1.7 & 1.5 \\ \hline
 Covariate & Sex & VigAct & Race & Age & Heart & Asset & Smoking & Education \\
 Probability$\%$  & 1.42 & 1.32 & 1.2 & 1.1 & 0.9 & 0.9 & 0.6 & 0.2 \\
   \hline
\end{tabular}
}
\label{tab:data_500:varincprob}
\end{table}

Table \ref{tab:data_500:varincprob} investigates the variable inclusion probabilities for the pyramid trees with this smaller training data.  
HLTHLM appears in almost every posterior sample $\mathcal{T}$ for this dataset, while the number of chronic conditions (CCs) and diabetes are included moderately often.
While SRHS was the most influential covariate in the moderate-sized sample, it plays a lesser role in this dataset. Instead, CCs and diabetes are more frequently included here, although they are also associated with SRHS.

\begin{figure}[t]
    \centering
    \includegraphics[width=1\linewidth]{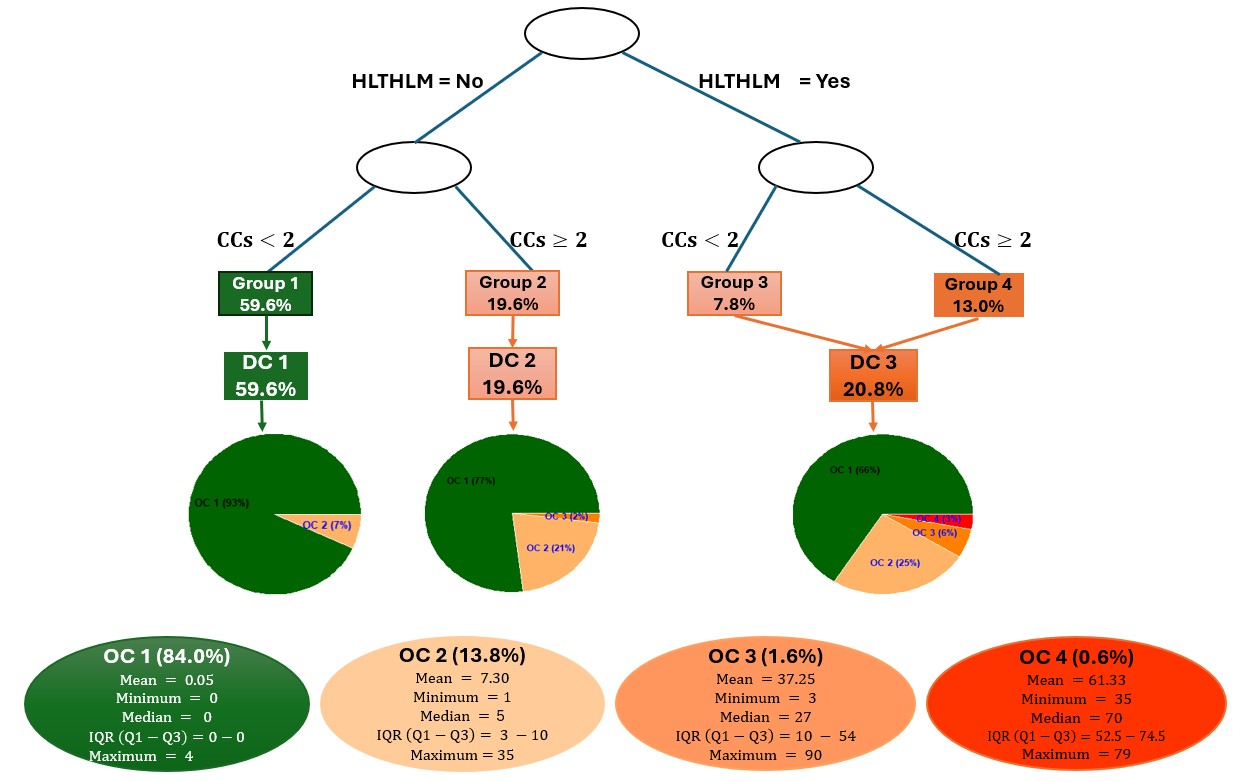}
    \caption{Estimated clustering structures for the small sample size training data.  The Dahl-estimated predictor groups with the corresponding pyramid tree are connected to three estimated distribution clusters.  For each distribution cluster, pie charts shows the probabilities that members are assigned to the three observational clusters.}
    \label{fig:capgm.500.est.clust.strctr}
\end{figure}
Similar to the moderate sample size,  we use the Dahl approach to obtain a posterior estimate of the OCs, DCs, and predictor groups, and we extract the tree $\mathcal{T}$ from the iteration corresponding to the estimated predictor group clustering.  These are represented in 
Figure \ref{fig:capgm.500.est.clust.strctr}.
The estimated pyramid tree uses the predictors HLTHLM and CCs, yielding four groups and 3 DCs.  
The resulting predictor group 1 is expected to be the healthiest, characterized by no HLTHLM and no more than one chronic condition, while group 4 appears to be at the highest risk.  
In the distribution clusters, predictor groups 3 and 4 (patients with mobilities restrictions regardless of number chronic conditions) are assigned into the same DC.  
For this sample size, we got four observational clusters.
For summary, the median (and range) of the observations assigned to each are 0 (0--4), 5 (1--35), 27 (3--90), and 70 (35--79).  The largest cluster contains 84\% of observations, the majority of whom have no ONHS. 
We see that DC 1 has 93\% probability of assigning members to OC1 and 0\% for OC 3 and OC 4. Most of the members in the highest rates of ONHS OCs 3 and 4 come from DC 3.

\begin{table}[t]
\caption{Estimated group characteristics. Group size ($n$) is shown under each group for small sample size.}
\centering
\begin{tabular}{lc|cccc}
\hline
& & Group 1 & Group 2 & Group 3 & Group 4 \\ 
Covariate & Category & $n=298$ &   $n=98$ &   $n=39$ &  $n=65$ \\ \hline

\multirow{2}{*}{SRHS (\%)} & SRHS Low  & 31 & 61 & 82 & 97 \\ 
                          & SRHS High & 69 & 39 & 18 & 3 \\
\hline
\multirow{2}{*}{HLTHLM (\%)} & No   & 100 & 100 & 0 & 0 \\ 
                             & Yes  & 0 & 0 & 100 & 100 \\
\hline
\multirow{2}{*}{Diabetes (\%)} & No  & 97 & 70 & 95 & 63 \\ 
                               & Yes & 3 & 30 & 5 & 37 \\ 
\hline
\multirow{2}{*}{CCs (\%)} & CCs $<2$   & 100 & 0 & 100 & 0 \\
                          & CCs $\geq 2$ & 0 & 100 & 0 & 100 \\ 
\hline
\multirow{2}{*}{Sex (\%)} & Female & 55 & 60 & 49 & 57 \\
                          & Male   & 45 & 40 & 51 & 43 \\
\hline
Age (mean) & &54.4 & 56.5 & 55.0 & 56.7 \\ \hline
\end{tabular}
\label{tab:group_est_500}
\end{table} 

The characteristics of the four estimated groups for this dataset are summarized in Table \ref{tab:group_est_500}. Observations in group 1 are generally more healthy, with no HLTHLM, fewer than two CCs, and predominantly high SRHS. In contrast, those in group 4 are the least healthy, with HLTHLM, more than two CCs, and lower SRHS. Observations in groups 2 and 3 are moderately healthy, with one risk factor. The average age is similar across all groups, and males and females are almost evenly distributed among them.

\begin{figure}[t]
    \centering
    \includegraphics[width=\linewidth]{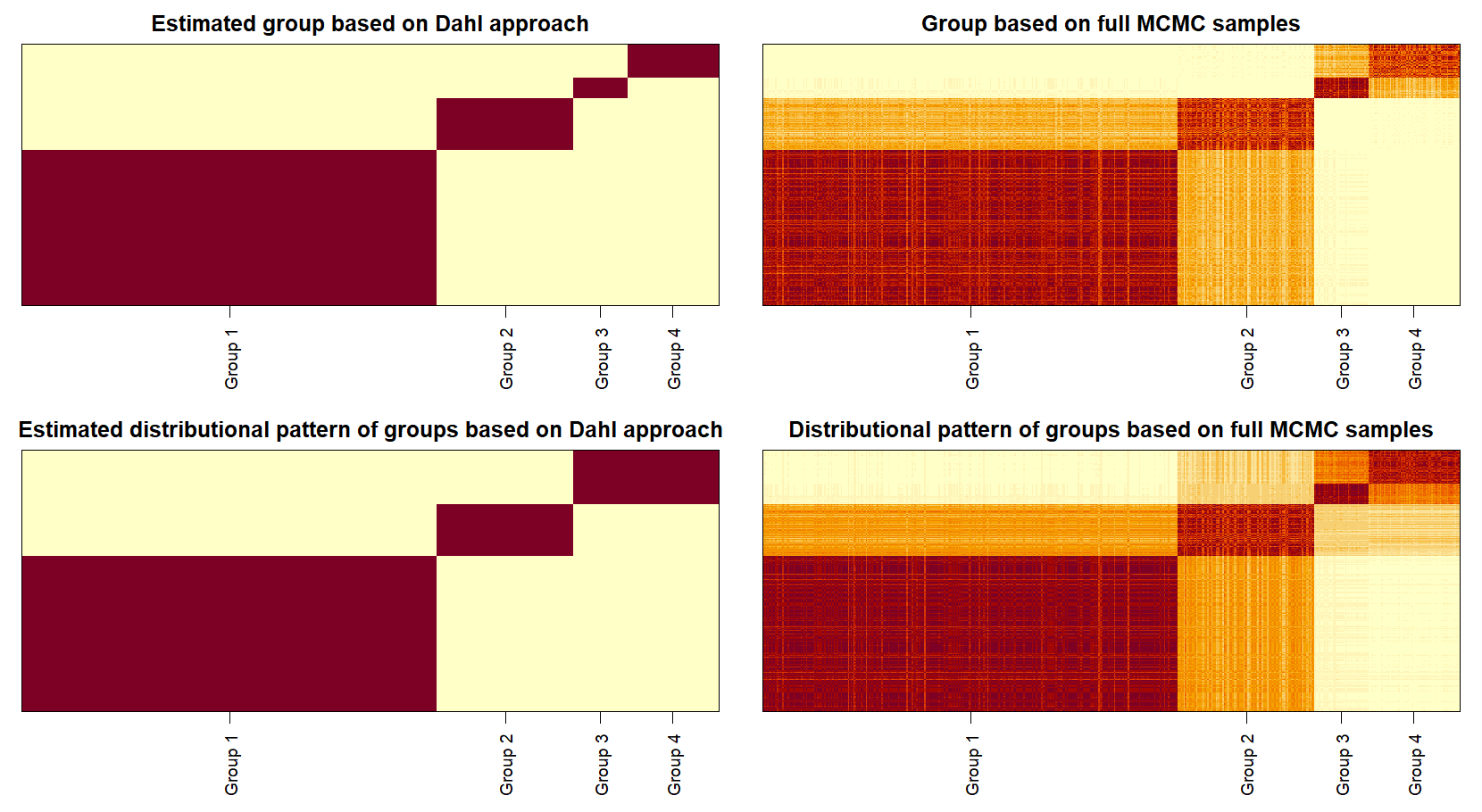}
    \caption{Pairwise co-clustering probabilities for observations in the same group and DC for small sample size: Dahl approach (left) vs. entire posterior samples (right)}
    \label{fig:data_500_est.group.dist.hm}   
\end{figure}

Figure \ref{fig:data_500_est.group.dist.hm} illustrates the grouping and distributional structure of observations based on co-grouping and co-distributional probabilities, indicating whether two observations $i$ and $i'$ share the same predictor group $G(\boldsymbol{X}_i) = G(\boldsymbol{X}_{i'})$ and distributional cluster $D_{G(\boldsymbol{X}_i)} = D_{G(\boldsymbol{X}_{i'})}$. The heatmaps (top-left and bottom-left) derived from the Dahl estimate show that observations with HLTHLM, regardless of the number of CCs (groups 3 and 4), tend to cluster together. Sorting observations by the estimated group and calculating pairwise probabilities that $G(\boldsymbol{X}_i) = G(\boldsymbol{X}_{i'})$ and $D_{G(\boldsymbol{X}_i)} = D_{G(\boldsymbol{X}_{i'})}$ across MCMC samples reveal grouping and distributional patterns consistent with the Dahl approach. Observations in groups 3 and 4 exhibit higher pairwise probabilities of clustering together, and in some MCMC samples, groups 2, 3, and 4 are observed to cluster together.

\subsection*{Results: Models Comparison}

\begin{table}[t]
\setlength{\tabcolsep}{2.5pt}

\caption{RMSPE and LPDS for training and test data from Health and Retirement Study with small sample size.} 
\centering
\begin{tabular}{lcccc}
\hline
&  \multicolumn{2}{c}{Within sample} &  \multicolumn{2}{c}{Out of Sample} \\
\cmidrule(lr){2-3} \cmidrule(lr){4-5}
Method    & RMSPE& LPDS & RMSPE & LPDS \\
\hline
CAPGM      & 8.25 & -525 & 10.35 & -10439\\
CAM (Demo) & 8.58 & -541 & 10.48 & -10576\\
CAM (CCs)  & 8.32 & -528 & 10.53 & -10607\\
CAM (Ins)  & 8.53 & -540 & 10.47 & -10552\\
DP         & 8.60 & -541 & 10.49 & -10565\\
PSBP (Reg) & 5.86 & -492 & 10.96 & -12015\\
PSBP (LGP) & 7.56 & -495 & 10.34 & -11140\\
\hline
\end{tabular}

\label{tab:data_500_rmspe.lpds}
\end{table}

The predictive accuracy and model fit for this training set, shown in Table \ref{tab:data_500_rmspe.lpds}, indicate similar performance trends as observed with the moderate sample of size $1000$. CAPGM outperforms all other methods on out-of-sample data in terms of both model fit and lower RMSPE. Within the sample, CAPGM demonstrates performance comparable to other methods, with the regression and LGP approaches providing the most accurate predictions. However, these approaches, particularly the regression method, perform poorly on out-of-sample data. Within sample the regression approach produces the largest (best) log-likelihood and the smallest RMSPE for estimating the mean; however, it performs worse on test data. This approach appears to overfit by creating a too many OCs leading to lower test performance. Consistent with findings from the sample of 1000, the CAM model based on demographic characteristics performs similarly to DP, while the CAM model based on chronic conditions does not perform as well, likely due to the smaller sample size.

\begin{table}[!t]
\caption{For small sample size, average number of non-empty OCs, observational clusters with at least 10 observations, distributional clusters, size of largest observational cluster, and their $95\%$ CI in the parenthesis.}
\centering
\begin{tabular}{lcccc}
\hline
Method     & non-empty OC & OC $\geq 1\%$ Obs & DC   & Largest cluster size \\ \hline
CAPGM     & 4.68 (3, 7) & 3.39 (2, 4) & 3.14 (2, 6) & 415 (327, 436)\\ 
CAM (Demo)& 4.2 (3, 7) & 3.1 (2, 4) & 1.3 (1, 3) & 420 (384 449) \\
CAM (CCs) & 3.89 (3, 6) & 3.17 (3, 4) & 2.83 (2, 4) & 426 (399, 447) \\
CAM (Ins) & 3.95 (3, 6) & 3.17 (2, 4) & 1.60 (1, 3) & 417 (374, 453) \\
DP        & 4.2 (3, 6) & 3.10 (2, 4)& & 413 (252, 440) \\
PSBP (Reg)&  9.3 (5, 13) & 4.40 (2, 7)& &  359 (191, 458) \\
PSBP (LGP)& 4.7 (3, 7) & 3.2 (2, 4)& & 413 (365, 436) \\
\hline
\end{tabular}
\label{tab:data_500:clustersize}
\end{table}

\begin{figure}[!t]
\centering

\begin{subfigure}{11cm}
    \centering
    \includegraphics[width=\linewidth]{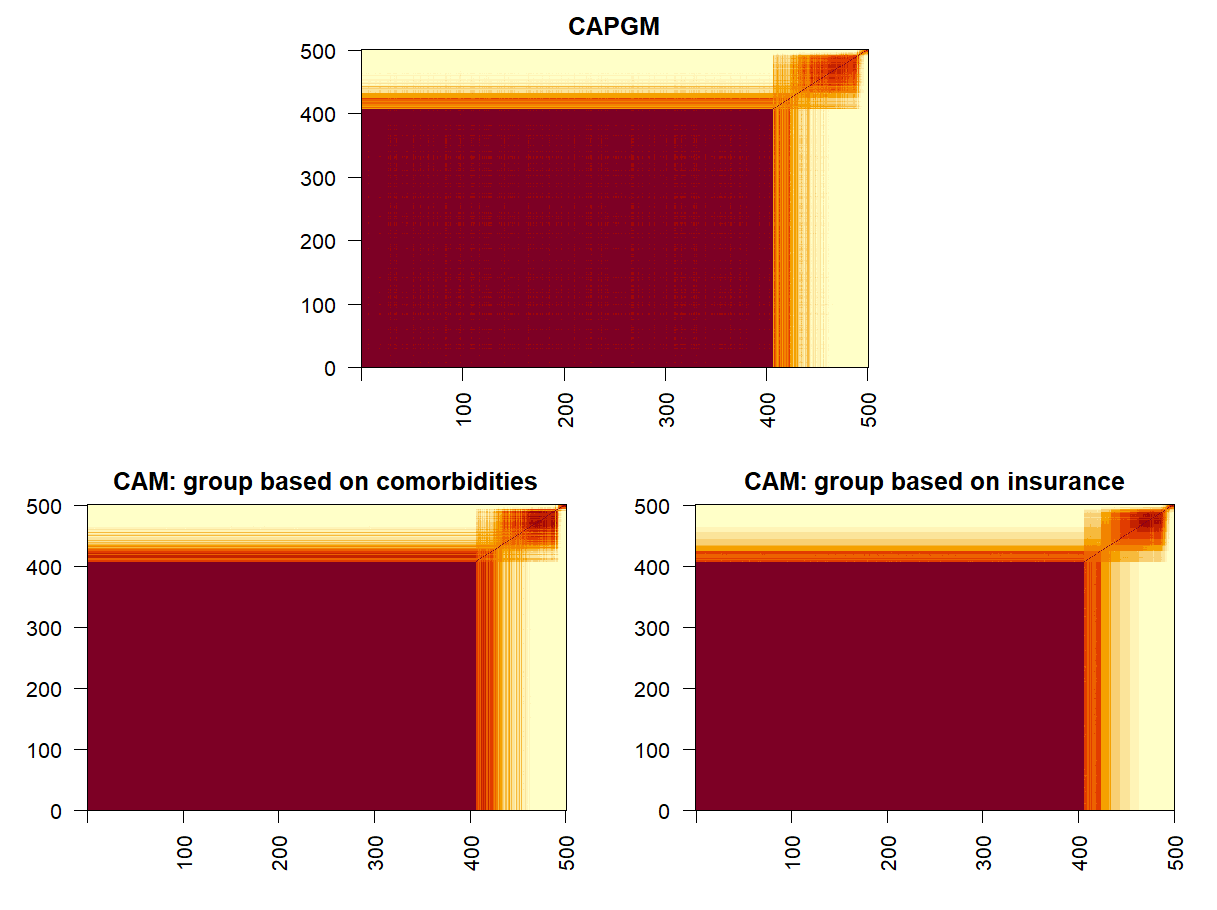}
\end{subfigure}

\begin{subfigure}{11cm}
    \centering
    \includegraphics[width=\linewidth]{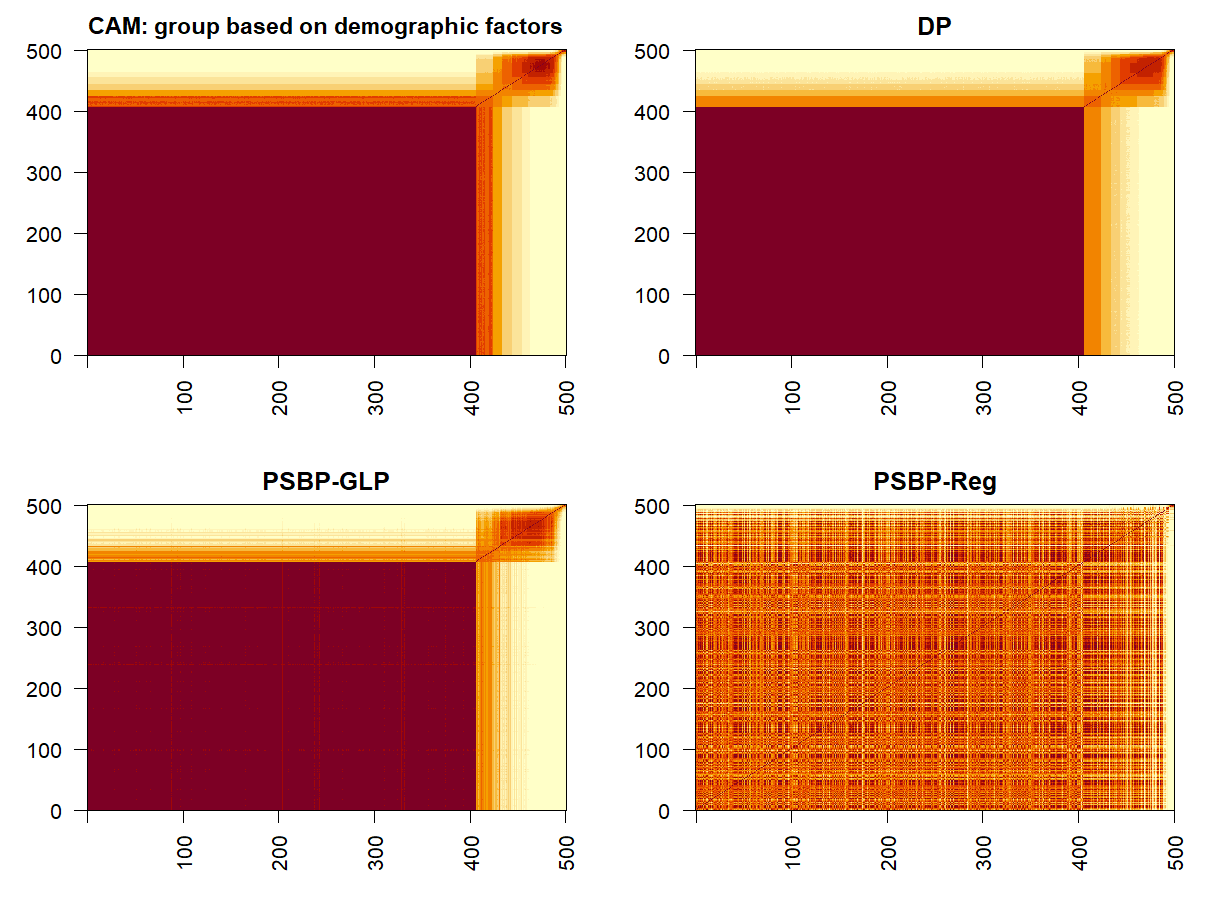}
\end{subfigure}

\caption{Comparison in the pairwise co-clustering probabilities of ONHS responses across models for small sample size. Observations are ordered in terms of increasing ONHS.}
\label{fig:data_500:heatmap}
\end{figure}

The observational clustering pattern illustrated in Table \ref{tab:data_500:clustersize} and Figure \ref{fig:data_500:heatmap}  resembles the results from  the moderate sample size data, although there is a common tendency towards fewer non-empty OCs across methods.
This is expected, as a smaller sample size tends to produce fewer clusters. CAPGM and LGP construct a similar number of OCs. DP and the demographic version of CAM generate a comparable number of OCs, which also reflects in the number of DCs. 
The demographic CAM version constructs an average number of DCs close to 1, meaning that it is frequently using the same observation-level cluster probabilities for all groups and ignoring the predictor information. 
The average number of DCs under CAPGM is slightly lower than in the moderate sample sizes.
The co-clustering probabilities for observational clusters are shown in Figure \ref{fig:data_500:heatmap}.  CAPGM and most of the methods clearly group the zeros together with very high probability, group most of the non-zero but small ONHS together, and have a small group in the top corner where individuals with very large counts are consistently co-clustered.  In contrast to the other approaches, the PSBP regression does not identify any structure consistent with values of ONHS (observations are ordered by increasing $y_i$),  possibly because this method is identify a clustering  primarily based on the predictors not the response.

\clearpage

\section{Results from HRS Data with Big Sample}
\label{S9_HRS_Data_Big_Sample}

\subsection*{Results: CAPGM}

\begin{table}[t]
\caption{Inclusion probability of covariates in $\mathcal{T}$, listed in descending order, for large sample size training data.}
\footnotesize
\centering
\resizebox{\textwidth}{!}{%
\begin{tabular}{rrrrrrrrr}  \hline
 Covariate& SRHS & HLTHLM & Lung & CCs & Diab & Cancer & Insurance & Stroke \\ \hline
 Probability$\%$ & 100.0 & 100.0 & 44.1 & 30.0 & 20.3 & 19.0 & 1.4 & 1.2 \\ \hline
 Covariate & VigAct & Psych & Asset & Heart & BMI & Age & Ever Smoke & Sex\\
Probability$\%$ &0.6 & 0.5 & 0.4 & 0.3 & 0.1 & 0.1 & 0.0 & 0.0 \\ \hline
 Covariate &  Race & Ethnicity & Education & Partner & HIBP & Arthr & Smoking & InPov \\
 Probability$\%$  & 0.0 & 0.0 & 0.0 & 0.0 & 0.0 & 0.0 & 0.0 & 0.0 \\
   \hline
\end{tabular}
}
\label{tab:data_2500:varincprob}
\end{table}
We now consider performance when the models are trained with a larger dataset with $n = 2500$ observations.
During MCMC sampling CAPGM constructs a higher number of non-empty OCs (5.76) and DCs (4.38) for the bigger sample size compared to smaller and moderate size sample.
The pyramid tree incorporates SRHS and HLTHLM with $100\%$ posterior probability as partitioning variables (see Table \ref{tab:data_2500:varincprob}), a pattern similar to moderate-sized sample where these two covariates were the most relevant. While the inclusion rates for lung disease status, diabetes, and cancer are moderate, other covariates have a lower likelihood of selection. 

\begin{figure}[t]
    \centering
    \includegraphics[width=1\linewidth]{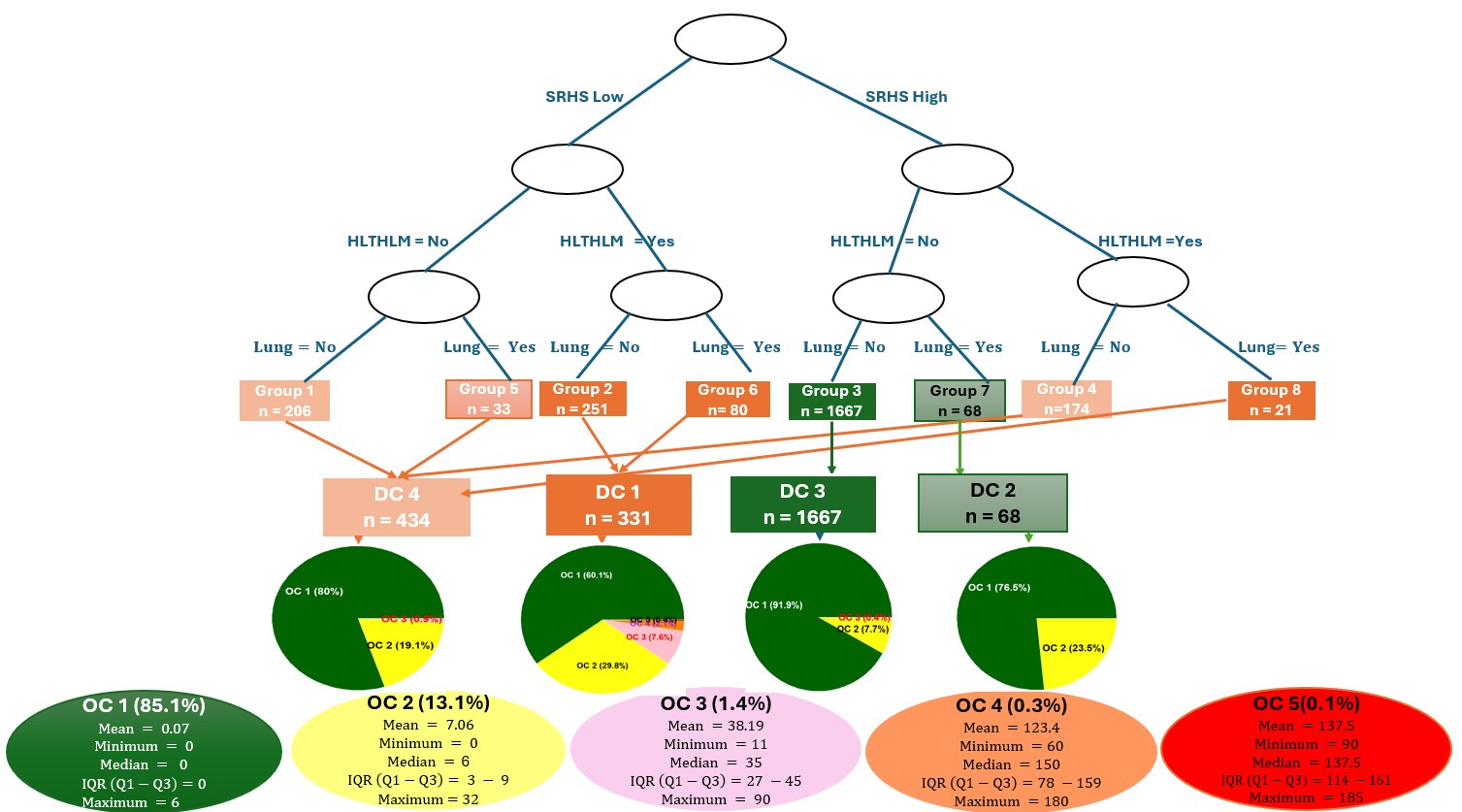}
    \caption{Estimated clustering structures for large sample training data.  The Dahl-estimated predictor groups with the corresponding pyramid tree are connected to four estimated distribution clusters.  For each distribution cluster, pie charts shows the probabilities that members are assigned to the three observational clusters.}
    \label{fig:capgm.2500.est.clust.strctr}
\end{figure}

As represented in Figure \ref{fig:capgm.2500.est.clust.strctr}, analysis with the larger training data yielded  five OCs, four DCs, and eight predictor groups based on the Dahl approach. As usual, we  extracted the tree $\mathcal{T}$ from the iteration corresponding to the estimated predictor group clustering.  
The estimated pyramid tree found eight predictors groups based on the top three most frequently selected predictors, SRHS, HLTHLM, and lung condition.
The resulting group 3 is expected to be the healthiest, characterized by higher SRHS, no HLTHLM, and no lung disease, while group 6 appears to be at the highest risk. 
The healthiest group 3 and group 7 with only the lung condition form correspond to their own DCs. Group 2 and 6 characterized with low SRHS and HLTHLM regardless of lung disease status assigned to the same DC 1. Moderate healthy groups 1, 5, 4, and 8 assigned to the DC 4.
The healthiest group (Group 3) and Group 7, which is defined by the presence of a lung condition, no HLTHLM, and high SRHS, each correspond to their own DCs. Groups 2 and 6, characterized by low SRHS and the presence of HLTHLM (regardless of lung disease status) are both assigned to DC 1. Moderately healthy groups, 1, 5, 4, and 8 are assigned to DC 4.

We obtain five observational clusters.
For summary, the median (and range) of observations assigned to each are 0 (0--6), 6 (0--32), 35 (11--90), 150 (60--180) and 137.5 (90--185), where this first cluster is contains 85\% of observations and most values are zero.
We noticed that all observations in OC 4 (0.3\%) and OC 5 (0.1\%), as well as the majority of observations in OC 3, come from the least healthy DC 1. 
Additionally,  the most  healthy DC 3 has the highest proportion of members associated with OC 1, while all DCs still assign over half to this cluster.
DCs 2 and 4 which represent separate groups at moderate risk include a large number of patients in OC2 with non-zero but generally small ONHS.

\begin{table}[t]
\small
\caption{Top 10 covariates combinations in $\mathcal{T}$ for large sample size.}
\centering
\begin{tabular}{lc}
\hline
Combination of $X$ & Percentage \\
\hline
SRHS, Lung and HLTHLM                       & 35.9 \\
SRHS, Diabetes and HLTHLM                   & 12.7 \\
SRSH, Comorbidities and  HLTHLM             & 12.1 \\
SRSH, Cancer, Comorbidities and  HLTHLM     & 9.8 \\
SRSH, Cancer and  HLTHLM                    & 7.2 \\
SRSH, and HLTHLM                            & 5.3 \\
SRSH, Diabetes, Comorbidities and  HLTHLM   & 3.7 \\
SRHS, Diabetes, Lung and  HLTHLM            & 2.6 \\
SRSH, Lung, Comorbidities and HLTHLM        & 1.4 \\
SRSH, , Comorbidities, HLTHLM and Insurance & 1.1 \\
\hline
\end{tabular}

\label{tab:data_2500:combX}
\end{table}
Table \ref{tab:data_2500:combX} provides additional insights into the pyramid tree structure, reporting the covariate combinations of the top 10 pyramid trees. All of the trees include the covariates SRHS and HLTHLM with the most common tree featuring SRHS, Lung, and HLTHLM, appearing $36\%$ of the time.

\begin{figure}[t]
    \centering
    \includegraphics[width=\linewidth]{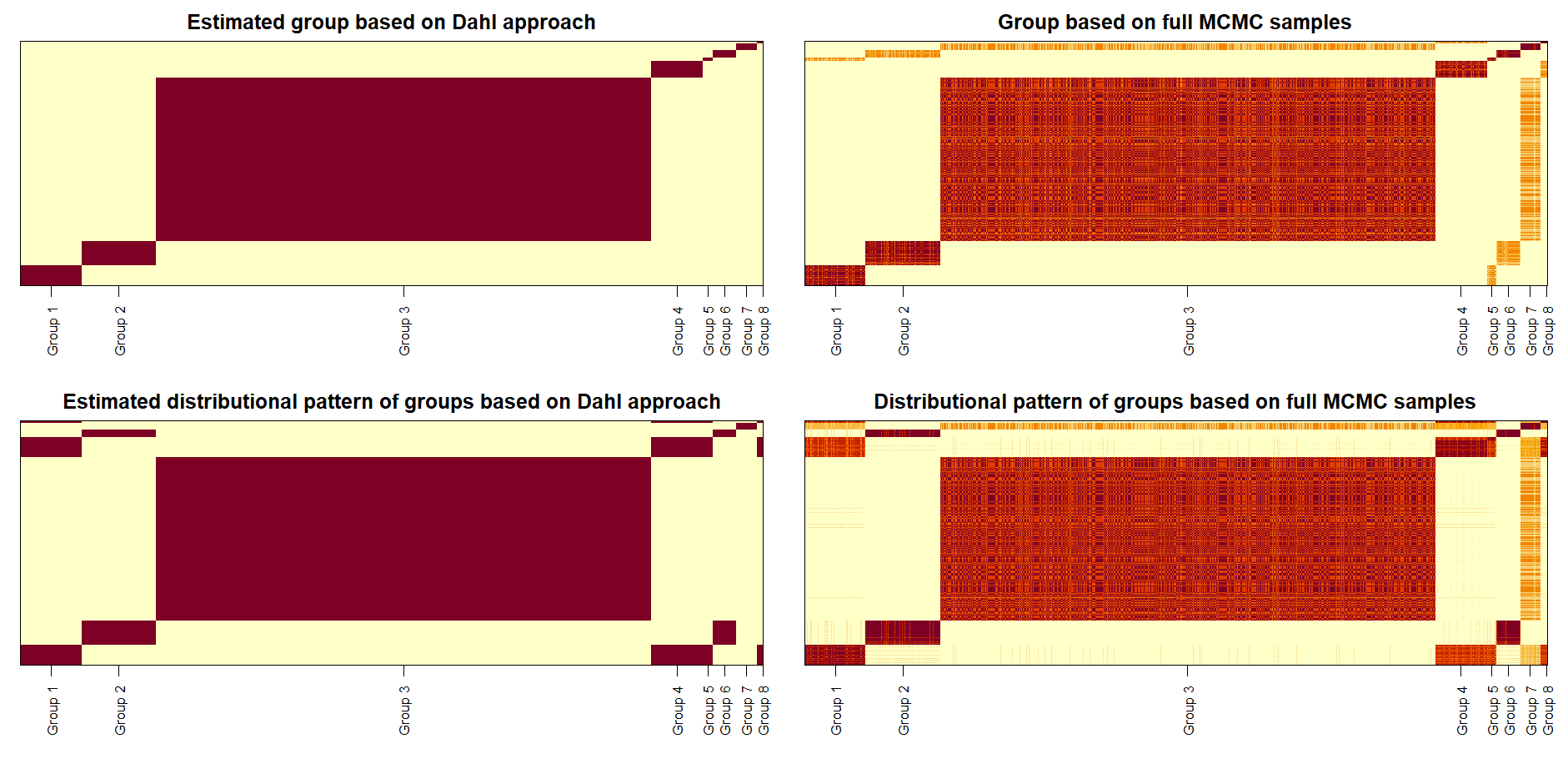}
    \caption{Pairwise co-clustering probabilities for observations in the same group and DC for large sample size: Dahl approach (left) vs. entire posterior samples (right) }
    \label{fig:data_2500_est.group.dist.hm}    
\end{figure}

Figure \ref{fig:data_2500_est.group.dist.hm} illustrates the grouping and distributional structure of observations based on co-grouping and co-distributional probabilities. The heatmaps (top-left and bottom-left) from the Dahl approach show that groups 1, 4, 5, and 8, representing moderate or less healthy individuals, cluster together. The healthiest group 3 remains distinct, while least healthiest groups 2 and 6, differing mainly by lung disease status, assign to the same DC and  individuals in these groups are more likely to hospital stays. When considering the pairwise probabilities across MCMC samples, we see much less uncertainty in the pyramid and distribution cluster structure as compared to the results from the small and moderate training data.  The co-clustering probabilities are highly  consistent with the Dahl estimate, except for groups 3 and 7, which are frequently sampled into the same DC during MCMC.

\subsection*{Results: Models Comparison}

\begin{table}[t]
\setlength{\tabcolsep}{2.5pt}
\caption{RMSPE and LPDS for training and test data from Health and Retirement Study with large sample size.} 
\centering
\begin{tabular}{lcccc}
\hline
&  \multicolumn{2}{c}{Within sample} &  \multicolumn{2}{c}{Out of Sample} \\
\cmidrule(lr){2-3} \cmidrule(lr){4-5}
Method    & RMSPE& LPDS & RMSPE & LPDS \\
\hline
CAPGM       & 9.60 & -2542 & 10.38 & -8174\\
CAM (Demo)  & 9.90 & -2632 & 10.53 & -8369\\
CAM (Com)   & 9.71 & -2573 & 10.45 & -8203\\
CAM (Ins)   & 9.82 & -2597 & 10.47 & -8312\\
DP          & 9.92 & -2640 & 10.54 & -8368\\
PSBP (Reg)  & 9.13 & -2487 & 10.54 & -8368\\
PSBP (LGP)  & 8.90 & -2391 & 10.32 & -8612\\
\hline
\end{tabular}
\label{tab:data_2500:RMSPE}
\end{table}

For this training set, CAPGM yields a lower RMSPE for mean estimation, based on the weighted average over clusters, than DP and various CAM versions, for both training and test data (see Table \ref{tab:data_2500:RMSPE}). In the test data, CAPGM’s RMSPE is slightly worse than the LGP predictions  but provides superior model fit, as evidenced by the highest LPDS. Although LGP achieves the lowest RMSPE across all methods for both training and test data, its model fit is poorest for the test data, as indicated by the lowest LPDS. 

\begin{figure}[t]
    \centering
    \includegraphics[width=1\linewidth]{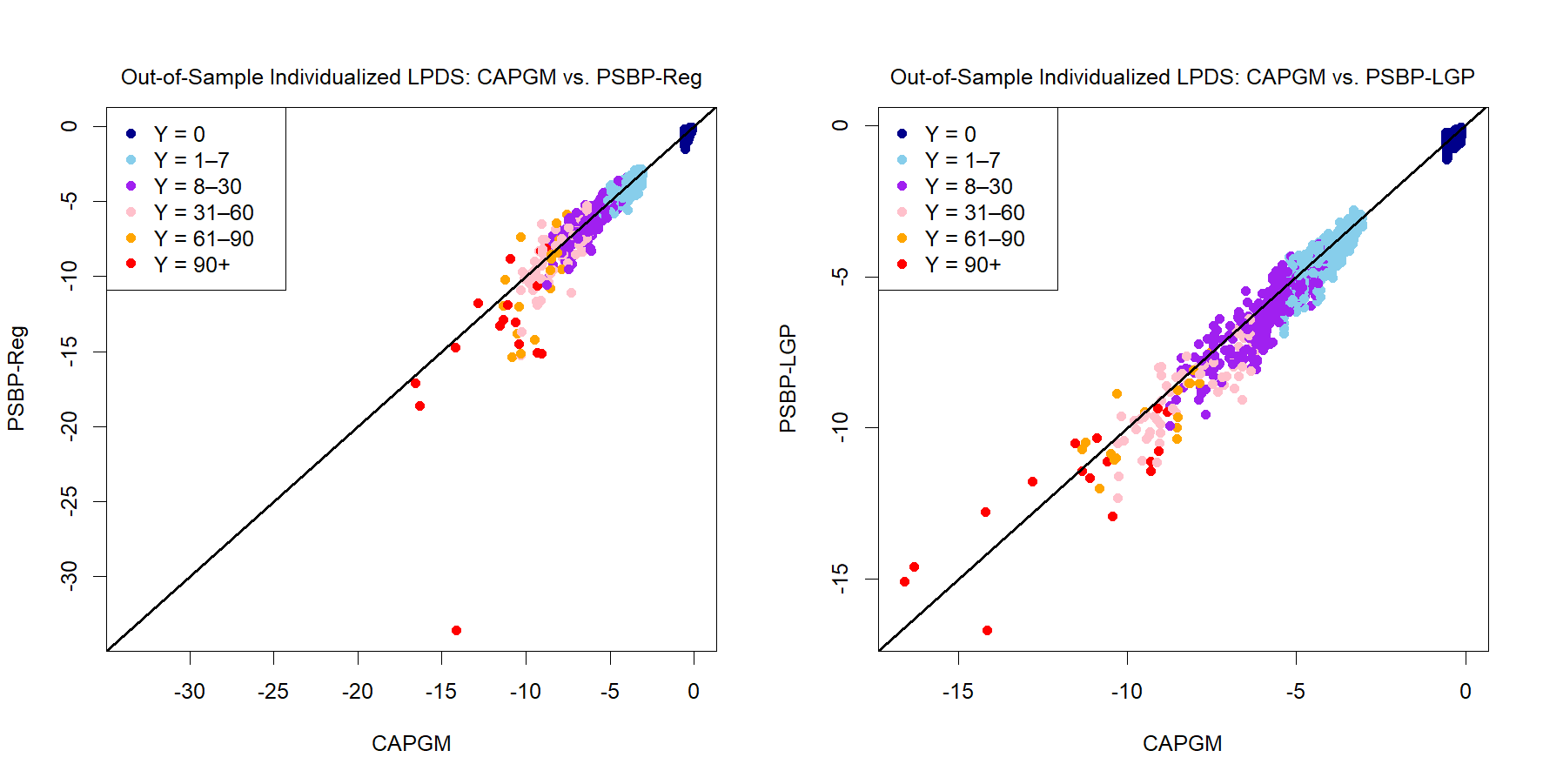}
    \caption{Comparison of model fit between CAPGM, PSBP-Reg and PSBP-LGP for large sample training data. Each point represents the estimate for $E_{post}\left\{ \log {f}(y_i\mid \boldsymbol{X}_i)\right\}$ for an observation in the out-of-sample test data, color-coded by the magnitude of response.}
    \label{fig:capgm.reg.lgp.2500.lpds}
\end{figure}

In training data, LGP produces the highest (best) LPDS but the lowest (worst) for test data. This is due to LGP forming additional smaller clusters with few observations for the training data (see Table \ref{tab:data_2500:clustersize}) that are not replicated in the out-of-sample test data.  PSBP-LGP samples an average number of 8.9 non-empty clusters during MCMC (Table \ref{tab:data_2500:clustersize}), but on average only 3.1 of these clusters have at least $1\%$ (25) observations. CAPGM places less weight on small clusters and is more robust in the tail of ONHS. Similar to the results presented in Figure 4 of the manuscript, LGP systematically yields smaller estimates of $E_{post}\left\{ \log f(y_i\mid\boldsymbol{X}_i)\right\}$ relative to CAPGM, as shown in Figure \ref{fig:capgm.reg.lgp.2500.lpds}.

\begin{table}[!t]
\caption{For large sample size, average number of non-empty OCs, observational clusters with at least 10 observations, distributional clusters, size of largest observational cluster, and their $95\%$ CI in the parenthesis.}
\centering
\begin{tabular}{lcccc}
\hline
Method    & non-empty OC & OC w $\geq 1\%$ Obs & DC   & Largest cluster size \\ \hline
CAPGM     & 5.76 (4, 9) & 3.06 (2, 4) & 4.38 (2, 7) & 2091 (1970, 2134) \\ 
CAM (Demo)& 7.9 (4, 15) & 4.0 (2, 7) & 2.4 (1, 4) & 1835 (984, 2081) \\
CAM (Com) & 7.02 (4, 11)& 3.52 (2, 5) & 4.50 (3, 6) & 2024 (1081, 2137) \\
CAM (Ins) & 6.40 (4, 10)& 3.23 (2, 4) & 3.18 (2, 4) & 2065 (1917, 2119) \\
DP        & 5.44 (4, 9) & 3.03 (2, 4) & &2042 (1881, 2104) \\ 
PSBP (Reg)& 10.5 (4, 16) & 5.1 (3, 6) & & 2068 (1916, 2124)\\
PSBP (LGP)& 8.91 (6, 12) & 3.13 (2, 4) & & 2086 (2012, 2138) \\
\hline
\end{tabular}
\label{tab:data_2500:clustersize}
\end{table}

\begin{figure}[!t]
\centering
\begin{subfigure}{11cm}
  \includegraphics[width=\linewidth]{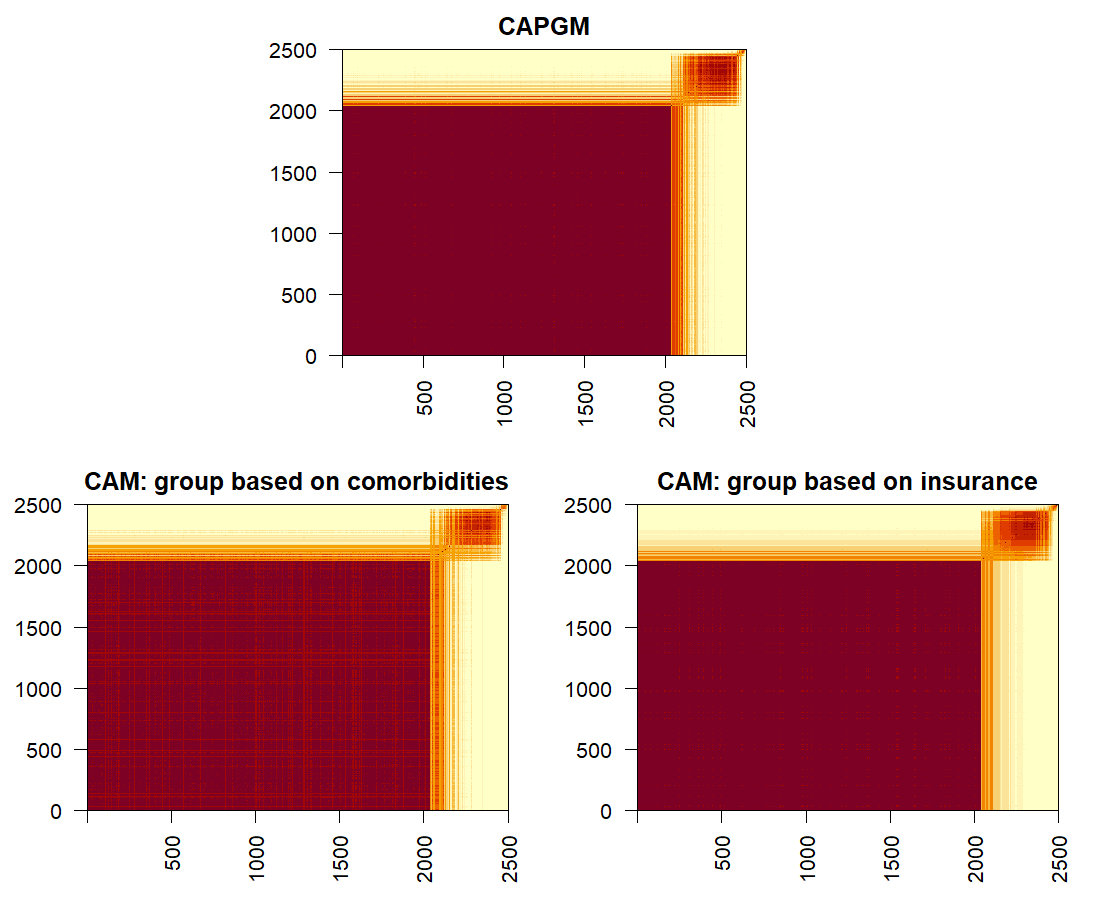}
\end{subfigure} 
\begin{subfigure}{11cm}
  \includegraphics[width=\linewidth]{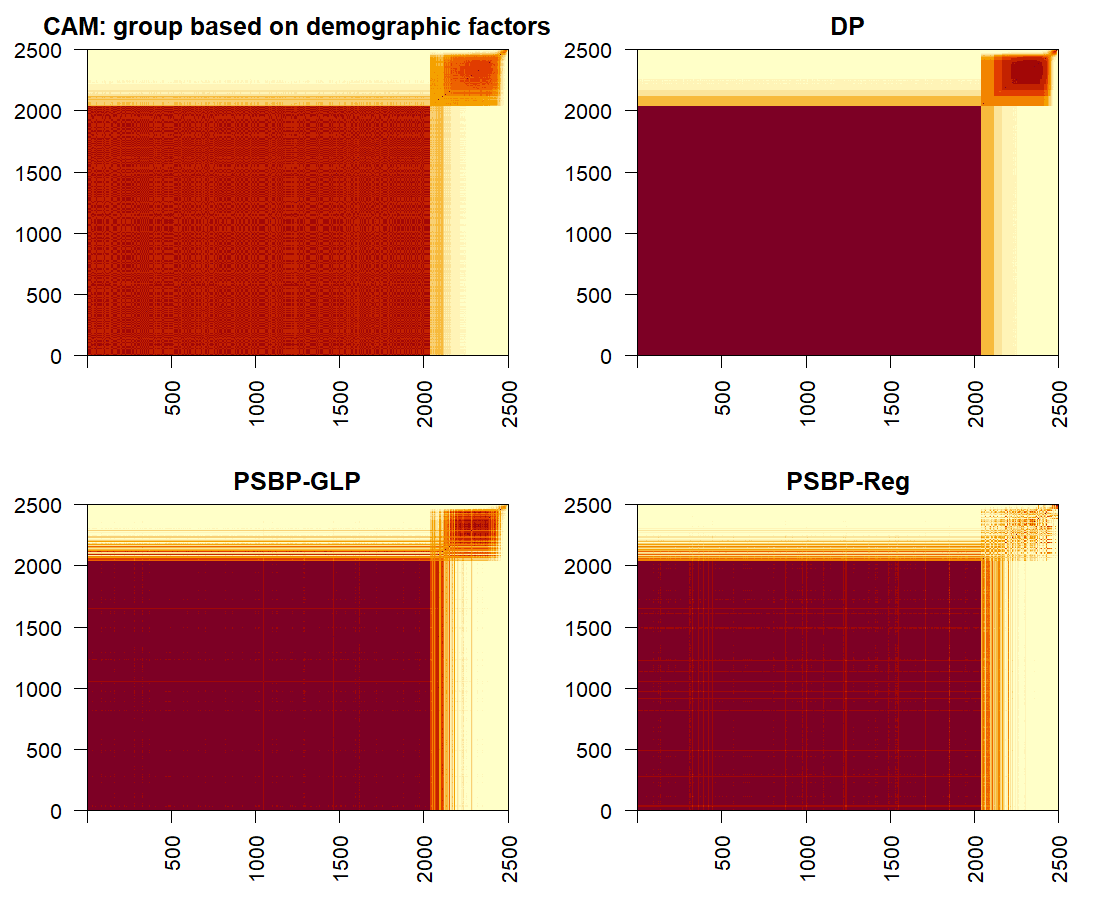}
\end{subfigure}
\caption{Comparison in the  pairwise co-clustering probabilities of ONHS responses across models for bigger sample size. Observations are ordered in terms of increasing ONHS.}
\label{fig:data_2500:heatmap}
\end{figure}

The heat map in Figure \ref{fig:data_2500:heatmap} reveals three distinct clusters identified by all methods. 
The first two correspond to our (Dahl-estimated) OCs 1 and 2, while there is less clear distinction in the larger and smaller OCs 3--5.
The largest cluster, located in the lower left corner and covered approximately $80\%$ of observations, represents patients who did not stay overnight in the hospital or had the shortest overnight hospital stays. This is consistent with the data, as $80.6\%$ of patients did not have any overnight hospital stays between the first and second waves. Patients with prolonged hospitalizations are clustered together in the upper right corner, while those with moderate hospital stays are situated between the largest and smallest clusters. Consistent with the other training data sample sizes, the co-clustering probability for the LGP and regression settings of PSBP is lower, as these methods tend to produce more clusters.  However, unlike the other two settings, PSBP-Reg is able to identify a single cluster containing the zero responses, whereas this structure was not discovered in the previous analyses.

\end{document}